\documentclass[twocolumn]{aastex631}

\usepackage{comment}
\usepackage{amssymb}
\usepackage{amsmath}
\usepackage{xspace}
\usepackage{enumitem}
\usepackage{graphicx}
\usepackage{rotating}
\usepackage{color}

\newcommand{\lya}        {Ly$\alpha$\xspace}
\newcommand{\hi}         {\ion{H}{1}\xspace}

\newcommand{\llya}       {$L({\rm Ly\alpha})$\xspace}
\newcommand{\NHI}        {\relax\ifmmode{N_{\rm HI}\xspace} \else {$N_{\rm HI}$}\expandafter\xspace\fi}
\newcommand{\NHIcl}      {\ifmmode{N_{\rm HI,cl}\xspace}\else{$N_{\rm HI,cl}\,$\xspace}\fi}
\newcommand{\unitcgssb}  {erg\,s$^{-1}$\,cm$^{-2}$\,arcsec$^{-2}$\xspace}

\newcommand{\unitcgslum} {erg\,s$^{-1}$\xspace}
\newcommand{\msun}       {M$_{\odot}$\xspace}
\newcommand{\kms}        {\ifmmode{\rm \,km\,s^{-1}}\else\,km\,s$^{-1}$\xspace\fi}
\newcommand{\unitNHI}    {\ifmmode{\rm \,cm^{-2}}\else\,cm$^{-2}$\xspace\fi}  
\newcommand{\vexp}       {\relax\ifmmode {v_{\rm exp}} \else {$v_{\rm exp}$}\expandafter\xspace\fi}
\newcommand{\sigsrc}     {\relax\ifmmode {\sigma_{\rm src}} \else {$\sigma_{\rm src}$}\expandafter\xspace\fi}
\newcommand{\dvpeak}     {$\Delta V_{\rm peak}$\xspace}
\newcommand{\SB}         {$\Sigma(R_p)$\xspace}
\newcommand{\pol}        {${\rm DoP}(R_p)$\xspace}
\newcommand{\spec}       {$F(\Delta V)$\xspace}
\definecolor{forestgreen}{rgb}{0.13, 0.55, 0.13}

\definecolor{myblue}{rgb}{0.13, 0.13, 0.55}
\newcommand\sj[1]{{\sf\color{myblue}{#1}}}

\shorttitle{\lya Radiative Transfer in Smooth and Clumpy Mediums}
\shortauthors{Chang et al.}
\begin{document}

\title{Radiative Transfer in Ly$\alpha$ Nebulae: \\
I. Modeling a Continuous or Clumpy Spherical Halo with a Central Source}

\author[0000-0002-0112-5900]{Seok-Jun Chang}
\affiliation{Max-Planck-Institut f\"{u}r Astrophysik, Karl-Schwarzschild-Stra$\beta$e 1, 85748 Garching b. M\"{u}nchen, Germany}
\affiliation{Korea Astronomy and Space Science Institute, 776 Daedeokdae-ro, Yuseong-gu, Daejeon 34055, Korea}

\author[0000-0003-3078-2763]{Yujin Yang}
\affiliation{Korea Astronomy and Space Science Institute, 776 Daedeokdae-ro, Yuseong-gu, Daejeon 34055, Korea}
\affiliation{University of Science and Technology, 217, Gajeong-ro, Yuseong-gu, Daejeon 34113, Korea}

\author[0000-0001-9561-8134]{Kwang-Il Seon}
\affiliation{Korea Astronomy and Space Science Institute, 776 Daedeokdae-ro, Yuseong-gu, Daejeon 34055, Korea}
\affiliation{University of Science and Technology, 217, Gajeong-ro, Yuseong-gu, Daejeon 34113, Korea}

\author[0000-0001-6047-8469]{Ann Zabludoff}
\affiliation{Steward Observatory, University of Arizona, 933 North Cherry Avenue, Tucson AZ 85721}

\author[0000-0002-1951-7953]{Hee-Won Lee}
\affiliation{Sejong University, 209 Neungdong-ro, Gwangjin-gu, Seoul 05006, Korea}




\begin{abstract}

%% The origin of \lya emitting nebulae, which extend over $\sim 100$\,kpc and may mark regions of early galaxy cluster formation, is a mystery. Possible mechanisms include photoionization by AGN, cooling radiation from cold accretion, shocks by galactic outflows, and scattering of \lya by the nebular gas. 
%%
To understand the mechanism behind high-$z$ \lya nebulae, we simulate the scattering of \lya in an \hi halo about a central \lya source. For the first time, we consider both smooth and clumpy distributions of halo gas, as well as a range of outflow speeds, total \hi column densities, \hi spatial concentrations, and central source galaxies (e.g, with \lya line widths corresponding to those typical of AGN or star-forming galaxies).
We compute the spatial-frequency diffusion and the polarization of the \lya photons scattered by atomic hydrogen. Our scattering-only model reproduces the typical size of \lya nebulae ($\sim 100\,$kpc) at total column densities $\NHI \geq 10^{20} \unitNHI$ and predicts a range of positive, flat, and negative polarization radial gradients.
We also find two general classes of \lya nebula morphologies: with and without bright cores. Cores are seen when \NHI is low, i.e., when the central source is directly visible, and are associated with a polarization jump, a steep increase in the polarization radial profile just outside the halo center. 
Of all the parameters tested in our smooth or clumpy medium model, \NHI dominates the trends. The radial behaviors of the \lya surface brightness, spectral line shape, and polarization in the clumpy model with covering factor $f_c \gtrsim 5$ approach those of the smooth model at the same \NHI.
A clumpy medium with high \NHI and low $f_c \lesssim 2$ generates \lya features via scattering that the smooth model cannot: a bright core, symmetric line profile, and polarization jump.

\end{abstract}



\section{Introduction} \label{sec:intro}

%% Ly$\alpha$ Emission
Hydrogen \lya is the most prominent emission line and thus a powerful tool for studying the early universe at $z > 2$. Narrowband imaging surveys have revealed various strong \lya-emitting sources: compact \lya emitters \citep[LAE;][]{gawiser07,ouchi08,sobral17,ouchi18}, \lya blobs \citep[LABs;][]{steidel00,matsuda04}, and enormous \lya nebulae \citep[ELANe;][]{hennawi15,cai17,fabrizio19}.
\lya blobs are typically extended over 50--100 kpc and have \lya luminosities of $\sim10^{44}$ \unitcgslum \citep{keel99,steidel00,matsuda04,dey05,yang09,yang10,travascio20}. They are believed to trace massive halos that will evolve into rich galaxy groups or even clusters today; as such, their embedded galaxies may be the progenitors of massive cluster galaxies
and their gas the precursor to the intracluster medium \cite[e.g., ][]{Badescu2017}. 

%% Host of LABs
Spatially extended \lya emission has long been associated with embedded sources---high-$z$ radio galaxies (HzRGs) are often surrounded by giant \lya halos \citep{heckman91,villar07,shukla22}. One of the archetypal \lya blobs, SSA22-LAB1, envelopes multiple small galaxies \citep{matsuda04, geach16, Umehata21}. Some \lya nebulae appear to be associated with  obscured active galactic nuclei (AGN) or starburst galaxies \citep{dey05, yang14a, cai17}. Recently, \cite{borisova16} and \cite{fabrizio19} found that nebulae with \lya emission extended over $\sim 50$ kpc scales are ubiquitous around bright radio-quiet QSOs.  

%% Energy source
The origin of this \lya emission is  controversial. Proposed power sources include photo-ionization by AGN \citep{steidel00,fabrizio19}, cooling radiation from cold-mode accretion \citep{trebitsch16,ao20,daddi20}, shocks due to fast outflows \citep{cabot16,travascio20},  and the scattering of \lya photons by the surrounding medium \citep{hayes11,li21}. Recently, \cite{li21} claimed that, based on  Ly$\alpha$/H$\beta$ line ratios, the extended \lya emission in SSA22-LAB1 originates from recombination in photo-ionized \ion{H}{2} regions and subsequent scattering by neutral gas. 

%% Polarization!!!
Polarimetric observations have emerged as a new tool to discriminate among these scenarios. \cite{hayes11} first observed a concentric polarization pattern around SSA22-LAB1, suggesting that \lya scattering from a central source was the most viable mechanism. Later, \cite{you17} and \cite{kim20} extended such polarization mapping, showing that polarized \lya emission is common among \lya blobs with various kinds of embedded sources and that the polarization morphologies are diverse. \cite{you17} found that the polarization vectors are aligned perpendicular to the major axis (also the direction of the jet) in B3~J2330+3927, a \lya blob around a radio galaxy at $z=3.087$. \cite{kim20} found an asymmetric polarization pattern where significant polarization was detected only toward the southeast of the \lya nebula LABd05.  This polarized \lya emission provides strong evidence that scattering by the neutral medium plays an important role in producing extended \lya emission. However, the interpretation of the polarization pattern and strength of the scattered \lya light is still challenging due to the small number of \lya polarization observations and, more critically, the lack of proper predictions for \lya polarization under various physical conditions.  

%% Ly$\alpha$ Radiative Transfer for LABs
Radiative transfer (RT) models are the essential tool to investigate the origin of the extended \lya emission. 
Previous RT work has concentrated mostly on the formation of the \lya line profile or the polarization under simple geometries: e.g., the classic double-peak solution in a static medium \citep{neufeld90}, 
the formation of \lya spectra in static or outflowing mediums without including polarization \citep{ahn02a, ahn03, zheng02, verhamme06}, and
the surface brightness and polarization profiles of the scattered \lya in a shell geometry \citep{ahn02,dijkstra08}. 
Using a spherical or ellipsoidal gas distribution on sub-kpc scales, \cite{eide18} investigate the polarization of scattered \lya in the context of compact \lya-emitting galaxies.
\citet{seon22} discuss the surface brightness and polarization profiles arising from smoothly varying mediums. To improve the RT models for LABs, simulations have to consider the large scale \hi distribution over several tens of kpc, broad \lya emission, and, most importantly, the clumpiness of the scattering medium. In addition, the RT calculation must carry comprehensive information about the scattered \lya photons, including spatial diffusion, spectral, and polarization properties.

%% While often used, the shell model is not adequate for \lya nebulae due to its physically thin \hi distribution. 
%% \citep[e.g][]{neufeld90, ahn02, zheng02, verhamme06}.
%% \cite{neufeld90} drove the analytic solution for double peaks structure in the static medium.
%% \cite{zheng02} and \cite{verhamme06} studies the formation of \lya spectra in static and outflow medium without polarization
%% \cite{ahn02} consider the polarization by the scattering to generate \lya spectra without spatial distribution.
%% \cite{dijkstra08} consider the surface brightness and spatial distributed polarization of the scattered \lya in shell model.

%% Clumpy CGM: observation
%% Observationally, it is found that the clumpiness of the gas in the CGM has to be taken into account --- but has been usually overlooked --- when investigating the gas properties.
There is evidence that LABs have clumpy gas distributions.
Several LABs have a \lya line peak in velocity space that coincides with the systemic velocity of the embedded galaxies \citep{prescott09,yang11,yang14a}. Given that \lya photons experiencing scattering in a {\it continuous} medium are always scattered off from the systemic velocity of the galaxies and gas (e.g., the double-peaked profile in the static medium or a redshifted profile in an outflowing medium), these \lya spectra are hard to reconcile with previous RT calculations. Through detailed photo-ionization modeling, \cite{hennawi13} and \cite{fabrizio15} suggest that the medium should be composed of numerous unresolved clumps with a \hi column density of $\lesssim 10^{20} \unitNHI$ and radius of $\lesssim 20$\,pc.

\lya RT calculations with a full treatment of the clumpy medium have not been fully explored, especially in the context of extended \lya emission. Previous studies concentrate on the \lya escape fraction from the clumpy medium of star-forming galaxies \citep{neufeld91,hansen06,duval14}. The main result is that the escape fraction increases in a clumpy medium thanks to surface scattering. When an optically thick photon encounters a clump, the photon is reflected through several scatterings at the surface. 
\cite{gronke16,gronke17} demonstrate that \lya spectra in clumpy mediums tend to be similar to those in continuous mediums when the covering factor of clumps is very high. \cite{trebitsch16} 
use a radiative hydrodynamic simulation to investigate the \lya polarization arising from cold gas accretion. They assume 100\%-polarized photon packets and adopt the method in \cite{rybicki99}, which is applicable only for Rayleigh scattering, to compute the polarization. 
This method is not suitable to describe the polarization behavior of photons resonantly scattered near the line center, because the polarization of a photon packet can decrease or even be negatively polarized after resonance scattering \citep[e.g.,][]{seon22}.

In this paper, we develop more realistic \lya radiative transfer simulations for LABs and consider the resulting behaviors of the observed \lya surface brightness, velocity, and polarization profiles.
To explore the physical parameter space, we present an extensive library of RT calculations for models in both smooth and clumpy mediums. 
Our simulations adopt a Monte-Carlo technique using ray-tracing in a grid-based geometry. To compute the polarization of scattered \lya accurately, we utilize a new method, including the effect of resonance scattering, developed in \cite{seon22}. A photon packet in our simulation carries multi-dimensional information, including wavelength, direction, position, and polarization state. We consider a geometry where a spherical scattering medium surrounds a point source. Our goal is to carry out a systematic study for LABs to examine if \lya scattering alone can explain the observed extended \lya emission.

This paper is organized as follows. 
In Section \ref{sec:rt}, we describe the algorithm used to generate our simulations.
In Section \ref{sec:model}, we explain the scattering geometry composed of a continuous or clumpy medium with a central point source.
In Section \ref{sec:result_a}, we present surface brightness profiles, polarization, and the integrated \lya spectra in the smooth medium. 
In Section \ref{sec:result_c}, we present those results for the clumpy medium and compare them with those for the continuous medium.
In Section \ref{sec:summary}, we summarize our conclusions and describe future work. 
This is the first in a series of papers focused on the scattering effect of \lya in continuous or clumpy spherical halos.

\section{{\lya} Radiative Transfer}
\label{sec:rt}

Our simulations are based on the 3D Monte Carlo code {\it LaRT}, standing for \lya Radiative Transfer, developed by \cite{seon20}. They used the {\it LaRT} code to investigate the Wouthuysen-Field effect by carefully dealing with the hyperfine structure of atomic hydrogen. In this work, we modify {\it LaRT} to deal with an emission line source with a broad line width that is embedded in the medium. In addition to a smooth medium, we also consider a clumpy halo with numerous \hi clumps.
In this section, we briefly describe the atomic physics related to the scattering of \lya adopted in {\it LaRT}. 

\subsection{Scattering Cross Section}

The scattering cross section of \lya is characterized by the oscillator strength $f_{12}=0.4162$.
In this work, no consideration is made for hyperfine structures, and \lya is a resonance doublet line associated with the fine structures $2P_{1/2}$ and $2P_{3/2}$ of the $2p$ level. We denote the transitions $(1S_{1/2} - 2P_{1/2})$ and $(1S_{1/2} - 2P_{3/2})$ by ``H''  and ``K,'' respectively.
The cross section of \lya is described by a sum of two Lorentzian functions near the H and K line centers in the rest frame of an atom.  Convolution with the local thermal motions of hydrogen atoms well described by a Gaussian distribution leads to the Voigt profile function representing the cross section in the local reference frame of medium. 

Explicitly, the scattering cross section of \lya as a function of the frequency is given by
\begin{equation}
    \sigma_\nu = {{\sqrt{\pi}e^2f_{12}} \over {\Delta \nu_D m_e c}} \left[ {1 \over 3} H(x_H,a) + {2 \over 3} H(x_K,a) \right].
\end{equation}
Here, $H(x,a)$ is the Voigt-Hjerting function given by
\begin{equation}
H(x,a) = {a \over \pi} \int^{\infty}_{-\infty} {e^{-y^2} \over {(x-y)^2 + a^2}}dy,
\end{equation}
where $a = \Gamma/(4\pi\Delta \nu_D)$ is the natural width parameter.
The damping constant is $\Gamma = 6.265 \times 10^{8} \rm\ s^{-1}$,
and thermal Doppler width is 
$\Delta \nu_D = \nu_{Ly\alpha}v_{\rm th}/c$, with 
thermal speed $v_{\rm th} = \sqrt{2k_BT/m_p}$.

The dimensionless frequency parameters corresponding to the H and K transitions are defined as
$x_H={{\nu - \nu_H} / {\Delta \nu_D}}$ and $x_K={{\nu - \nu_K} / {\Delta \nu_D}}$, respectively.
The central frequency of \lya is $\nu_{Ly\alpha} = 2.466 \times 10^{15}\, \rm Hz$, and
the frequency difference between the H and K transitions is $\Delta \nu_{HK} = 1.08 \times 10^{10}\, \rm Hz$.
In this work, the temperature of the scattering medium is fixed to $10^4\rm \, K$ so that
the velocity difference of the two lines, $c\nu_{HK}/\nu_{Ly\alpha}$, is much smaller than $v_{\rm th}$.
This, in turn, leads to a total cross section that is well described by a single Voigt profile \citep{ahn02,seon20}.

\subsection{Polarization}

No rigorous distinction between resonance and Rayleigh scattering can be made in the scattering process of a \lya photon. A commonly accepted usage of the term ``resonance scattering'' refers to a scattering process that occurs in a frequency range within a few $\Gamma$ from the line center in the rest frame of the scattering atom. In contrast, a scattering process occurring far from the line center may be called Rayleigh scattering. 
Note that resonance and Rayleigh scatterings are referred to as ``core'' and ``wing'' scatterings, respectively, throughout the paper.

In the case of \lya, Rayleigh scattering is more effective at yielding linearly polarized radiation than resonance scattering
because the $1S_{1/2} - 2P_{1/2}$ transition of resonance scattering results in completely unpolarized radiation, whereas resonance scattering associated with the $1S_{1/2} - 2P_{3/2}$ transition produces weak polarization.

% In particular, it is fascinating that resonance scattering due to the $1S_{1/2} - 2P_{1/2}$ transition results in completely unpolarized radiation, whereas resonance scattering associated with the $1S_{1/2} - 2P_{3/2}$ transition produces weak polarization.
%To describe the polarization state of \lya,
%\cite{rybicki99} adopts the classical method where 
%the single photon with 100\% polarized vector for one ray in the simulation \citep[e.g.][]{dijkstra08,dijkstra12,trebitsch16}.
%When the initial photon is generated in this classical method, 
%the polarization vector randomly originates, and the degree of polarization is 100\%.
%During \lya photon is traveling in the scattering medium, the direction of polarization is changed, but the degree of polarization is always 100\%.
%Hence, this method cannot include the polarization behavior by resonance scattering.

\cite{ahn02} investigate polarized radiative transfer using photon packets carrying the polarization information incorporated into a Hermitian 2$\times2$ density matrix \cite[e.g.,][]{eide18,chang20}.
Here, a photon packet represents an ensemble of numerous photons. 
%They adopt the photon packet to describe the polarization after resonance scattering.
In this formalism, the polarization state of an initial photon packet is chosen to be unpolarized.
The density matrix is renewed to assign the polarization information at each time of scattering in the observer's frame.
Depending on whether it is Rayleigh or resonance scattering, two different update schemes are applied separately to the density matrix.
The scattering type is determined in a probabilistic way after an appropriate assessment of the occurrence probabilities of Rayleigh and resonance scatterings as a function of $\nu$.

For our simulation, we adopt the Stokes vector of a photon packet to represent the polarization state of \lya, as described in \cite{seon22}.
The Stokes vector is represented as a column vector:
\begin{equation}
{\bf S} = 
\begin{pmatrix}
I \\ Q \\ U \\ V
\end{pmatrix}
.
\end{equation}
Here, the Stokes parameters are defined by
\begin{eqnarray}
\nonumber
I &=& \left< E_m E_m^* + E_n E_n^* \right>; \\
\nonumber
Q &=& \left< E_m E_m^* - E_n E_n^* \right>; \\
\nonumber
U &=& \left< E_m E_n^* + E_m^* E_n \right>; \\
V &=& i \left< E_m E_n^* - E_m^* E_n \right>,
\end{eqnarray}
where $E_m$ and $E_n$ are the electric field along the polarization basis vectors $\bf \hat m$ and $\bf \hat n$, respectively. 
The transverse nature of the electromagnetic waves requires that the polarization basis 
vectors are orthogonal to the wavevector $\bf \hat k$.

The renewed Stokes vector ${\bf S}'$ is determined by the polar scattering angle $\theta = \bf \hat k \cdot \bf \hat k'$, where $\bf \hat k'$ is the wavevector of the scattered radiation.
%The Stokes vector is defined by the basis vectors in the photon's frame.
 \cite{seon22} introduce the two matrices ${\bf M}(\theta)$ and ${\bf L}(\phi)$  
 to obtain ${\bf S}'$:
 %given by
\begin{equation}
{\bf S'} = {\bf M}(\theta){\bf L}(\phi){\bf S},
\end{equation} where $\phi$ is the azimuth scattering angle. Here, ${\bf M}$ denotes the scattering matrix and ${\bf L}$ the rotation matrix of the Stokes vector.

Use is made of the explicit expressions of ${\bf M}(\theta)$ and ${\bf L}(\phi)$ to yield
${\bf S}'$ as follows:
\begin{eqnarray}\label{eq:polarization}
\nonumber
I' &=& [{3 \over 4}E_1 (\cos^2 \theta + 1) + E_2]I \\
\nonumber
   &+& {3 \over 4}E_1 (\cos^2 \theta - 1)(Q \cos 2\phi + U \sin 2\phi); \\
\nonumber
Q' &=& {3 \over 4}E_1 (\cos^2 \theta - 1)I \\
\nonumber
   &+& {3 \over 4}E_1 (\cos^2 \theta + 1)(Q \cos 2\phi + U \sin 2\phi); \\
\nonumber
U' &=& {3 \over 2}E_1 \cos \theta(-Q \sin 2\phi + U \cos 2\phi); \\
V' &=& {3 \over 2}E_3 \cos \theta V.
\end{eqnarray}
Here, the parameters $E_{1}$, $E_{2}$, and $E_{3}$ are given by 
\begin{equation}
    E_1 = {{2x_K x_H + x_H^2} \over {x_K^2 + 2x_H^2}}, \,
    E_2 = 1 - E_1, \,
    E_3 = {1 \over 3}(E_1 +2),
\end{equation} as functions of the frequency.
The scattering angles $\theta$ and $\phi$ must be specified before one can carry out the computation of ${\bf S}'$ using Eq.~(\ref{eq:polarization}).
The polar angle $\theta$ is randomly selected from the marginal probability density of $\theta$, which is proportional to $I'/I$ integrated over $\phi$,
\begin{equation}
    P (\cos \theta) =  {3 \over 4}E_1 (\cos^2 \theta + 1) + E_2.
\end{equation}
The azimuth angle $\phi$ is randomly chosen, using a rejection method, to follow Equation (28) in \citet{seon22} in a range between 0 and 2$\pi$.
%Because the polar scattering angle $\theta$ is selected from the probability function, 
We normalize the Stokes vector ${\bf S}'$ by dividing by $I'$ so that the intensity component of $S'$ is fixed to be unity.
After the nomalization,
%Then 
the new Stokes vector ${\bf S}'$ is 
%given by, 
\begin{equation}
{\bf S'} = 
\begin{pmatrix}
1 \\ Q'/I' \\ U'/I' \\ V/I'
\end{pmatrix}
.
\end{equation}
The \lya emission source is assumed to be isotropic and unpolarized, and therefore the initial Stokes vector is given by
\begin{equation}
{\bf S_i} =
    \begin{pmatrix}
    1 \\ 0 \\ 0 \\ 0
    \end{pmatrix}.
\end{equation}

\section{Modeling a spherical halo with a central point source} \label{sec:model}
\begin{figure*}[ht!]
\centering
\includegraphics[width=\textwidth]{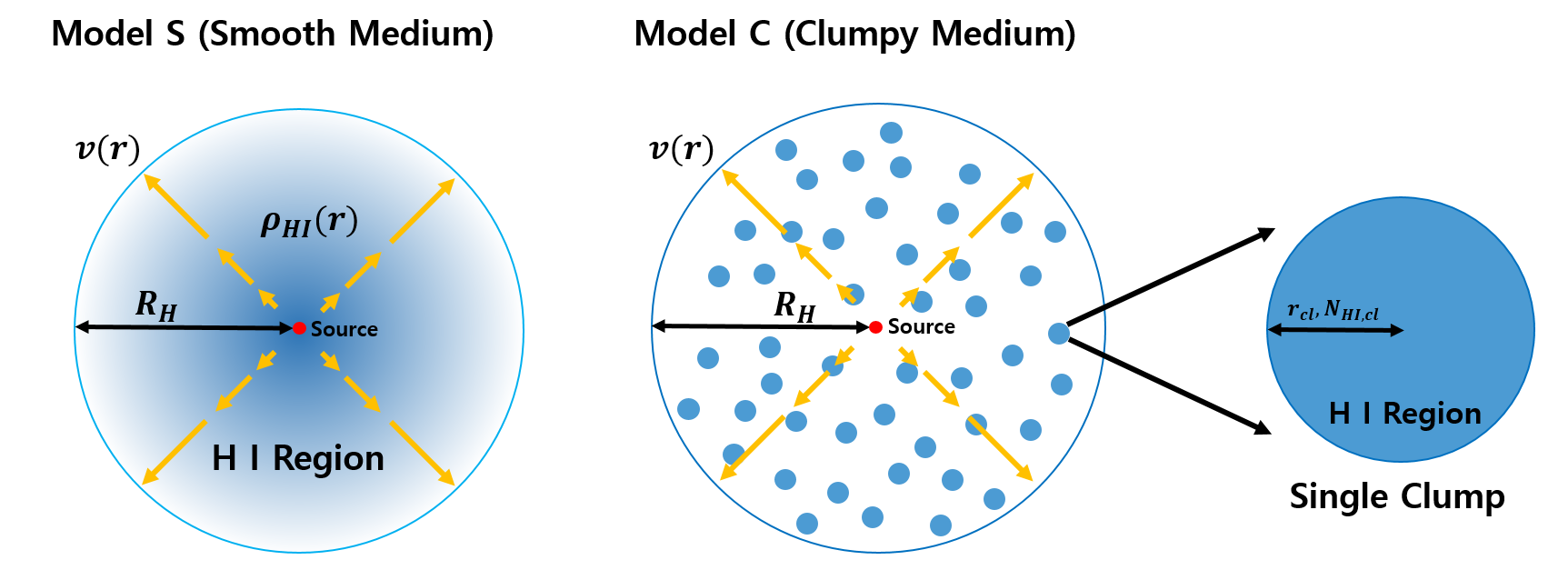}
\caption{
Schematic illustration for Model S (smooth medium; left) and Model C (clumpy medium; middle). 
The red dot represents the central \lya source. Orange arrows represent the expansion velocity, which is proportional to the radius.
In Model C, the clumps are uniformly distributed in the halo, and the density of each clump is uniform (right).
}
\label{fig:scheme}
\end{figure*}

To simulate \lya halos produced mainly by \lya scattering, we consider a spherical \hi halo with a central \lya point source.
The radius of the scattering medium is fixed to $100$~kpc, which is comparable to the typical sizes of LABs or ELANe. We consider two types of models: Model S (``smooth'' medium) %model) 
and Model C (``clumpy'' medium) %model) 
depending on the distribution of neutral \hi gas in the halo.
Figure \ref{fig:scheme} shows the schematic illustration for the two models. The \hi medium in Model S is continuously distributed.
In Model C, the scattering medium is composed of numerous spherical clumps; the intra-clump region is set to be empty.
The source at the center emits \lya photons with an initial spectrum approximated by a Gaussian function with a width of several hundred \kms. Throughout the paper, we assume that the model halo is located at $z = 3$ with an angular scale of 7.855 kpc arcsec$^{-1}$ and a Ly$\alpha$ luminosity of $10^{44}$ \unitcgslum. The number of the photon packets emitted from the source is $5 \times 10^5$. In the following, we describe the detailed physical conditions of the two models and the central point source.

\subsection{Smooth Medium (Model S)}

In Figure \ref{fig:scheme} (left), we show that for Model S, the \hi number density of the spherical \hi region declines exponentially as a function of the distance from the central source. We adopt an expanding velocity field where the outflow speed is proportional to the distance\footnote{While we do not explicitly explore inflows here, they can be modeled by switching \vexp to a negative value.
The resulting surface brightness profiles and  polarization behavior are identical to those for an outflow due to the symmetry of the \lya\ scattering cross-section.
The inflow spectrum will be a mirror image of the outflow spectrum.
}. The \hi number density $n_{\rm HI}(r)$ and the radial velocity $v(r)$ are given by
\begin{equation}
n_{\rm HI}(r) = n_{0} ~ e^{-{r / {R_e}}}, 
\quad
v(r) = \vexp ~ {r \over R_H} ~, 
\label{eq:velocity}
\end{equation}
where $R_e$ is the effective radius of the exponential profile, and $r$ is the radius from the central source. 
The parameter $R_H$ is the halo radius and fixed to $100\, \rm kpc$.
The outflow velocity characterizing the kinematics of the scattering region is an expansion velocity \vexp.
We explore a range of \vexp = 0, 100, 200, and 400 \kms\
and halo gas concentrations  $R_e/R_H$ = 0.3, 0.5, 1, and $\infty$ (i.e., uniform distribution).
For these sets of parameters, the total \hi column density \NHI, which characterizes the optical thickness of the \hi region, is given by
\begin{equation}
    \NHI = \int^{R_H}_{0} n_{\rm HI}(r) dr.
\end{equation}
We set the range of \NHI to $10^{18-21} \unitNHI$.
Then, the total neutral \hi mass of the spherical halo is 
%given by
\begin{equation}
    M_{\rm HI} 
    \approx 2.5 \times 10^{8} M_{\odot} {\left( R_e \over R_H \right)^4}{\left( N_{\rm HI} \over {10^{18} \rm cm^{-2}}\right) }.
\end{equation}
In the case of the uniform medium ($R_e = \infty$),
\begin{equation}
M_{\rm HI} 
\sim 3.4 \times 10^{8} M_{\odot} {\left( N_{\rm HI} \over {10^{18} \rm cm^{-2}}\right) }, 
\end{equation}
where the range of \NHI = $10^{18-21}\, \rm cm^{-2}$ corresponds to $M_{\rm HI} \sim 10^{8-11}$\,\msun.

\subsection{Clumpy Medium (Model C)}

Figure \ref{fig:scheme}\,(middle and right) illustrates Model C, where the halo gas consists of numerous small clumps.  We adopt the same halo size of $R_H$ = 100 $\rm kpc$ as for Model S.  All neutral \hi is confined within these clumps, and clumps do not overlap each other.
In the smooth model (Model S), we vary $R_e/R_H$ to explore the variation of number density profiles. However, due to the technical limitations of Model C, we consider only a uniform number distribution of clumps. 
%% In other words, the clumps are uniformly distributed over the halo in Model C. 
In a clump, the \hi medium is static and uniform; the \hi number density is $\NHIcl / r_{cl}$, where \NHIcl is the clump \hi column density.
With photo-ionization modeling of the giant \lya nebula around a bright QSO (UM287), \cite{fabrizio15} found a clump hydrogen column density of \NHIcl $\lesssim 10^{20} \unitNHI$ and a clump size of $r_{cl}$ $\lesssim$ 19\,pc. We allow $r_{cl}$ to range from 10\,pc to 1\,kpc ($r_{cl} \ll R_H$). As a result, the clumps are not spatially resolved at $z>2$, which is consistent with observations.
Note that \cite{gronke17} considered clumps with radii of $10^{-3} - 10^{-2}$\,pc in a 100\,pc halo to simulate the \lya RT in the ISM. The ratios of the clump and halo sizes are similar to our assumption.

The total column density of the clumpy halo (\NHI) is given by
\begin{equation}
    \NHI = f_c \NHIcl,
\end{equation}
where the covering factor $f_c$ represents the number of clumps in the line of sight between the central source and an observer.
We explore the ranges $f_c$ =  1 -- 100 and $\NHIcl$ = $10^{16-19}\, \rm cm^{-2}$.
To describe outflows in the \hi halo, we set the radial velocities of individual clumps proportional to the distance between the central source and each clump center, following Eq.~\ref{eq:velocity}.
The total \hi mass of the clumpy halo is given by
\begin{equation}
    M_{\rm HI}  
    \sim 4.5 \times 10^8 M_{\odot} {\left( f_c \NHIcl \over {10^{18}\, \rm cm^{-2}}\right) }. 
\end{equation}
The mass of Model C is greater by $\sim 30\%$ than that of the uniform medium Model S at the same total column density \NHI.

\subsection{Central Point Source}

In our simulation, we assume that the emission from the central source is isotropic. 
To predict the linewidth of \lya emission, we consider the spectrum of the input source to follow a Gaussian function $F(\lambda_i)$,
\begin{equation}
    F(\lambda_i) = {1 \over \sqrt{2 \pi} \sigma_{\lambda, {\rm src}}} \exp \left[-{(\lambda_i-\lambda_{Ly\alpha})^2 \over {2 \sigma_{\lambda, {\rm src}}^2}}\right],
\end{equation}
where $\lambda_i$ is the wavelength of an initial photon, and $\sigma_{\rm \lambda, src}$ is the intrinsic \lya linewidth
of the input source. Hereafter, we adopt the velocity width $\sigsrc = c \sigma_{\lambda, {\rm src}}/\lambda_{\rm Ly\alpha}$ for $\sigma_{\rm \lambda, src}$.
Depending on the nature of the galaxies embedded in the \lya halo, we consider \sigsrc from 100\,\kms (as in a star forming galaxy (SFG)) to 400 \kms (as associated with an AGN).

\begin{table*}[]
\centering
\begin{tabular}{llll}
\hline
        & Parameter   & Range                                                     & Note \\ \hline
Model S & \NHI        & $10^{18}, 10^{19}, 10^{20}, 10^{21} \unitNHI$             & total \ion{H}{1} column density \\
        & $R_e$       & 0.1, 0.3, 0.5, 1.0\,$R_H$,\, $\infty$ (uniform density)   & effective radius   \\
        & \vexp       & $0, 100, 200, 400 \kms$                                   & expansion velocity \\
        & \sigsrc     & $100 \,(\rm SFG) , 200, 400 \,(\rm AGN) \kms$             & \lya source velocity width  \\[0.2em]\hline
Model C &  \NHIcl     & $10^{16-21} \unitNHI$                                     & clump column density  \\
        &  $r_{cl}$   & $0.01, 0.1, -1\,$kpc                                      & clump radius  \\
        &  $f_c$      &  $1, 2, 5, 10, 100$                                       & covering factor  \\
        &  \vexp      & $0, 100, 200, 400 \kms$                                   & expansion velocity   \\
        &  \sigsrc    & $100 \,(\rm SFG) , 200, 400 \,(\rm AGN) \kms$             & \lya source velocity width  \\[0.2em]
\hline
\end{tabular}
\caption{Parameters of Models S and C.}
\label{table:model_s}
\end{table*}

\begin{table*}[]
\centering
\begin{tabular}{rl}
\hline
%    Figure Number    & Contents  \\ \hline
 & For spatial diffusion as a function of projected radius $R_p$ \\ \hline
Figure~\ref{fig:sb_NH_s} & Surface brightness profiles \SB for \NHI  $= 10^{18-21} \unitNHI$
\\
\ref{fig:sb_vexp_s} & \SB for \vexp = 0 -- 400 \kms 
\\
\ref{fig:sb_A_s} & \SB for $R_e/R_H = 0.3-1,$ and $\infty$ (uniform \hi density)
\\
\ref{fig:size_s} & Observable radius $R_{obs}$
\\ \hline
 & For polarization as a function of projected radius $R_p$ \\ \hline
Figure~\ref{fig:pol_NH_s} & Degree of polarization profiles \pol for \NHI  $= 10^{18-21} \unitNHI$
\\
\ref{fig:pol_vexp_s} & \pol for \vexp = 0 -- 400 \kms 
\\
\ref{fig:pol_A_s} & \pol for $R_e/R_H = 0.3-1$ and $\infty$
\\
\ref{fig:Pobs_s} & Polarization at $R_{obs}$ ($P_{obs}$)
\\ \hline
 & For frequency diffusion as a function of Doppler shift $\Delta V$ \\ \hline
Figure~\ref{fig:spec_NH_s} & Integrated \lya spectra \spec for $\NHI = 10^{18-21} \unitNHI$
\\
\ref{fig:spec_vexp_s} & \spec for \vexp = 0 -- 400 \kms
\\
\ref{fig:spec_A_s} &  \spec for $R_e/R_H = 0.3-1$ and $\infty$
\\
\ref{fig:peak_s} &  Peak shift of \lya spectra $\Delta V_{peak}$
\\[0.2em]\hline
\end{tabular}
 	\caption{Figures showing results for Model S.}
	\label{table:figures_s}
\end{table*}

\section{Smooth Medium (Model S) Results}\label{sec:result_a}

\begin{figure*}[ht!]
	\includegraphics[width=\textwidth]{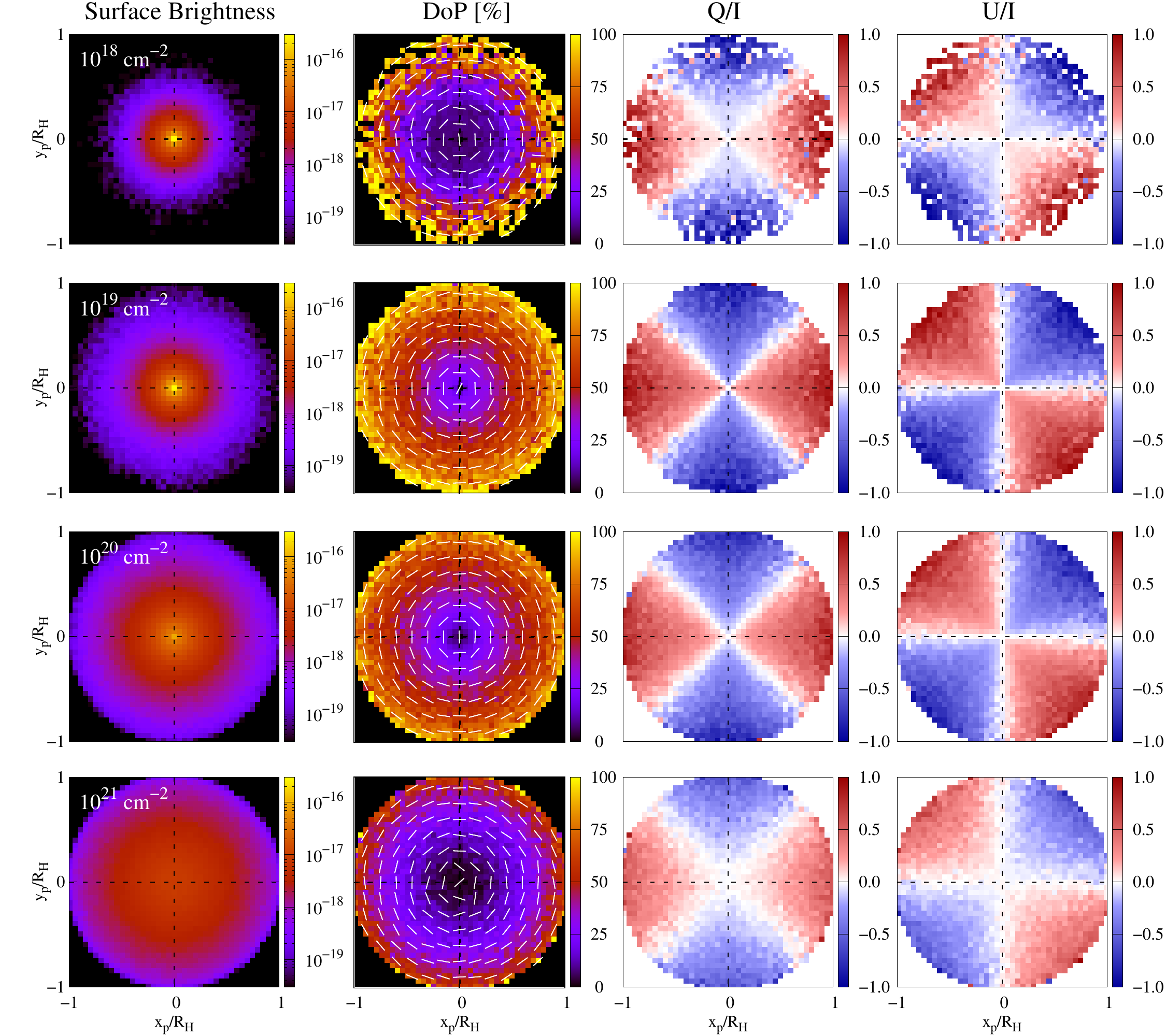}
	\caption{
		Projected surface brightness and polarization maps of Model S for \hi column density $\NHI = 10^{18} \unitNHI$ (first), $10^{19} \unitNHI$ (second), $10^{20} \unitNHI$ (third), and $10^{21} \unitNHI$ (fourth row) with a expansion velocity $\vexp = 400$ \kms, a \hi halo effective radius $R_e/R_H=\infty$ (uniform \hi density), and a \lya source velocity width $\sigsrc = 100$ \kms.
		The left panels are the surface brightness in logarithmic scale.
		The three right panels represent the polarization information: the degree of polarization and the Stokes parameters $Q/I$ and $U/I$.
		The white solid lines show the direction of the polarization.
		At the center, the surface brightness is highest, and the polarization is zero.
		As \NHI increases, the surface brightness profile becomes more extended, and the central bright core disappears at \NHI$=10^{21} \unitNHI$.
		At \NHI = $10^{19-20}$ \unitNHI (second and third rows), the polarization steeply increases near the center.
		We refer this dramatic increase as a ``polarization jump.''
		The \NHI$=10^{19} \unitNHI$ case (second row) shows the highest degree of polarization. Note that the overall degree of polarization does not behave monotonically as a function of \NHI.
		This non-monotonic behavior originates from the varying contributions of the three \lya scattering processes illustrated in Figure~\ref{fig:single_multi}. 
	}
	\label{fig:image}
\end{figure*}

To investigate the behavior of \lya in the smooth medium, we produce simulated images of observables such as \lya intensity and polarization.
In Figure \ref{fig:image}, we show examples of Model S: surface brightness distributions, degrees and orientations of polarization, and Stokes parameters ($Q/I$ and $U/I$) for four column densities, $\NHI= 10^{18}$, $10^{19}$, $10^{20}$, and $10^{21}\, \rm cm^{-2}$, respectively. In the figure, we fix the expansion velocity \vexp  = 400\kms and the \lya source velocity width \sigsrc = $100\, \kms$. 

Here, we briefly describe the general trends of Model S. We discuss the results as functions of various parameters in the following sections.
First, we find that the surface brightness profiles become more extended as the total \hi column density (\NHI) increases (Figure \ref{fig:image}, left column). 
Second, the polarization patterns are concentric due to the spherical symmetry and increase radially outward (Figure \ref{fig:image}, second column). These predictions are consistent with previous findings \citep{dijkstra08,eide18}.
Third, the degree of polarization (DoP) does {\it not} behave monotonically as a function of \NHI. The overall DoP peaks at $\NHI = 10^{19}\unitNHI$ relative to $\NHI = 10^{18}$, $10^{20}$, and $10^{21}\unitNHI$.
Lastly, at $\NHI = 10^{19} \unitNHI$, the degree of polarization increases steeply from nearly 0\% at the center of the halo to 20\%.
Throughout the paper, we will refer to this behavior as a ``polarization jump.'' This discontinuous DoP profile is one of the most surprising results from our simulations.

In the following sections, we describe the \lya halos in the parameter space defined in Table \ref{table:model_s}.
Table~\ref{table:figures_s} summarizes the figures showing the Model S results.
We present the predicted surface brightness profile and degree of polarization as a function of projected radius in Sections \ref{sec:result_sb_s} and \ref{sec:result_dop_s}, respectively.
In Section~\ref{sec:result_sb_s}, we determine the  observable sizes of the model \lya halos ($R_{obs}$), thereby testing whether \lya photons scattered from a central point source can produce realistic LABs with typical sizes of 100\,kpc ($\sim$10\arcsec) at high redshift.
In Section~\ref{sec:result_dop_s}, we explain the origin of the polarization jump in detail and compute the polarization at $R_{obs}$ ($P_{obs}$). 
In Section~\ref{sec:result_spec_s}, we present the integrated \lya spectra and explore the velocity offsets generated by scattering in the expanding medium with the Hubble flow-like velocity field. 
%% We measure the velocity offset of the line peak to investigate the relation between the offset and the kinematics of the scattering medium.

For the \lya source spectrum, we consider a range of \lya velocity widths to represent different source types: $\sigsrc = 100\, \kms$ for typical SFGs and $\sigsrc = 400\, \kms$ for AGN. We add a third, intermediate value of  $\sigsrc = 200\, \kms$ to approximate a star forming galaxy with broader emission. We will refer to total column densities of \NHI = $10^{19}$ and $10^{21}$\unitNHI as low- and high-\NHI cases, respectively.

%------------------------------------------------------------------------------------------
\subsection{Surface Brightness}\label{sec:result_sb_s}

\begin{figure*}[ht!]
\centering
    \includegraphics[width=\textwidth]{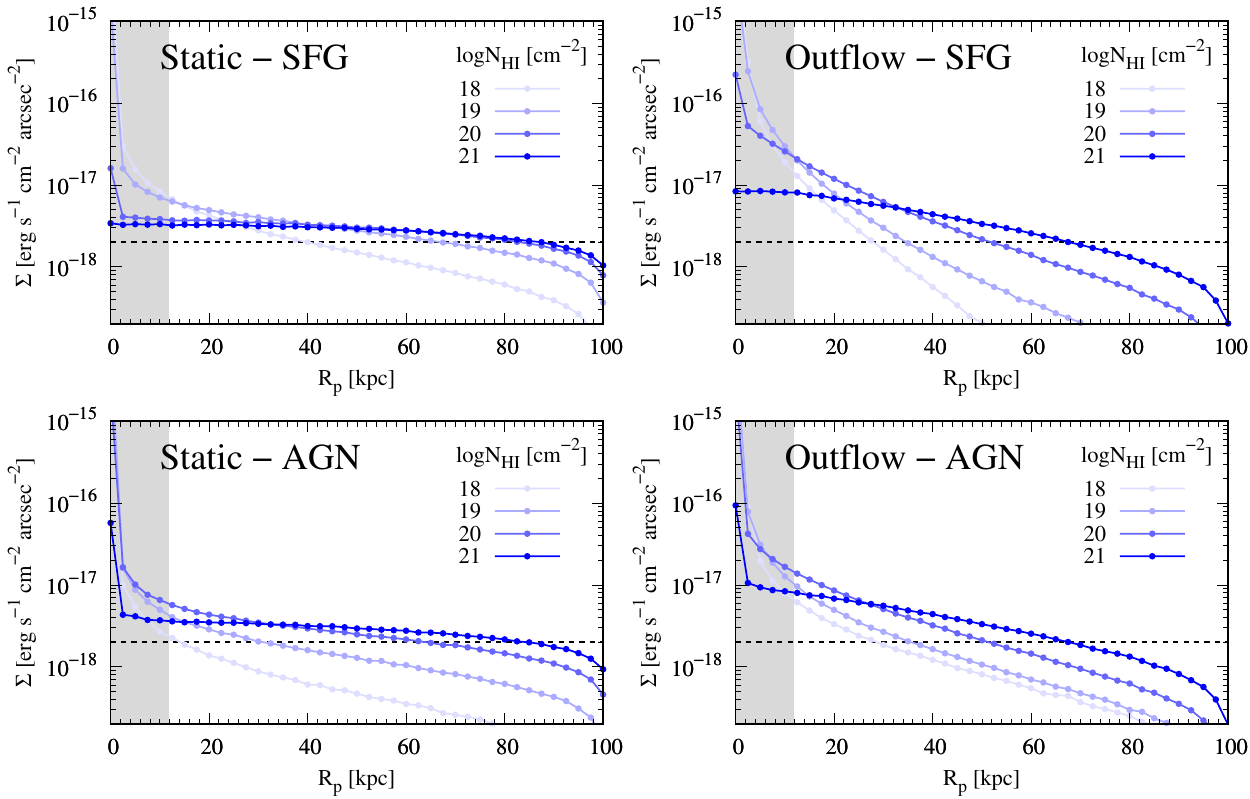}
	\caption{
		Radial surface brightness profiles of Model S for %the \hi column density
		$\NHI= 10^{18-21} \unitNHI$.
		The $x$-axis is the projected radius $R_p = \sqrt{x_p^2 + y_p^2}$.
            The gray shaded region with $R_p \lesssim 12\, \rm kpc$ indicates where the data would be affected by ground-based seeing.
            The $y$-axis is the surface brightness in cgs units.
		The H~I density is uniform (i.e., $R_e = \infty$).
		The top and bottom panels represent %SFG and AGN cases, respectively.
		cases where the central source is a star-forming galaxy (SFG, \sigsrc = 100 \kms) and active galactic nucleus (AGN, \sigsrc = 400 \kms), respectively.
		``Static" (left panels) and ``Outflow" (right) models have $\vexp = 0\, \kms$ and $400\, \kms$, respectively.
		The horizontal dashed line is the observational threshold: $2 \times 10^{-18}$ \unitcgssb.
		The surface brightness profiles in the top right panel are 
		%the radial profile of projected surface brightness in
		from the first column of Figure~\ref{fig:image}. 
		When \NHI increases, the surface brightness profile becomes more extended and flattened due to the increasing number of scatterings. For $\NHI \leq 10^{19} \unitNHI$, the profiles are composed of two components: a bright core from directly escaping photons and a diffuse halo due to scattering. At $\NHI = 10^{21} \unitNHI$, the bright core disappears, because numerous scatterings prevent us from seeing the central source.
	}
	\label{fig:sb_NH_s}
\end{figure*}

\begin{figure*}[ht!]
\centering
	\includegraphics[width=\textwidth]{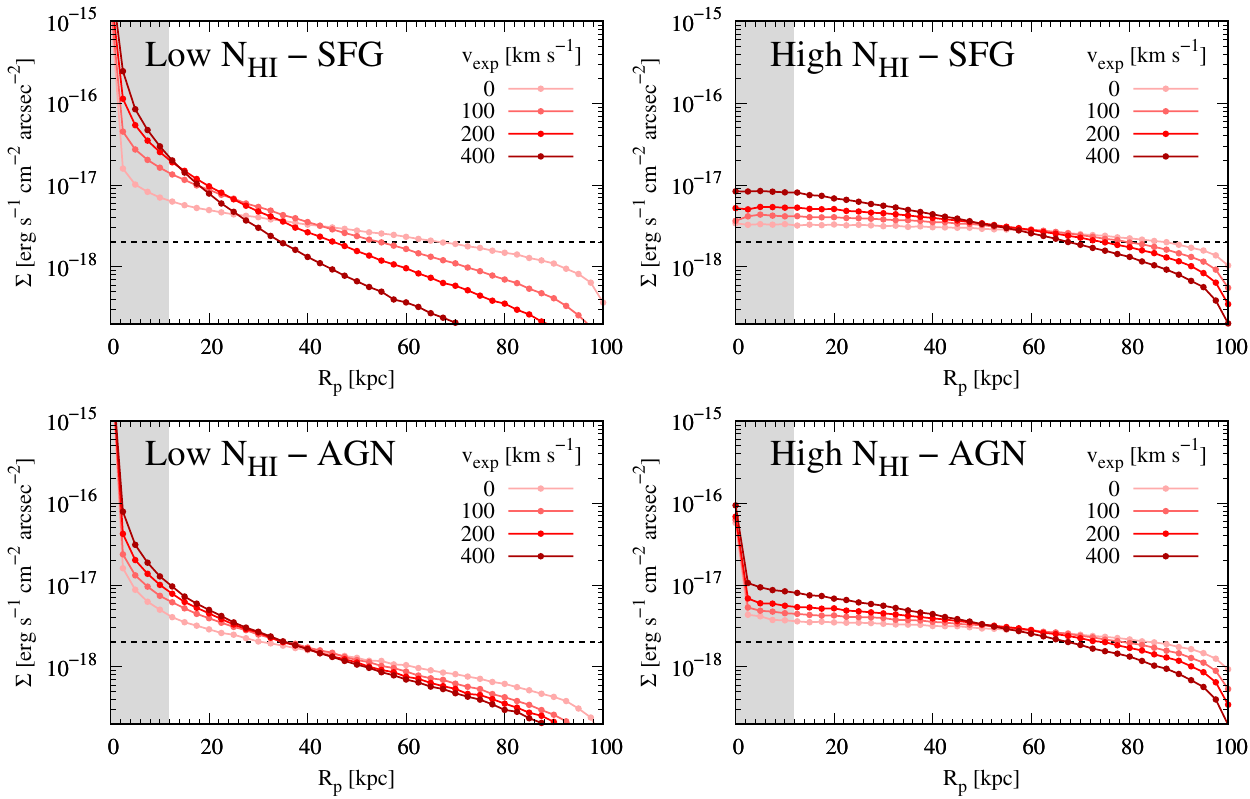}
	\caption{
	Surface brightness profiles of Model S for $\vexp = 0 - 400 \kms$.
	The structure of this figure is similar that of Figure \ref{fig:sb_NH_s}. ``Low \NHI'' (left) and ``High \NHI'' (right panels) show the results for $\NHI = 10^{19}$ and $10^{21} \unitNHI$, respectively.
	The surface brightness profile becomes more extended with decreasing \vexp. 
	For low \NHI cases (left panels), the input source (\sigsrc) strongly affects the dependence of \SB on \vexp: the low \NHI--SFG case profiles are sensitive to the variation of
	\vexp, of which the value is comparable to or larger than \sigsrc. On the other hand, the surface brightness profiles in the high \NHI case do not strongly depend on \sigsrc, although bright cores can exist in the AGN cases.
	}
	\label{fig:sb_vexp_s}
\end{figure*}

\begin{figure*}[ht!]
\centering
	\includegraphics[width=\textwidth]{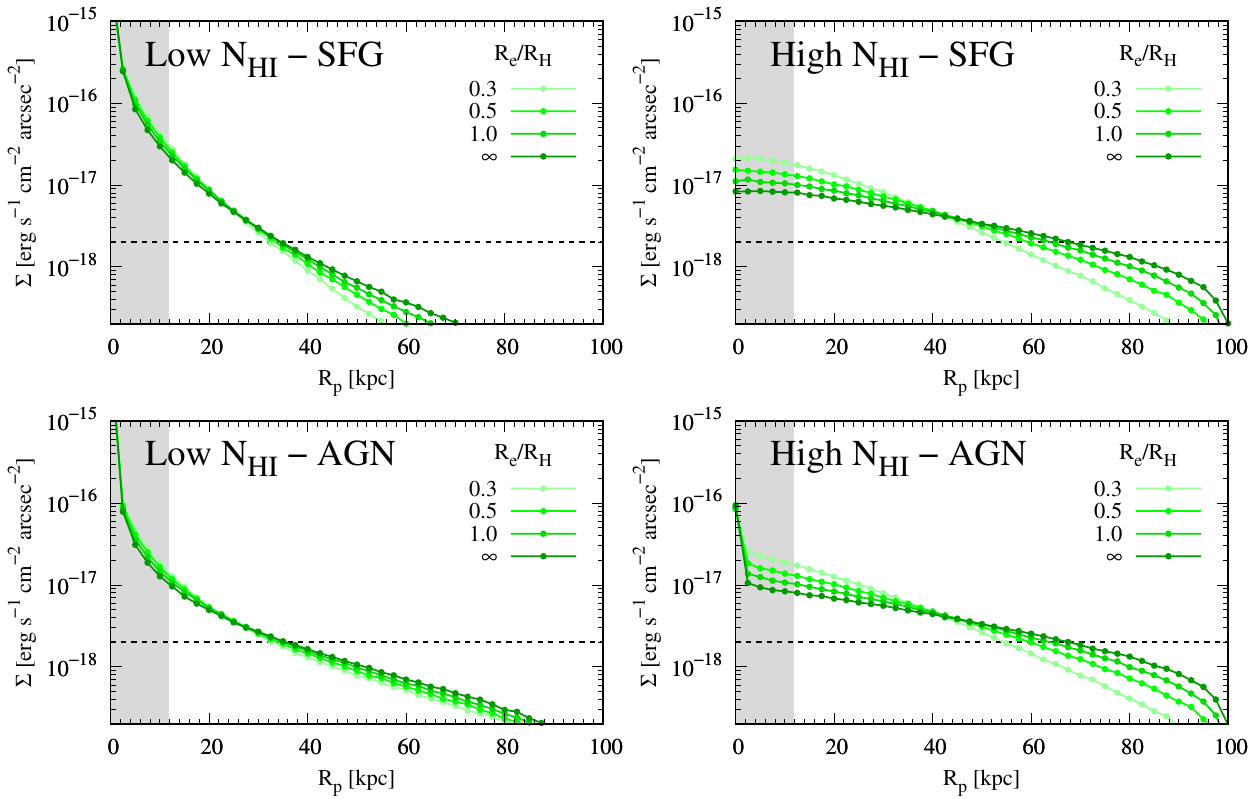}
	\caption{
            Surface brightness profiles of Model S for an \hi halo effective radius $R_e = 0.3 R_H$ to $\infty$ (uniform \hi density) and a strong outflow, $\vexp = 400\kms$.
            The dependence on $R_e/R_H$ of low \NHI cases is minimal compared to other parameters ,
            because the number of \lya scatterings in the inner region dominates the formation of the \lya halo. For high \NHI, there is more dependence on $R_e/R_H$, because \lya scattering can occur everywhere in the \hi halo.
        }
	\label{fig:sb_A_s}
\end{figure*}

\begin{figure*}[ht!]
\centering
	\includegraphics[width=\textwidth]{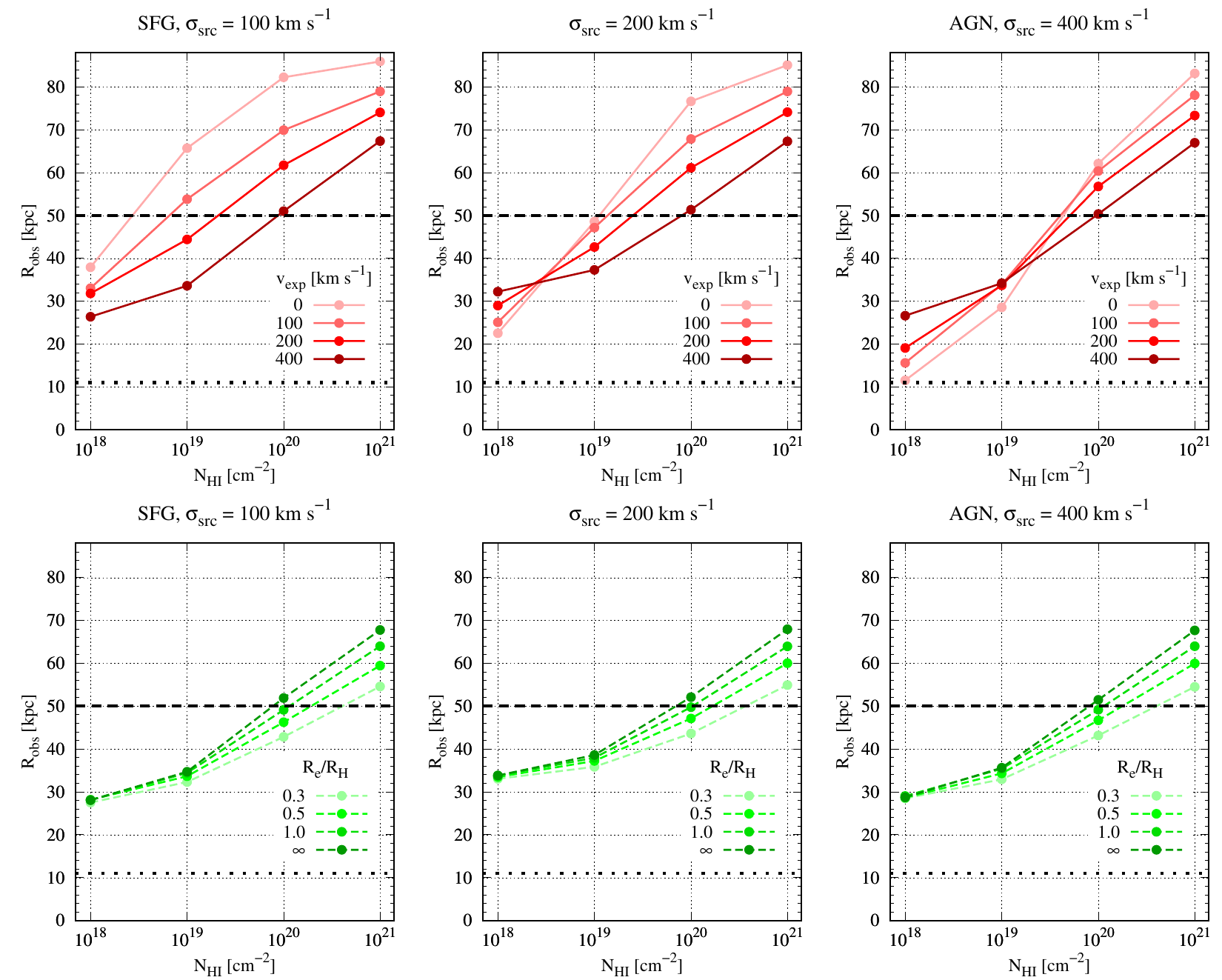}
	\caption{
	Model S \lya halo size ($R_{obs}$) 
	as a function of \NHI for $\sigsrc = 100 \kms$ (SFG), 200 \kms, and 400 \kms (AGN). 
	The red {\bf solid} and green {\bf dashed} lines of different shades in the top and bottom panels represent $\vexp$ and $R_e/R_H$, respectively.
	The top panels are for various \vexp, with $R_e/R_H$ fixed to $\infty$ (i.e., a uniform \hi density).
	The bottom panels are for various $R_e/R_H$, with \vexp fixed to 400 \kms.
	The horizontal dotted lines represent the radius of the \lya halo considering only seeing effects without scattering ($R_{obs} = 12\,$kpc).
	The black dashed lines represent the typical size of giant \lya nebulae ($\sim 100\,$kpc, $R_{obs} = 50\,$kpc).
	Increasing \NHI leads to larger $R_{obs}$.
	When \NHI $\geq 10^{20} \unitNHI$, a \lya point source within a smooth scattering medium will be observed as a giant \lya nebula (i.e., diameter $\sim 100\,$kpc; $R_{obs} \gtrsim 50\,$kpc), regardless of other parameters.
	In this high \NHI regime, lower \vexp leads to larger $R_{obs}$.
	%% $R_{obs}$ in AGN cases tends to be smaller than the size in SFG cases except $\NHI = 10^{21} \unitNHI$ cases.
	The dependence on $R_e/R_H$ is weaker than for other parameters, although not negligible at $\NHI \geq 10^{20} \unitNHI$ (see Section~\ref{sec:Re_s} and Figure~\ref{fig:sb_A_s}).
	}
	\label{fig:size_s}
\end{figure*}

In our model, \lya photons diffuse outward in both the spatial and frequency domains until they can escape the system. The spatial diffusion renders the central \lya point source into an extended \lya halo. Naturally, the more \lya photons experience scattering, the more extended and flattened surface brightness distribution emerges. To investigate the properties of \lya halos resulting from Model S, we extract radial surface brightness profiles \SB as a function of the projected distance from the center, $R_p = \sqrt{x_p^2 + y_p^2}$. 
Our findings are:

\begin{enumerate}[leftmargin=+0.5cm,itemsep=0pt]

\item[$\bullet$] 
The surface brightness profile becomes more extended with increasing \NHI and with decreasing \vexp.
The most dominant parameter is \NHI. (see \S~\ref{sec:sb_NH_s} and \S~\ref{sec:sb_vexp_s})

\item[$\bullet$] 
In the static medium, the SFG cases produce more extended halos than the AGN cases. (\S~\ref{sec:sb_input_s})

\item[$\bullet$] 
At high column density ($\NHI = 10^{21} \unitNHI$), the surface brightness profiles become very flat and do not depend on the source types (SFG vs.~AGN). (\S~\ref{sec:sb_input_s})

\item[$\bullet$] 
The \lya halo in Model S extends over $\sim$100 kpc as long as $\NHI \geq 10^{20} \unitNHI$. (\S~\ref{sec:LAH_size})

\end{enumerate}

\subsubsection{Dependence on Column Density (\NHI)}
\label{sec:sb_NH_s}

%% Dependence on N(HI)
The surface brightness profile \SB becomes more extended with increasing column density (\NHI) due to the increased number of scatterings. Figure \ref{fig:sb_NH_s} shows the change in \SB for \NHI $=10^{18}$ to $10^{21}$ \unitNHI and for four different combinations of outflow speeds and input \lya sources. As \NHI increases, all profiles become more extended, and the resulting halos will be observed to be larger. 

%% Morphology of Lya halos: probably more to different paragraph?
Depending on how much the input \lya photons are scattered off the sightline, the morphology of \lya halos can vary. 
%% Core
In the SFG cases with $\NHI \leq 10^{20}$\unitNHI, \lya photons from the central source can directly escape the system without much scattering; therefore, the spatial profile near $R_p = 0$ is sharply peaked, like a point-source. In other words, observers can see the input source directly through the scattering medium. This is also true for both the (Static--AGN) and (Outflow--AGN) cases because \lya photons in the large velocity wings are optically thin. When observed, these bright point sources will appear as bright cores due to the seeing or point spread functions (shaded gray regions at $R_p \lesssim 12$ kpc; $\sim$1.5\arcsec; Figure \ref{fig:sb_NH_s}).

%% Halo
At high column density (\NHI$=10^{21} \unitNHI$), it takes more scatterings for photons to escape the system, and \SB becomes more extended and flatter. In these scattering-dominated cases, one cannot see the input source directly through the gas halo, but only the scattered photons as a diffuse halo. In analogy, one can only see the scattered light from a flashlight in a thick fog.
%% Preview for polarization jump + core
The aforementioned polarization jump originates because these directly escaping photons dilute any polarized signal near the center of the halo. In Section~\ref{sec:result_dop_s}, we show that the bright core and polarization jump occur at the same time.

\subsubsection{Dependence on Outflow Speed (\vexp)}
\label{sec:sb_vexp_s}

\SB becomes more extended and flattened with decreasing \vexp.
Figure \ref{fig:sb_vexp_s} shows the variation of the surface brightness profile as a function of \vexp (0, 100, 200, 400\,\kms) for four different combinations of column densities (high/low \NHI case for $10^{19}$/$10^{21} \unitNHI$) and input \lya sources (SFG and AGN).
This dependence can be understood as follows. 
When the scattering occurs in an outflowing medium compared to a static one, photons can escape the system more easily due to large changes in the wavelength after scattering. Thus the photons see smaller optical depth, experience a smaller number of scatterings, and escape the system at a distance closer to the center. Therefore, \lya halos become more compact with increasing \vexp.

This dependence of the spatial extent on \vexp can be different depending on the width of the input \lya emission (\sigsrc). We find that \SB shows the largest dependence on \vexp in the case with low \NHI $(\leq 10^{19} \unitNHI$)--SFG where \vexp is larger than \sigsrc (see the low \NHI panels in Figure \ref{fig:sb_vexp_s}). 
For low \NHI--AGN case, the profiles (left bottom panel) are almost indistinguishable. 

%% Furthermore, the dependence on \vexp can be even inverted at \NHI = $10^{18} \unitNHI$, as we will discuss in Section \ref{sec:LAH_size}.

\subsubsection{Dependence on Concentration ($R_e/R_H$)}\label{sec:Re_s}

\SB becomes more extended with increasing effective radius ($R_e$), with a uniform halo being most extended. In other words, the \lya halo looks more extended for a more extended scattering medium. However, this dependence is significant only for large \NHI.
Figure \ref{fig:sb_A_s} shows \SB for $R_e/R_H$ = 0.3, 0.5, 1.0, and $\infty$ for the four \NHI--\sigsrc combinations. Note that a uniform distribution corresponds to the limiting case of $R_e/R_H$ $\rightarrow$ $\infty$. In the low column density regime (\NHI = $10^{19}$\unitNHI), the dependence on $R_e/R_H$ is negligible. 
For high column density ($\NHI = 10^{21} \unitNHI$), \SB shows a significant variation with increasing $R_e/R_H$, because \NHI is large enough to scatter photons even at the outer radii.

\subsubsection{Dependence on Input Source (\sigsrc)}
\label{sec:sb_input_s}
Given that the dependence on input source (\lya source velocity width \sigsrc) is more subtle, whenever possible we contrast two extreme cases (SFG and AGN) while other parameters fixed in Figures \ref{fig:sb_NH_s} -- \ref{fig:sb_A_s}.

In general, \SB for the SFG case is more extended than for the AGN case at the same \NHI (e.g., the Static cases in Figure \ref{fig:sb_NH_s}, left panels). In the AGN case (\sigsrc = 400\,\kms), there could be photons with wavelengths much further from the line center of the scattering medium; therefore, these photons easily escape due to the smaller optical depth. 
But this dependence on \sigsrc is minor compared to the trends with \NHI and \vexp, and there is also an exception in the case of large \vexp and high \sigsrc (e.g., the low-\NHI cases in Figure~\ref{fig:sb_vexp_s}, left panels).

When the outflow speed is high enough so that photons in the velocity wings of the AGN are effectively scattered by the medium, \SB for the AGN case can be more extended than for the SFG case.
At $\NHI \leq 10^{19} \unitNHI$, the fast-moving outer halo can cause scattering of the initial photons in the wavelength blueward from 0 to $-\vexp$. 
In the right panels of Figure \ref{fig:sb_NH_s}, for which $\vexp = 400 \kms$, \SB for the AGN case with $\NHI \leq 10^{19} \unitNHI$ becomes more flattened than the SFG case. We will show that the blueward photons of the AGN case are more likely to be scattered by the fast-moving halo by examining the \lya spectrum in Section \ref{sec:result_spec_s}.

The surface brightness profile arising from high column density gas (\NHI$\sim 10^{21} \unitNHI$) does not depend on \sigsrc (darkest lines in Figure \ref{fig:sb_NH_s}); 
the strong contribution of multiply scattered photons at this high column density erases the information about the intrinsic \lya emission. 
In the right panels of Figures \ref{fig:sb_vexp_s} and \ref{fig:sb_A_s}, the SFG and AGN cases show almost identical extended profiles, except for the small central bright core in the AGN case.

\subsubsection{Can Large \lya Halos be Produced through Scattering Alone?}
\label{sec:LAH_size}

To investigate when scattering by the neutral gas around galaxies or AGN can produce large \lya halos such as \lya blobs or ELANe \cite[e.g.,][]{steidel00, yang14a,yang14b, fabrizio19}, we measure the size of each \lya halo from the model library.  We define an observed halo radius $R_{obs}$ as the distance from the center to a fixed observational threshold ($\Sigma(R_{obs})$ = $2 \times 10^{-18}$ \unitcgssb), which corresponds to the horizontal dashed line in Figures \ref{fig:sb_NH_s} -- \ref{fig:sb_A_s}. 

Figure \ref{fig:size_s} shows $R_{obs}$ as a function of \NHI for the three types of sources, $\sigsrc = 100$ (SFG), 200, and 400 \kms (AGN). We also show the dependence of $R_{obs}$ on \vexp (red lines) and $R_e/R_H$ (green lines) in the upper and lower panels, respectively. If this radius is at least 50 kpc in a model, we regard that model as producing a LAB or ELAN.

As discussed above, the most dominant factor in determining the size of the scattering halo is \NHI. In SFG cases with $\NHI = 10^{18-19} \unitNHI$ (left top panel), $R_{obs}$ also shows strong dependence on \vexp: the smaller \vexp, the larger the \lya halo.  The concentration of the scattering medium ($R_e/R_H$) has less effect on the halo size (bottom panels).

For $\NHI \geq 10^{20} \unitNHI$, the profiles always extend out to at least $R_{obs}$ $\sim$ 50\,kpc, no matter the source or the expansion velocity. These systems will be observed as typical \lya blobs at $z=3$ as long as the central source has \llya = $10^{44}$ \unitcgslum as we assumed in Section \ref{sec:model}.
Somewhat lower 
\NHI value ($\sim 10^{19} \unitNHI$) still produce a large enough halo
when the 
source is SFG-like (\sigsrc = 100--200\,\kms) {\it and} the outflow speed is weak (\vexp $\sim$ 0-100\,\kms).
These results demonstrate that scattering alone can produce realistic \lya halos.

\subsection{Polarization}\label{sec:result_dop_s}

\begin{figure*}[ht!]
\centering
\includegraphics[width=150mm]{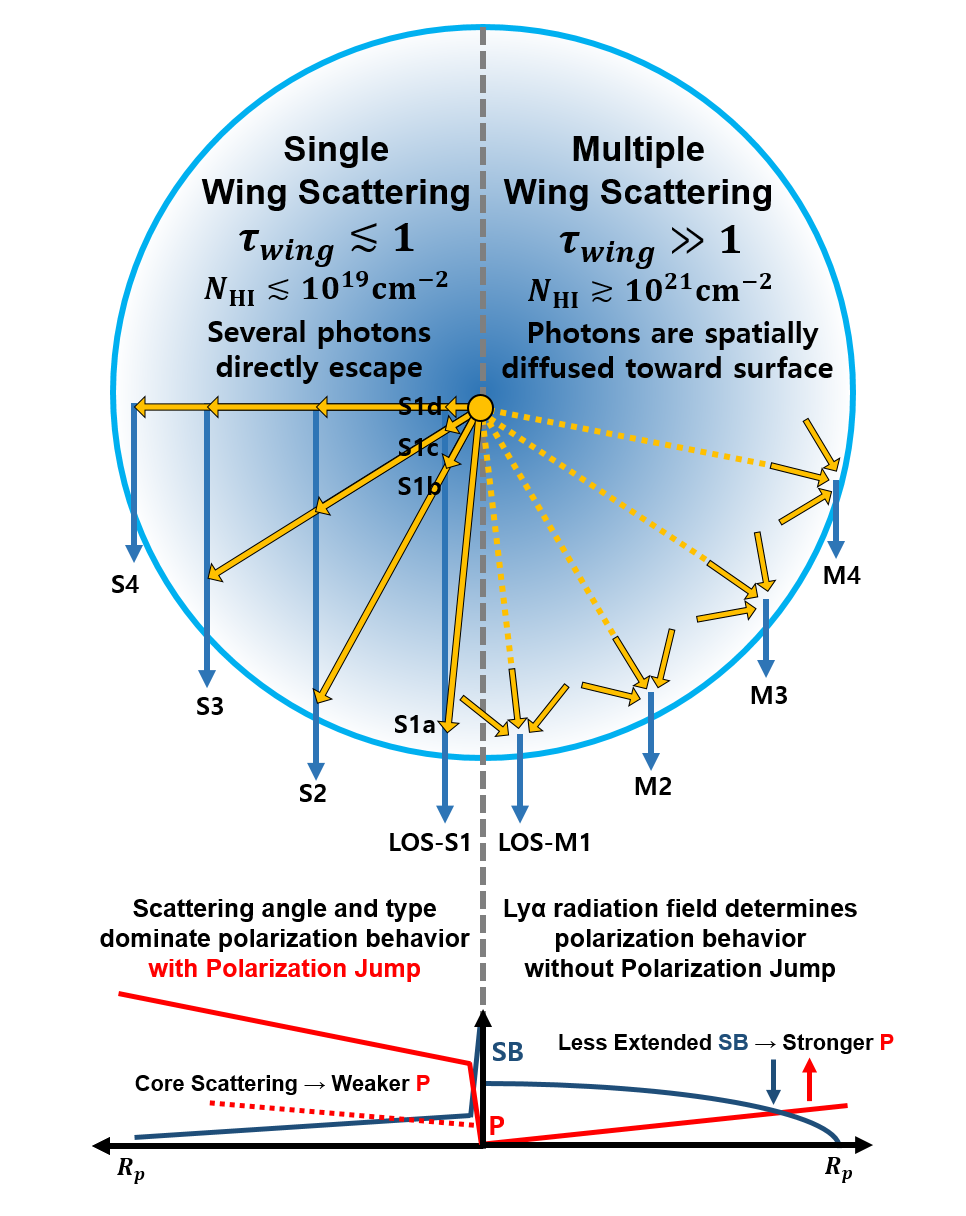}
	\caption{
    Schematic illustration of the smooth model for the cases dominated by a single wing scattering (left side) and by multiple wing scatterings (right side).
	The blue arrows represent the direction of the photon escaping from the \hi halo to observers. The yellow arrows represent the incident direction of the photon at last scattering.
	In the {\it single wing scattering case}, photons from the wing region in Figure~\ref{fig:cross_a} (the optical depth as a function of Doppler factor) can escape with a large scattering angle from deep inside the halo, because $\tau_{init} \lesssim 1$. Furthermore, photons in the far wing region can directly escape without any scattering. Polarization is induced mostly by photon packets with scattering angles close to $90^\circ$ (e.g., photons scattered at LOS S1c and S1d).
	The bright core and ``polarization jump" are produced in this case.
	When \NHI is very small ($\NHI \leq 10^{18} \unitNHI$), even photons from the core region in Figure~\ref{fig:cross_a} can escape through core scattering, decreasing the degree of polarization (red dotted line). 
	In the {\it multiple wing scattering case}, photons are continuously diffused toward the surface of the halo, resulting in
	a flattened surface brightness profile and a gradually increasing polarization profile.
	Polarization in this case is produced by the anisotropy of the radiation field of multiply scattered photons near the surface.
	The symmetry of the incident radiation fields at the last scattering points is gradually broken from LOS-M1 to M4; thus, the polarization gradually increases outward.
	}
	\label{fig:single_multi}
\end{figure*}

\begin{figure}[ht!]
%	\plotone{depth.pdf}
\includegraphics[width=85mm]{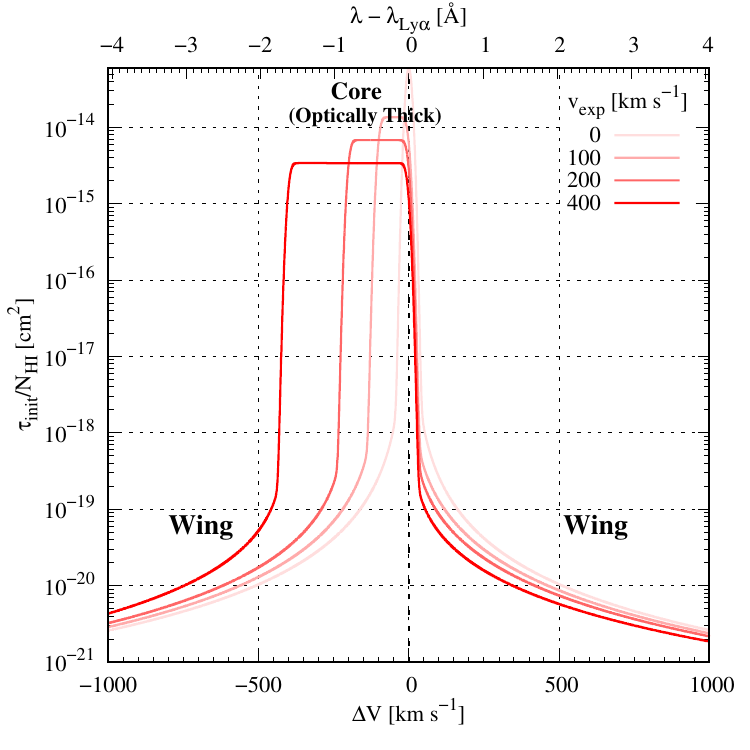}
	\caption{
		Optical depth of the Model S uniform density halo, $\tau_{init}$, given in Equation~\ref{tau_init} as a function of $\lambda - \lambda_{\rm Ly \alpha}$ (top axis) and Doppler factor $\Delta V$ (bottom axis).
		Given that $\tau_{init}$ is proportional to \NHI in %a uniform density halo
		this case (i.e., $R_e/R_H = \infty$), normalized optical depth ($\tau_{init}/\NHI$) is plotted on $y$-axis on a logarithmic scale. 
		The colors of the solid lines represent various \vexp.
		The core ($\Delta V$ = [$-$\vexp, 0]) and the wing regions (outside the core) are represented as Gaussian- and Lorentzian-like cross sections, respectively. At the boundaries of the two regions, $\tau_{init}$ decreases dramatically. The core region is always optically thick for the explored range of \NHI, while the wing regions can be either optically thin at $\NHI = 10^{18-19} \unitNHI$ or optically thick at $\NHI = 10^{21} \unitNHI$.
	}
	\label{fig:cross_a}
\end{figure}

\begin{figure*}[ht!]
\centering
\includegraphics[width=\textwidth]{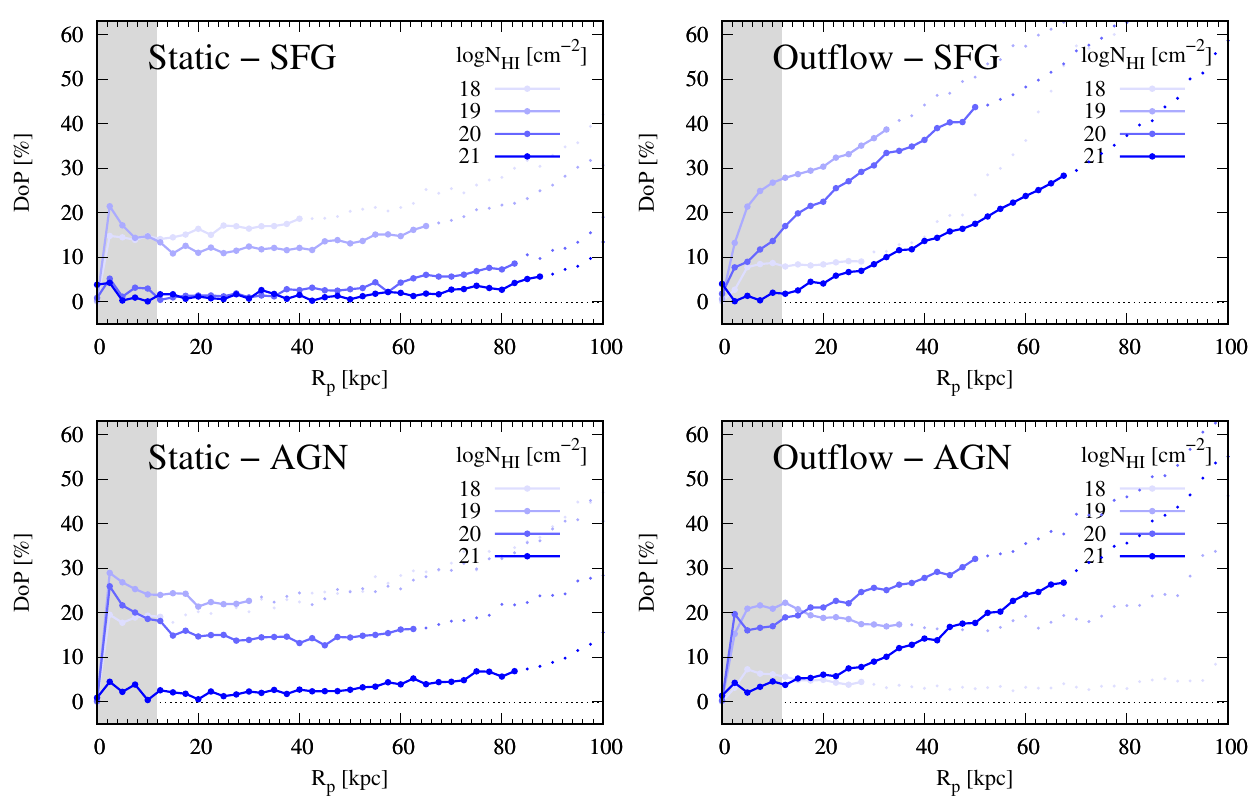}
	\caption{
        Model S degree of polarization profiles, \pol, for $\NHI= 10^{18-21} \unitNHI$.
        The $y$-axis is the degree of polarization, ${\rm DoP} = \sqrt{Q^2 + U^2}/I$.
        The dots with/without solid lines represent the surface brightness over/under the observational thresholds 
        (horizontal dashed lines in Figure \ref{fig:sb_NH_s}).
        The profiles in the top right panel correspond to the second column of Figure~\ref{fig:image}. 
        \pol for the static cases is more flattened than for the outflow cases, because the frequency diffusion by the outflowing medium causes
        photons with angles close to 90$^\circ$ to escape from deeper regions of the halo
        (e.g., LOS-S1c and S1d in the schematic illustration of Figure~\ref{fig:single_multi}). 
        The variation of \pol is not monotonic with \NHI, because the contributions of the three scattering mechanisms (core, single-wing, and multiple-wing scattering) strongly depend on \NHI. 
	}
	\label{fig:pol_NH_s}
\end{figure*}

\begin{figure*}[ht!]
\centering
\includegraphics[width=\textwidth]{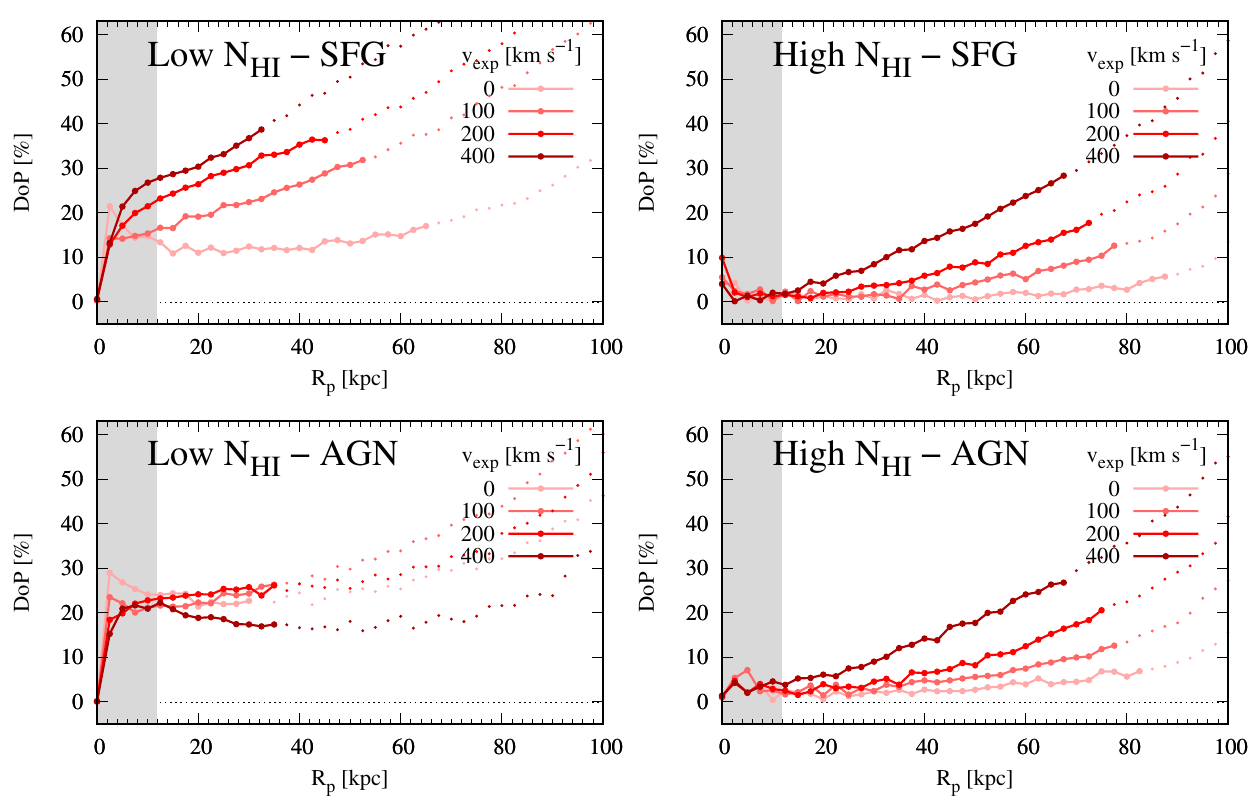}
	\caption{		
	Model S degree of polarization profiles, \pol, for $\vexp = 0 - 400\,\kms$. The parameters in each panel are identical to Figure \ref{fig:sb_vexp_s}.
	Polarization jumps occur at low \NHI due to single wing scatterings.	
	In the left panels,
	\pol for the low \NHI--AGN case is weaker than for the low \NHI--SFG case due to core scattering.
	For High \NHI cases (right panels) where multiple wing scatterings dominate, \pol gradually increases without a polarization jump. The overall polarization increases with increasing \vexp, but \pol does not depend on the input line width (\sigsrc; SFG vs.~AGN). Because \SB in high \NHI cases becomes less extended with increasing \vexp, the more concentrated \SB tends to show stronger polarization. 
	}
	\label{fig:pol_vexp_s}
\end{figure*}

\begin{figure*}[ht!]
\centering
	\includegraphics[width=\textwidth]{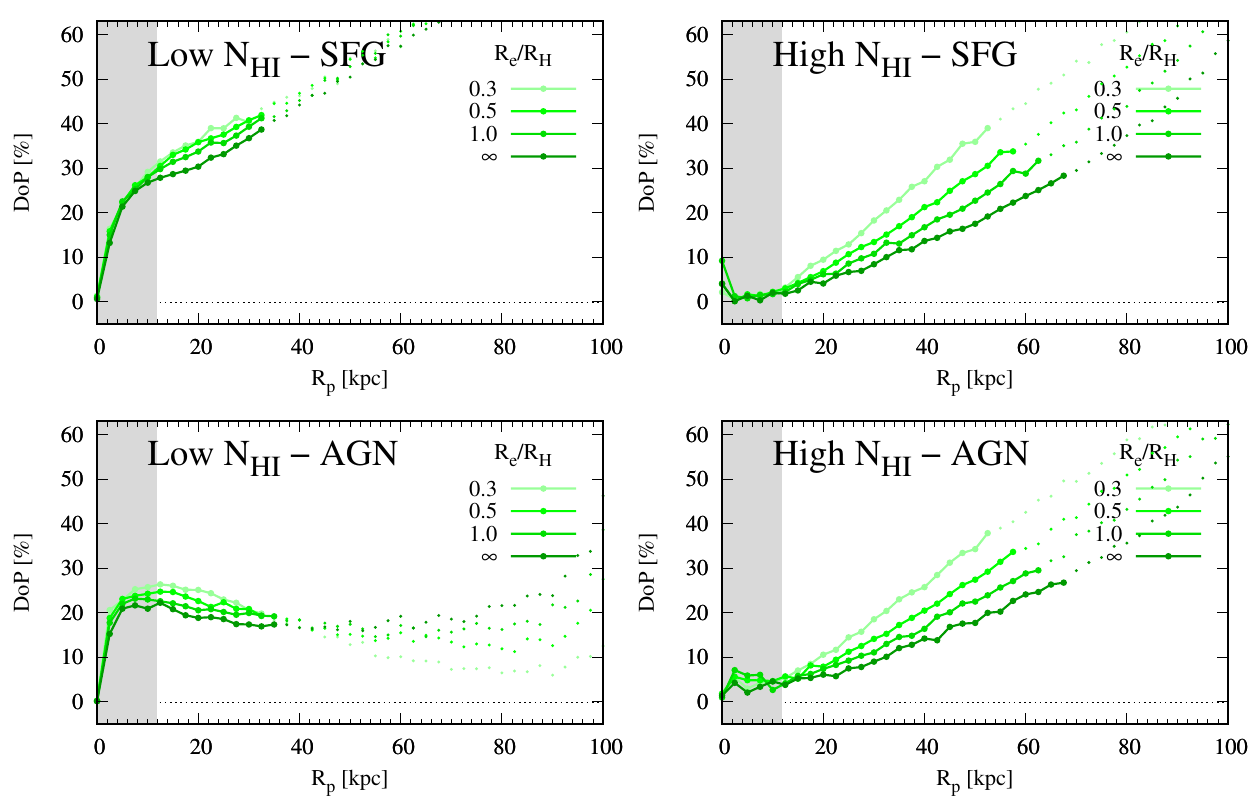}
	\caption{
	Model S degree of polarization profiles \pol for $R_e = 0.3R_H - \infty$ at $\vexp = 400 \kms$.
	The parameters in each panel are identical to Figure \ref{fig:sb_A_s}.
	The overall degree of polarization increases with decreasing $R_e/R_H$, especially at high \NHI.
	However, this dependence 
	on $R_e/R_H$ is much weaker than the dependence on \NHI (Figure~\ref{fig:pol_NH_s}) and \vexp (Figure~\ref{fig:pol_vexp_s}).
	}
	\label{fig:pol_A_s}
\end{figure*}

\begin{figure*}[ht!]
\centering
	\includegraphics[width=\textwidth]{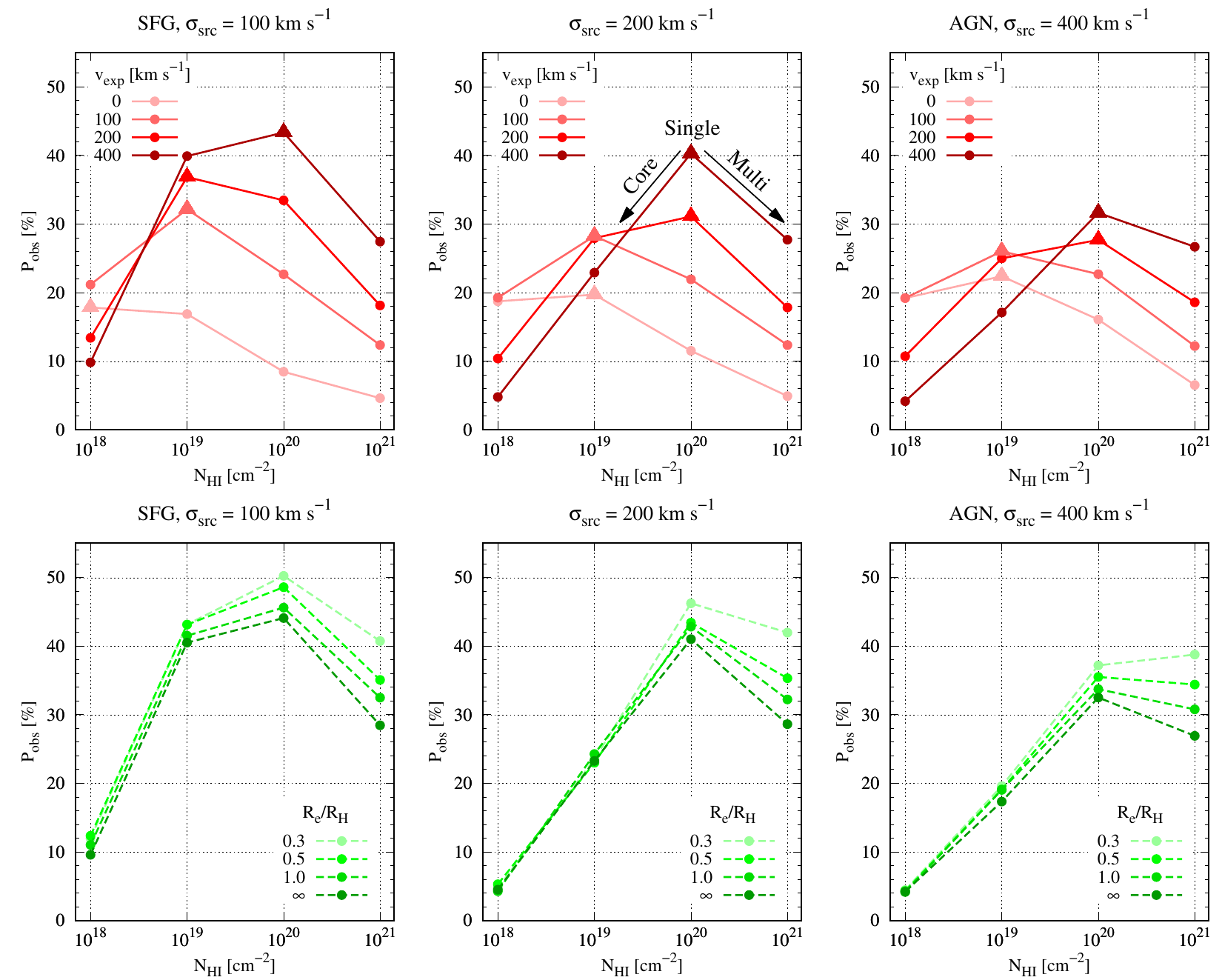}
	\caption{
        Degree of polarization at the observable radius $R_{obs}$, $P_{obs}$ (i.e., ${\rm DoP}(R_{obs})$), for Model S with $\sigsrc = 100 \kms$ (SFG), 200 \kms, and 400 \kms (AGN).
        The red solid and green dashed lines represent 
        the variation of
        $\vexp$ and $R_e/R_H$, respectively, with $R_e/R_H$ = $\infty$ fixed in the upper panels and \vexp = 400 \kms fixed in the bottom panels.
        The triangles in the upper panels indicate when the polarization reaches a maximum for a given outflow speed.
        %% NHI dependence
        $P_{obs}$ is not monotonic as a function of \NHI, a behavior dominated by core scattering at $\NHI = 10^{18} \unitNHI$ and by multiple wing-scattering at $10^{21} \unitNHI$. 
        At $\NHI = 10^{18} \unitNHI$, $P_{obs}$ decreases with increasing \vexp and \sigsrc, because the effect of core scattering becomes stronger.
        At $\NHI = 10^{21} \unitNHI$, $P_{obs}$ does not depend on \sigsrc, because multiple wing-scattering dominates and erases the information of the input photons; larger \vexp produces higher $P_{obs}$.
        At $\NHI = 10^{19-20} \unitNHI$, $P_{obs}$ reaches a maximum due to the dominance of single-wing scattering; the critical \NHI values corresponding to the $P_{obs}$ peak are determined by a complex function of \NHI, \vexp, and \sigsrc.
        In the bottom panels, the dependence on $R_e/R_H$ is not strong enough to change the trends with \NHI, \vexp, and \sigsrc. 
	}
	\label{fig:Pobs_s}
\end{figure*}

\lya scattering processes include three types of scatterings: single-wing, multiple-wing, and core scattering. The polarization in our models can be explained by the relative contributions of these scattering types. 
We show the predicted degree of polarization as a function of projected radius $R_p$, \pol, in Figures~\ref{fig:pol_NH_s}--\ref{fig:Pobs_s}. Here we summarize the findings that are discussed in more detail in the following sections:

\begin{enumerate}[leftmargin=+0.5cm,itemsep=0pt]

\item[$\bullet$]
A polarization jump occurs when single-wing scattering dominates the escape of \lya photons: at $\NHI \lesssim 10^{19} \unitNHI$ in the SFG case and at $\lesssim 10^{20} \unitNHI$ in the AGN case, respectively. (\S~\ref{sec:polarization_jump})

\item[$\bullet$]
Due to the mixing of the three scattering types, the resulting \pol is complex and does not behave monotonically with \NHI, \vexp, or \sigsrc. (\S~\ref{sec:polarization_jump} -- \S~\ref{sec:multiple_scattering})

\item[$\bullet$]
The overall normalization of \pol increases with \NHI, peaks, and then declines. The peak occurs at a characteristic \NHI value, which depends on \vexp and \sigsrc.
(\S~\ref{sec:pol_parameter_s})

\item[$\bullet$]
%$P_{obs}$ (${\rm DoP}$ at $R_{obs}$)
The overall normalization of \pol
increases with increasing \vexp, except for the AGN case at $\NHI \sim 10^{18} \unitNHI$.
(\S~\ref{sec:pol_parameter_s})

\item[$\bullet$]
At \NHI $\sim 10^{21} \unitNHI$, the polarization profile is dominated by \vexp, regardless of \sigsrc. (\S~\ref{sec:pol_parameter_s})

\end{enumerate}

\subsubsection{Polarization of Scattered \lya Photons}

Before we describe the polarizations in Model S in detail, we first explain the behavior of the single- and multiple-wing scattering cases.
We define $\tau_{wing}$ as the optical depth for wing (or Rayleigh) scattering. Figure~\ref{fig:single_multi} shows a schematic illustration for the cases where single- ($\tau_{wing} \lesssim 1$) and multiple- ($\tau_{wing} \gg 1$) wing scattering dominates. 
The solid blue and yellow arrows represent directions of the incident and escaping photons, respectively. The dotted yellow lines indicate the spatial diffusion that photons experience through multiple scatterings prior to the last scattering point.

In the case of  single-wing scattering, when the wavelength of a photon is far from the line center and $\tau_{wing} \lesssim 1$ arises from low \NHI ($\leq$ $10^{19}$\unitNHI), the photon can escape the \hi halo through a single scattering or even without any scattering. In this case, the scattering angle determines the amount of polarization produced by this singly-scattered photon. When the angle is closer to 90\arcdeg, the scattered photon is more strongly polarized.

In the regime determined by multiple-wing scattering, a photon can be wing-scattered numerous times due to the large optical depth ($\tau_{wing} \gg 1$), and it is diffused to the surface of the halo before the last scattering. In this case, the \lya radiation field at the last scattering surface determines the amount of polarization carried by the escaping photon. As the radiation field becomes more isotropic, the escaping photons are more weakly polarized.

The initial wavelength of the photon emitted by the central source ultimately determines the subsequent journey in the radiative process, especially at low column densities. 
To illustrate this, in Figure \ref{fig:cross_a} we show the optical depth of an initial photon $\tau_{init}$ in a uniform halo as a function of wavelength/velocity for different expansion velocities:
\begin{equation}
\label{tau_init}
\tau_{init}(\Delta V) =  \int^{R_H} n_{\rm H I}(r) \sigma_{\nu(r)} dr = {\NHI \over R_H} \int^{R_H} \sigma_{\nu(r)} dr, 
\end{equation}
where $\nu (r)$ is the frequency of \lya considering the radial velocity $v(r)$.
We choose to plot $\tau_{init}/\NHI$ on the $y$-axis, because $\tau_{init}$ is proportional to \NHI. The \lya cross-section of atomic hydrogen follows the Voigt profile.

When the velocity offset ($\Delta V$) is measured from the systemic velocity ($\Delta V = 0 \kms$), $\tau_{init}$ is extremely high and flat in the velocity range, $-\vexp < \Delta V < 0 \kms$.
Outside this region, $\tau_{init}$ dramatically decreases and becomes the Lorentzian function. The two regions, the flat and Lorentzian-like regions, are referred to as ``core'' and ``wing'' regions, respectively. At low column density ($\NHI \lesssim 10^{19} \unitNHI$), the photon in the wing region is optically thin, and single scattering dominates. The photon in the core region has to random-walk to the wing region through several scatterings to escape the \hi halo. At high column density ($\NHI \gtrsim 10^{21} \unitNHI$), both the core and wing region are optically thick. The photons must experience multiple wing scatterings before they escape.

%% The two scattering types are resonance (core) and Rayleigh (wing) scattering.
The type and angle of scattering determine the polarization state of the scattered \lya photon packet \citep{ahn02,chang17, eide18}. 
The photon after wing (Rayleigh) scattering maintains the degree of polarization in forward and backward scattering cases or develops strong polarization if the scattering angle is close to 90$^\circ$. 
Core (resonance) scattering produces unpolarized light ($1s_{1/2} - 2p_{1/2}$ transition) or weak polarization ($1s_{1/2} - 2p_{3/2}$ transition) \citep{stenflo80,lee94}. 
If $\tau_{wing} \ll 1$ ($\NHI \leq 10^{18} \unitNHI$), escaping photons experience only resonance scattering and show weaker polarization (the red dashed line in Figure~\ref{fig:single_multi}).
We will explain the effect of resonance scattering through simulated results in \S~\ref{sec:core_scattering}.

The single- and multiple-wing scattering cases are crucial to understanding the polarization behavior of \lya.
If single-wing scattering dominates in the model, the scattering angle is the key parameter that determines the overall polarization \citep{chang17,seon22}. But the contribution from the core scattering must also be considered for accurate calculation of polarization, particularly when $\NHI \lesssim 10^{19} \unitNHI$.
On the other hand, if the photon experiences multiple wing scatterings, most individual photon packets are entirely polarized ($\sim 100 \%$). In this case, the anisotropy of the \lya radiation field determines the polarization \citep{seon22}.

%% In the case that the core scattering dominates, the photon packets can be weakly polarized (singly scattered) or unpolarized (multiple scattering) depending on the number of scatterings. 

\subsubsection{Polarization Jump from Single Wing Scattering}
\label{sec:polarization_jump}

We find that the {\it polarization jump} originates from the singly wing-scattered photons. On the left side of Figure \ref{fig:single_multi}, we schematically illustrate how the polarization jump develops when  single-wing scattering dominates at low column density.
The photons projected at $R_p \sim 0$ are either directly escaping and unpolarized or scattered by the medium into the line-of-sight. Due to spherical symmetry, the degree of polarization at the center should be zero, even if the line-of-sight includes polarized scattered photons. 

When the line-of-sight diverges from the center (LOS-S1 in Fig.~\ref{fig:single_multi}), the symmetry is broken, and the observed photons are those last-scattered at inside locations of LOS-S1 (e.g., S1a--S1d in Fig.~\ref{fig:single_multi}). Due to the low $\tau_{wing}$, photons emitted from the central source propagate directly to this location, are scattered, and escape the system.
Because photons scattered at an angle close to $90^\circ$ (e.g., S1d) are strongly polarized and can enter the sightline, the degree of polarization steeply increases immediately outside the center. From there, the degree of polarization increases radially outward, because the fraction of photons scattered at angles near 90$^\circ$ increases.

%%
%Because photons directly escaping from the central source are the main reason for the polarization jump, the single scattering case produces a surface brightness profile combining a bright core and faint halo. Our prediction of a polarization jump is interesting, but testing it might be challenging due to several factors: 1) observational seeing would diminish any steep gradient, and 2) spherical symmetry is easily broken in a realistic environment.

The polarization jump is strongest at $\NHI \sim 10^{19} \unitNHI$, regardless of \sigsrc, and disappears at $\NHI \ge 10^{20}$ and \NHI $\ge$ $10^{21} \unitNHI$ in the SFG and AGN cases, respectively (Figure \ref{fig:pol_NH_s}). 
This is because the polarization jump originates from a large contribution of single-wing scattered photons. Figures \ref{fig:pol_vexp_s} and \ref{fig:pol_A_s} show how the polarization profile changes for a range of $\vexp = 0 - 400$\,\kms and $R_e/R_H = 0.3 - 1$, $\infty$, respectively. In the low \NHI cases ($10^{19} \unitNHI$; left panels), the polarization jump can be as high as 15\%--30\%.
As shown in the low \NHI--AGN case (bottom left of Figure \ref{fig:pol_vexp_s}), the polarization jump decreases with increasing \vexp because the contribution of core scattering increases.
%%
%% The level of the polarization jump decreases with increasing \vexp due to increasing contribution of core scattering.
%%

While a polarization jump is not generally expected at high \NHI, a small jump can still exist if the input \lya spectrum has enough photons in the wing region. In the high \NHI--AGN cases (bottom right panels of Figures \ref{fig:pol_vexp_s} and \ref{fig:pol_A_s}), we find small polarization jumps, at the $\sim$5\% level, which would be hard to observe. As shown in Figures \ref{fig:sb_vexp_s} and \ref{fig:sb_A_s}, the small bright cores are still visible in this case, confirming that the core + halo morphology and the polarization jump occur at the same time. 

The polarization jump is an excellent diagnostic to verify the dominance of the single-wing scattering process. Because photons directly escaping from the central source are the main reason for the polarization jump, the single scattering case produces a surface brightness profile combining a bright core and faint halo.
However, the observation of this characteristic feature might be challenging, because the ground-based seeing could erase any steep gradient, as indicated by the gray shaded regions in Figures \ref{fig:pol_NH_s} -- \ref{fig:pol_A_s} ($R_p$ $\lesssim$ 12\,kpc; $\sim$1.5\arcsec). In addition, realistic \hi halos near galaxies do not have completely spherical symmetry. 

\subsubsection{Effect of Core Scattering}
\label{sec:core_scattering}

One of the key features of our \lya RT work is its consistent treatment of core scattering, which must be considered to calculate the polarization correctly. If core scattering is not included in the RT calculation, the overall degree of polarization must increase with decreasing \NHI, because single-wing scattering occurs more frequently as \NHI decreases.
However, our inclusion of core scattering leads to a different behavior---Figure \ref{fig:image} shows that the overall degree of polarization {\it diminishes} from $\NHI=10^{19} \unitNHI$ to $10^{18} \unitNHI$---which likely arises from the many photons that escape through only core scattering due to the small optical depth. Because the core scattering occurs near the line center of scattering atom, blue photons in the core region can experience core scattering.

In Figure \ref{fig:cross_a}, if $\NHI \ge 10^{19}\unitNHI$, $\tau_{init}$ in the wing region near the core region is $\gtrsim 1$. Thus, the initial photons in the core region go through numerous core scatterings, move to the wing region, and escape the system through wing scattering eventually.
However, at $\NHI = 10^{18}\unitNHI$, $\tau_{init}$ is smaller than unity near the core region and $\ll 1$ in the wing region. 
In this case, the photons in the core region escape after going through only several core scatterings, while the photons initially in the wing region can directly escape. Because core-scattered photon is unpolarized or weakly polarized, the polarization becomes weaker due to the contribution from core-scattering. 

Core scattering affects the polarization behavior significantly at low column density ($\NHI \lesssim 10^{19} \unitNHI$). In the right panels of Figure \ref{fig:pol_NH_s}, we find that resonantly escaping photons reduce the overall \pol at $\NHI = 10^{18} \unitNHI$ when the condition for escape through core scattering is met. 
Similarly, if the strong outflow medium surrounds an input source with broad emission (\sigsrc = 400\,\kms), a reasonable fraction of core-escaping photons can induce a decrease in \pol. In the bottom left panel of Figure \ref{fig:pol_vexp_s}, the overall \pol at low column density ($\NHI = 10^{19} \unitNHI$) weakens with increasing \vexp. The effect of core scattering becomes negligible at $\NHI = 10^{20-21} \unitNHI$. In this high \NHI regime, photons can not escape through only core scattering, because the photons have to go through wing scattering to escape from the \hi halo.

\subsubsection{Polarization Profile from Multiple Wing Scattering}
\label{sec:multiple_scattering}

In an optically thick case where multiple wing scattering dominates, a gradually increasing polarization pattern emerges without a polarization jump. 
Figure \ref{fig:pol_NH_s} shows \pol for $\NHI = 10^{18-21} \unitNHI$ and the combinations of \vexp and \sigsrc. The solid lines represent the region within the observable halo radius $R_{obs}$ above the surface brightness threshold, while the unconnected dots indicate the area beyond $R_{obs}$ where observations of the surface brightness and polarization would be extremely challenging.
At high column density ($\NHI=10^{21} \unitNHI$), all profiles gradually and monotonically increase from 0\% at the center to 10--30\% at $R_{obs}$.

In an extremely optically thick case like the static halo with high column density ($\NHI=10^{21} \unitNHI$), most \lya photons are spatially diffused toward the halo's surface through multiple wing scatterings, then scattered for the last time before reaching the observer. In other words, the last scattering surface of the observed photons is close to the halo boundary.
In this case, the surface brightness profile becomes very flat due to the large number of scatterings, and we can only see the surface of the halo. At this last scattering surface, the radiation field of multiply scattered photons becomes anisotropic, because there are very few \lya photons incident from the outside.
%% because there are no \lya photons beyond this radius of the surface.
The symmetry about a radial direction of the radiation field is slowly broken as photons propagate from deep inside the halo radially outward. As a result, this gradual variation of the radiation field causes the degree of polarization to slowly increase with radius.

The anisotropy of the radiation in the vicinity of the halo boundary and the breaking of spherical symmetry determine the degree of polarization in the multiple-wing scattering case.
This polarization behavior can be understood from Figure \ref{fig:single_multi} (right) and is different than for the single-wing scattering case (left side of Figure \ref{fig:single_multi} and \S \ref{sec:polarization_jump}).
In the single-wing scattering case, most incident radiation at the last scattering comes directly from the central source. The spherical symmetry is suddenly broken if the sightline diverges from the center, producing the polarization jump.
However, in the multiple wing scattering case, as the sightline diverges from the center, the spherical symmetry is gradually broken due to the incident radiation from various directions. The average scattering angle of the photons is also more likely to be close to 0\arcdeg\ at LOS-M1. Therefore, the degree of polarization profile shows a gradual increase without a polarization jump. As $R_p$ increases (e.g., LOS-M1 $\rightarrow$ M4), the scattering angle tends to be close to 90\arcdeg; thus, \pol increases radially outward.

\subsubsection{Polarization Dependence on Model Parameters}
\label{sec:pol_parameter_s}

Here we describe the dependence of polarization features on various model parameters: 

\paragraph{Non-monotonic Polarization Profile.}

\pol does not always increase monotonically as a function of projected radius ($R_p$), excluding in the case dominated by multiple wing scatterings ($\NHI \geq 10^{21} \unitNHI)$. This is because the relative contribution of core scattering and the single- and multiple-wing scattering varies over the projected radius.
If \lya photons go through mostly one type of scattering mechanism (core- vs.~single- vs.~multiple-wing scattering), \pol would always increase radially outward. However, in our simulation, the relative contribution of three scattering mechanisms determines the radial polarization profile due to large \sigsrc.

For example, in the static--SFG case with $\NHI=10^{19} \unitNHI$ (top left panel of Figure \ref{fig:pol_NH_s}), the degree of polarization decreases to $\sim$10\% after the high polarization jump (20\%) and then increases gradually at large $R_p$. This non-monotonic behavior is also observed in the static--AGN case with \NHI = $10^{20} \unitNHI$. In these cases, the polarization is dominated by single-wing scattering of the initial photons in the wing region in the inner halo, while multiple-wing scattering dominates in the outer halo. 

In the outflow--AGN case with $\NHI=10^{19} \unitNHI$ (bottom right panel of Figure \ref{fig:pol_NH_s}), the polarization profile is not monotonic due to the contribution of core scattering. 
In the outer halo, blue photons in the core region in the initial source spectrum can escape with only core scattering because their wavelengths are close to the line center of expanding \hi halo. This imprint clear features in the blueward of \lya spectra as will be discussed in Section~\ref{sec:result_spec_s}.
The relative contribution of core- and single-wing scattering becomes more complex with increasing \sigsrc, because the subsequent journey of a photon, i.e., which scatterings a photon will experience in the halo, is mainly determined by the photon's initial wavelength. The broader the width of the source \lya emission, the more different scattering processes can play a role.
As a result, a more complex polarization pattern emerges.

\paragraph{Dependence on \vexp.}

The dependence of the overall degree of polarization ($P_{obs}$) on \vexp is complex and depends on \NHI. In Figure \ref{fig:Pobs_s}, we measure the degree of polarization at the observable halo radius ($R_{obs}$), $P_{obs}$, for three input sources, \sigsrc = 100, 200, and 400\,\kms.
At \NHI $\ge$ $10^{20}$ \unitNHI, the overall degree of polarization tends to increase with \vexp for all three input sources. For example, $P_{obs}$ for the SFG case at $\NHI = 10^{21} \unitNHI$ (also shown in the top right panel of Figure \ref{fig:pol_vexp_s}) increases from $P_{obs}$ $\sim$ 5\% to 28\% when \vexp increases from 0 to 400\,\kms.
At large \vexp, photons can easily escape due to large changes in their wavelengths induced by the fast-moving medium. As shown in Figure \ref{fig:cross_a}, $\tau_{init}$ in the wing region decreases as the outflow becomes stronger. This decrease allows more photons to escape through single-wing scattering, thus increasing the overall polarization strength.

However, this dependence of $P_{obs}$ on \vexp inverts for \NHI = $10^{18} \unitNHI$ due to the increased contribution of photons escaping through core scatterings. In the upper panels of Figure~\ref{fig:Pobs_s}, $P_{obs}$ at $\NHI = 10^{18} \unitNHI$ decreases with increasing \vexp.
At $\NHI \sim 10^{19} \unitNHI$, $P_{obs}$ has a complex dependence on \vexp, because the relative contributions of three scattering mechanisms change depending on \vexp and \sigsrc. Note that the exact \NHI values at which this decrease of $P_{obs}$ by core scattering occurs are different for different input sources (\sigsrc). 
In the low \NHI--AGN case ($\NHI = 10^{19} \unitNHI$; the bottom left panel of Figure~\ref{fig:pol_vexp_s}), \pol for $\vexp = 400 \kms$ is about $\sim10\%$ smaller than for other \vexp's. In this outflow medium, the broad input \lya emission increases the contribution of core escaping photons.

\paragraph{Dependence on $R_e/R_H$.}

Similar to the case of strong outflows, a high \hi density concentration allows the photons to escape from the more inner \hi halo. Note that a higher concentration (small $R_e/R_H$) at fixed column density or mass implies that the density declines faster than for halos with low concentration (large $R_e/R_H$). The radiation field is more anisotropic when more photons escape from the inner halo. In Figure~\ref{fig:pol_A_s}, the overall \pol increases with decreasing $R_e/R_H$.
In Figure~\ref{fig:Pobs_s}, the dependence of the polarization on $R_e/R_H$ in the high \NHI case is stronger than in the low \NHI case (see the bottom panels of Figure~\ref{fig:size_s}).
Unlike \vexp (upper panels), the dependence of $P_{obs}$ on $R_e/R_H$ (lower panels) is not strong enough to change the behavior of $P_{obs}$ with \NHI.

%----------------------------------------------------------------------
\subsection{\lya Spectrum}\label{sec:result_spec_s}

\begin{figure*}[ht!]
\centering
	\includegraphics[width=\textwidth]{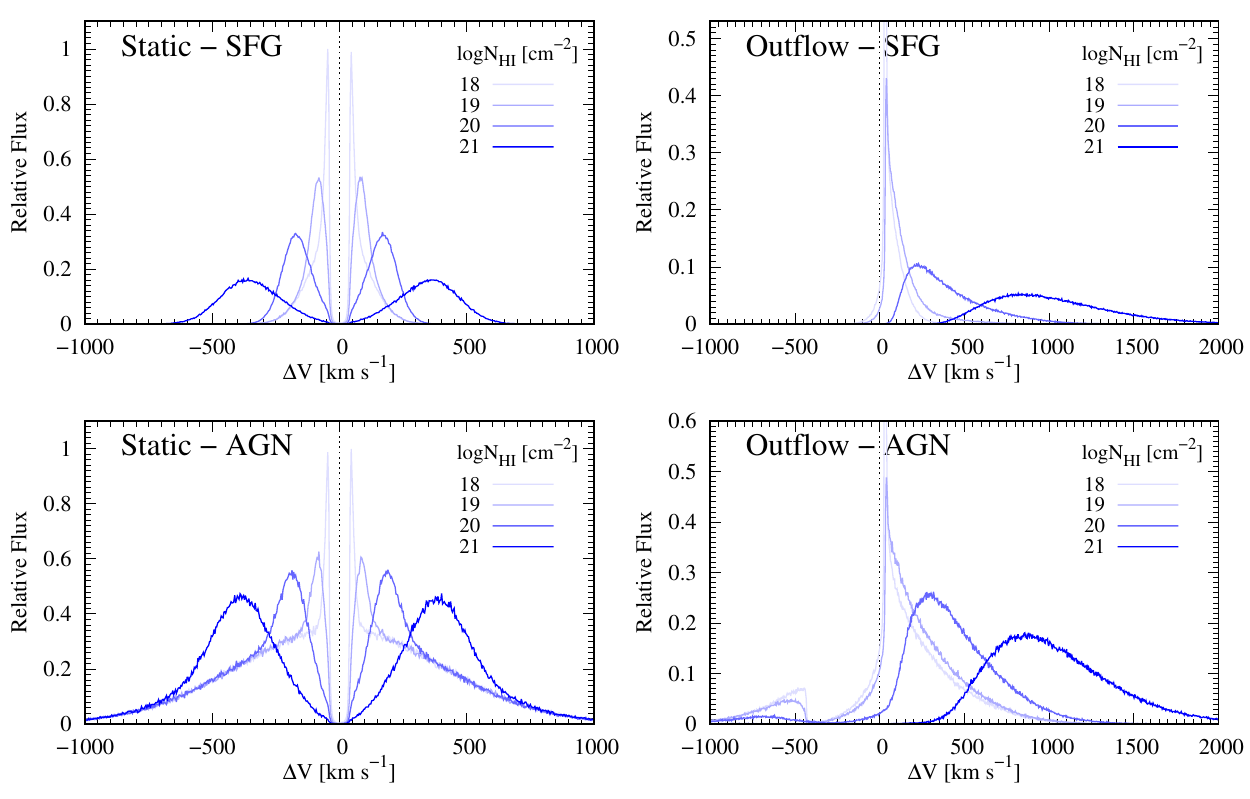}
	\caption{
		Total integrated \lya spectra \spec for Model S with $\NHI = 10^{18-21} \unitNHI$.
		The $y$-axis is the relative flux normalized to the peak value of the $\NHI = 10^{18} \unitNHI$ case.
		The $x$-axis is the Doppler shift $\Delta V$ from the systemic velocity.
		The parameters in each panel are identical to Figure \ref{fig:sb_NH_s}.
		Static and Outflow cases show double peaks and red enhanced peaks, respectively.
		As \NHI increases, the spectral peaks shift more from the systemic velocity. 
		The spectral profiles also depend on the type of source, e.g., the broad wing over $\pm$ 500\,\kms in Static--AGN case originates from the direct escape of input photons.
	}
	\label{fig:spec_NH_s}
\end{figure*}

\begin{figure*}[ht!]
\centering
	\includegraphics[width=\textwidth]{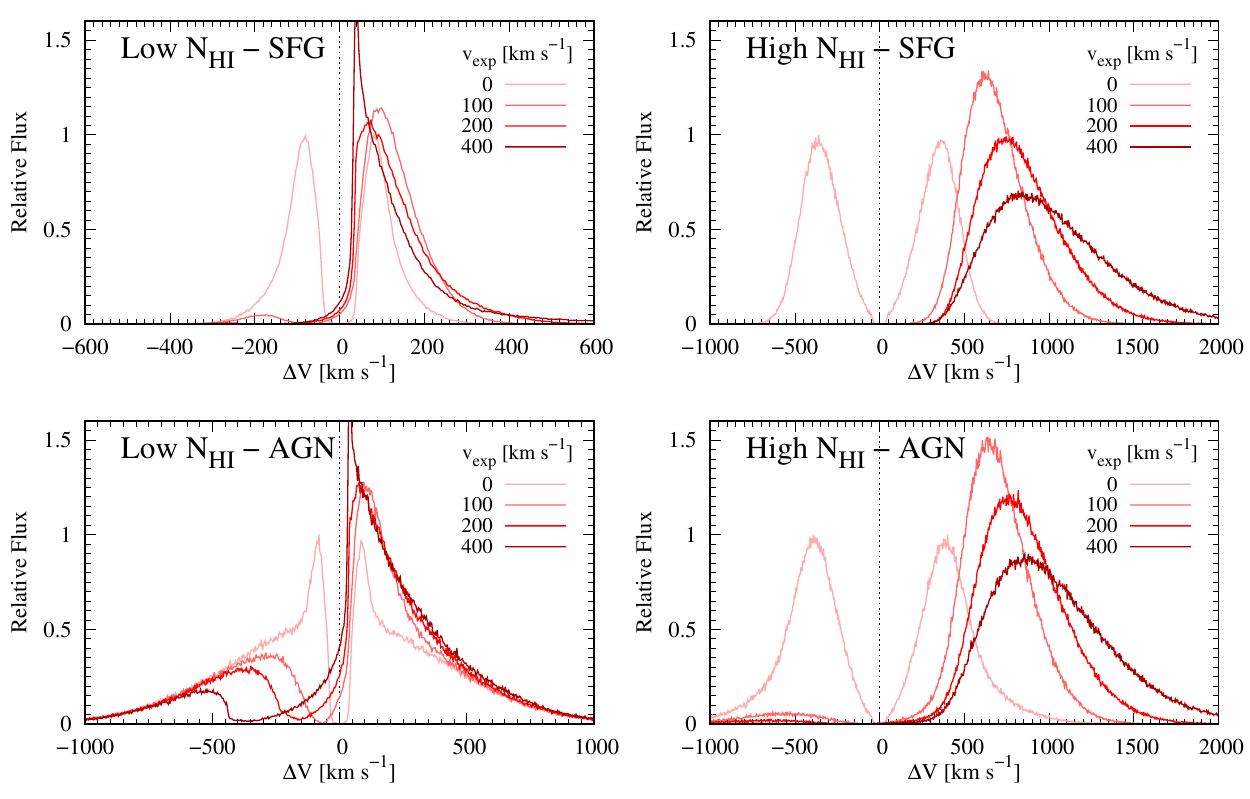}
	\caption{	
	    Total integrated \lya spectra \spec for Model S with \vexp = 0 -- 400 \kms. 
	    The parameters in each panel are identical to Figure \ref{fig:sb_vexp_s}.
	    For High \NHI, the profiles do not depend on the input source (\sigsrc), much as \SB and \pol are insensitive to \sigsrc (Figures~\ref{fig:sb_vexp_s} and \ref{fig:pol_vexp_s}). This insensitivity arises because multiple wing scatterings destroy the information associated with the input source.
	    On the other hand, the blueward spectra in the low \NHI cases depend on the input source types.
	    Low \NHI--AGN case shows absorption features at velocity ranges corresponding to the optically thick core region in Figure~\ref{fig:cross_a}. These absorption features broaden with increasing \vexp.
	}
	\label{fig:spec_vexp_s}
\end{figure*}

\begin{figure*}[ht!]
\centering
	\includegraphics[width=\textwidth]{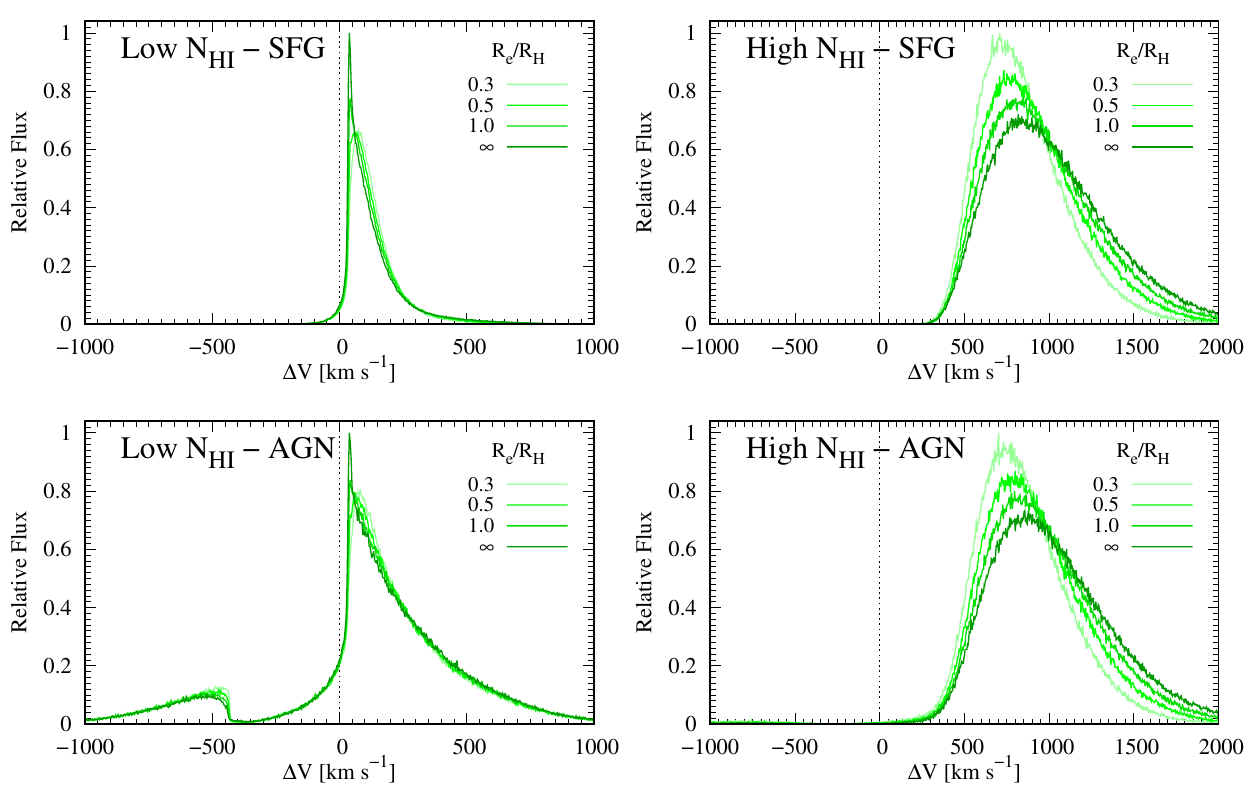}
	\caption{
	    Total integrated \lya spectra \spec for Model S with $R_e/R_H = 0.3 - \infty$.
	    The parameters in each panel are identical to Figure \ref{fig:sb_A_s}.
	    The dependence on $R_e/R_H$ is negligible for low \NHI.
	    For high \NHI (right panels), the \lya peak becomes more extended and redshifted with increasing $R_e/R_H$. 
	    A higher $R_e/R_H$ means a higher \hi number density in the outer region, therefore increasing the number of photons scattered by faster outflowing medium occurs causing more diffusion in frequency space.
	    However, this dependence is much weaker than the dependence on \NHI and \vexp (Figures~\ref{fig:spec_NH_s} and \ref{fig:spec_vexp_s}).
	}
	\label{fig:spec_A_s}
\end{figure*}

\begin{figure*}[ht!]
\centering
	\includegraphics[width=\textwidth]{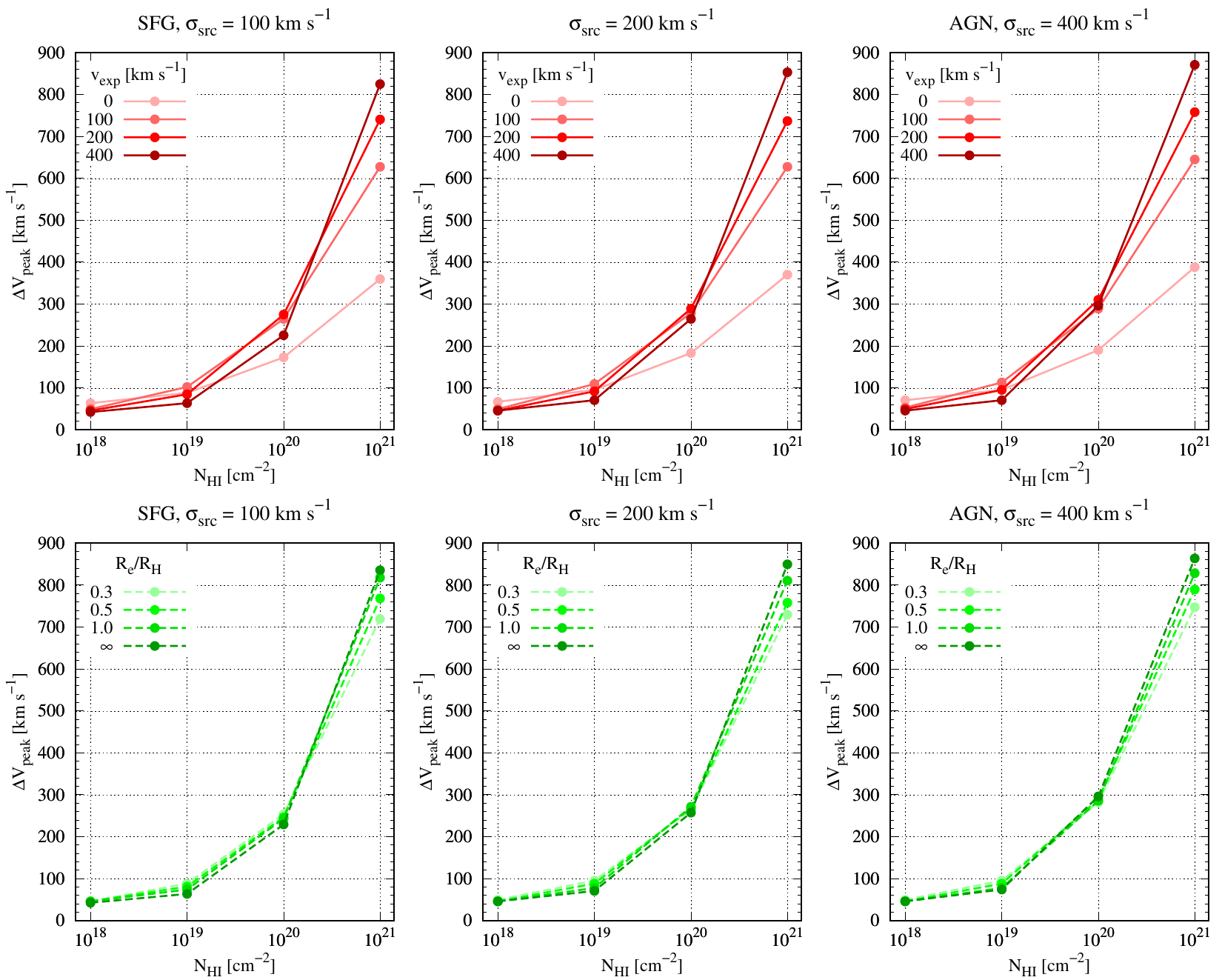}
	\caption{	
            Velocity offset of the spectral peak \dvpeak for Model S with $\sigsrc$ = 100 \kms (SFG), 200 \kms, and 400 \kms (AGN) from the left to the right columns.
            The red solid and green dashed lines represent various $\vexp$ and $R_e$, respectively. \NHI is the most dominant parameter; 
        %such that
        \dvpeak increases with increasing \NHI.
            In general, \dvpeak also increases with increasing \vexp. The dependence on \vexp is negligible at \NHI $\leq 10^{20} \unitNHI$ ({\it top panels}), but becomes stronger at $\NHI = 10^{21} \unitNHI$, where the multiple wing scattering process dominates. \dvpeak is insensitive to \sigsrc and $R_e/R_H$ in all cases.
	}
	\label{fig:peak_s}
\end{figure*}

\subsubsection{Origin}

Before we proceed to describe the \lya spectra in detail, we provide a brief summary of \lya line formation. 
Scatterings cause not only spatial diffusion, but also a broadening and shift of \lya emission lines. \lya line photons are transferred through diffusion in both frequency space and real space. In a static medium, a typical frequency shift resulting from each scattering is comparable to the thermal motion. Because escape is made through diffusion into the wing regions in frequency space after a large number of local core scatterings, the resultant profile is characterized by two peaks symmetric about the line center. The peak separation and peak widths
increase as the scattering optical depth increases.

In a static medium, the scattered \lya photon wavelength changes by random amounts from one scattering to another. The width of these changes is determined by the thermal motion of atoms. The spectrum from the static medium shows a characteristic symmetric double peak. The width of each peak and the separation between them become larger with increasing optical depth \cite[e.g.,][]{neufeld90,ahn02}.

In an expanding medium, the diffusion process in frequency space becomes asymmetric, systematically enhancing redward frequency shift and suppressing blueward shift, which leads to formation of \lya line profiles characterized by a weak blue peak and strong red peak \citep{zheng02,verhamme06,dijkstra08}.
Figure \ref{fig:spec_NH_s} shows the integrated \lya spectra for $\NHI = 10^{18-21} \unitNHI$ and the combinations of \vexp and \sigsrc. We confirm that the spectra are broadened with increasing \NHI at the same \vexp. The static (left panels) and outflow medium (right panels) show double-peaked and red asymmetric profiles, respectively.

In our model, we adopt a Hubble-flow-like velocity field such that the outflow velocity is proportional to the distance from the central source. In this case, unlike the thin shell geometry often studied in the past \citep{dijkstra08}, the scattered photons are always redshifted, as explained below. 
The scattering medium is always expanding at any position toward any line of sight. Thus, the optical depth profile follows a profile similar to $\tau_{init}$ in Figure \ref{fig:cross_a}, consisting of a flat core and Lorentzian-like wing region. When an incident photon of wavelength $\lambda_{in}$ is scattered by a hydrogen atom, the wavelength of the scattered photon $\lambda_{s}$ is given by
\begin{equation}\label{eq:lambda}
\lambda_{s} = \lambda_{in} [1 + {{v_{rel}} \over c}{(1 - {\bf k_{in}} \cdot {\bf k_s})}],
\end{equation}
where ${\bf k_{in}}$ and ${\bf k_s}$ are the wavevectors of the incident and scattered photon, respectively. Here $v_{rel}$ is the relative velocity between the current and previous scattering position. In the Hubble-flow-like outflow, $v_{rel}$ is always positive, because ${\bf k_{in}}$ and the relative velocity are along the same direction. The forward scattering does not change the wavelength, while the backward scattering causes a Doppler shift toward $+2v_{rel}$. Therefore, the wavelength becomes longer as long as the photon is scattered in the outflowing medium.

\subsubsection{Velocity Offset of \lya Line Peak}

In this section, we carry out a systematic study of \lya spectra for the various parameters in Model S. Figures \ref{fig:spec_NH_s} -- \ref{fig:spec_A_s} show the total integrated \lya spectra \spec that we produce by summing all of the photons from the central source. In Figure \ref{fig:peak_s}, we show how the velocity offset (\dvpeak) of the line peak varies as a function of \vexp, \sigsrc, and $R_e/R_H$. Note that \dvpeak should vary as a function of distance from the center, but, for simplicity, we adopt a single value of \dvpeak measured from the integrated spectrum. Our findings are:

\begin{enumerate}[leftmargin=+0.5cm,itemsep=0pt]

\item[$\bullet$]
The total column density (\NHI) is the dominant parameter affecting \dvpeak. The velocity offset of the line peak increases with increasing \NHI. 

\item[$\bullet$]
At $\NHI \sim 10^{21} \unitNHI$, the line peak moves redward with increasing \vexp and $R_e/R_H$.

\item[$\bullet$]
At $\NHI \lesssim 10^{19} \unitNHI$, the dependence of \dvpeak on \vexp is negligible. 
%% At $\NHI \lesssim 10^{19} \unitNHI$, the dependence on \vexp is the opposite, although negligible. As \vexp increases, the line peak becomes closer to the line center.

\item[$\bullet$]
In the AGN case with $\NHI \leq 10^{19} \unitNHI$, the absorption feature appears in the blue region at $\Delta V = -\vexp$ due to the outflow. 

\item[$\bullet$]
The spectral width of the central source (SFG vs. AGN) does not affect the velocity offset of the line peak.

\end{enumerate}

\paragraph{Dependence on \NHI.}
First, we confirm the basic trend of \lya RT that \lya lines become broader, and the line peaks are shifted further in velocity space, as \NHI increases. \NHI is the most dominant parameter affecting \dvpeak of the resulting \lya profiles.
In the static medium (the left panels of Figure \ref{fig:spec_NH_s}), the separations between the double peaks increase, while both red- and blue-side profiles broaden with increasing \NHI. Note that \spec in the AGN case has much broader wings up to $\sim\pm$1000\,\kms than the SFG case due to directly escaping photons at $\Delta V > 500 \kms$. However, the separations between the double peaks in both cases are similar at a given \NHI. 
In the presence of outflows (right panels), \lya lines are redshifted. For example, the line peaks appear at $\Delta V \sim 800 \kms$ for $\NHI = 10^{21} \unitNHI$.

\paragraph{In the high column density regime.}
At $\NHI \gtrsim 10^{21} \unitNHI$, where multiple wing-scattering dominates, the \lya line peaks become broader and more extended to the red with larger velocity offsets as \vexp and $R_e/R_H$ increase.
At $\NHI \sim 10^{21} \unitNHI$, $\tau_{init}$ in the wing region (Figure \ref{fig:cross_a}) is large enough to cause additional wing scattering. Although the scattered photon's wavelength is in the wing region, multiple scatterings are required for the photons to escape the system. Hence, the \lya spectrum in the outflow medium shows an asymmetric profile with a single red peak.

In Figure \ref{fig:spec_vexp_s}, we show how \lya spectra vary with $\vexp = 0-400 \kms$ for the combinations of \NHI and \sigsrc. In the right panels ($\NHI = 10^{21} \unitNHI$), \spec becomes broader and more extended to the redward with increasing \vexp. When the photons are multiply scattered in the wing region, the strong outflow causes large velocity changes in each scattering. 

Figure \ref{fig:spec_A_s} is the same plot, but shows the dependence of the \lya spectra on gas concentration ($R_e/R_H$ = 0.3, 0.5, 1, and $\infty$). In the right panels ($\NHI = 10^{21} \unitNHI$), the line peak is shifted more redward with increasing $R_e/R_H$ because the larger $R_e/R_H$ implies more neutral gas at large distance; thus, the photons are more likely to be scattered by the faster medium in the outer halo.
We find that the dependence of \dvpeak on $R_e/R_H$ is weaker than the trend with \vexp.

\paragraph{In the low column density regime.}
At $\NHI \lesssim 10^{19} \unitNHI$ (left panels in Figure \ref{fig:spec_vexp_s} and \ref{fig:spec_A_s}), where most of the photons can escape through single wing scattering, the resulting spectra have very small offsets (\dvpeak = 0 -- 100\,\kms) and become almost indistinguishable over a range of \vexp and $R_e/R_H$, especially redward.
For example, in the right panels in Figure \ref{fig:spec_NH_s}, the spectra with $\NHI = 10^{18-19} \unitNHI$ have velocity offsets close to zero.

We note the tendency of the red peak to move closer to the systemic velocity with increasing \vexp (left panels of Fig.~\ref{fig:spec_vexp_s}). Although this trend is too weak to be observed, this behavior is the opposite of that of the high column density ($\NHI = 10^{21} \unitNHI$) case and provides insight into the details of the \lya RT. 

This opposite \dvpeak -- \vexp trend arises because the photons in the red wing can directly escape from the system at low column density.
As shown in Figure \ref{fig:cross_a}, the optical depth of initial red photons $\tau_{init}$ ($\Delta V$ $>$ 0\kms) is smaller than unity in the low column density regime. 
Because of this small optical depth, most of \lya photons in the red wing of the input spectrum can directly escape, leading to the observed asymmetric profiles with small velocity offsets. As \vexp increases, the overall optical depth in this red wing decreases (Figure~\ref{fig:cross_a}), and the line peak gets closer to the systemic velocity and also sharper. 
While photons escaping through back-scattering with large scattering angles ($>$90$^{\circ}$) appear at $\Delta V \sim \vexp$ in the spectra, there are too few to
affect the \dvpeak of the integrated spectrum.

At low column density, $R_e/R_H$ affects the only sharpness of the line peak. Figure \ref{fig:spec_A_s} shows \lya spectra for $R_e/R_H = 0.3$ to $\infty$, respectively. The left panels indicate that the peaks become sharper with decreasing $R_e/R_H$.

%% Absorption features in AGN case
\paragraph{Absorption features in AGN case.}
In the AGN case with low column density ($\NHI \lesssim 10^{19} \unitNHI$), the outflows imprint blue absorption features on the \lya spectra.
In the bottom left panel of Figure \ref{fig:spec_vexp_s}, this blue absorption feature becomes broader and more blueshifted with increasing \vexp. This feature originates from the optically thick core region, i.e., the central flat part in Figure \ref{fig:cross_a} that is stretched from $\Delta V$ = $-$\vexp to 0\,\kms.

This increasing strength of blue absorption as a function of $\vexp$ can explain why the surface brightness profiles in the low \NHI--AGN cases of Figure~\ref{fig:sb_vexp_s} are more extended in the strong outflow. 
Because the outflowing medium is optically thick to the initial photons in $\Delta V$ = [$-$\vexp, 0], these photons are scattered into the outer part of the halo. Given that most of the scattering occurs near the $\Delta V$ = 0\kms of the atom's rest frame, the scattering probability near the surface of the \hi halo increases when the photon wavelength approaches $-\vexp$.
This behavior is in contrast to the SFG case (top left panel of Figure \ref{fig:spec_vexp_s}), where blue absorption features are not produced; because \sigsrc is smaller than $\vexp$, the initial \lya spectrum does not cover the optically thick core region.

%% Summary & \sigsrc
\paragraph{Dependence of \dvpeak on other parameters}
In Figure \ref{fig:peak_s}, we summarize how the velocity offset \dvpeak depends on \vexp and $R_e / R_H$ for \sigsrc = 100 (SFG), 200, and 400 (AGN)\,\kms. We find that (1) \dvpeak increases with increasing \NHI, (2) \dvpeak increases with increasing \vexp and $R_e / R_H$ at high column density ($\NHI \sim 10^{21} \unitNHI$), (3) \dvpeak is  insensitive to \vexp and $R_e / R_H$ at $\NHI \lesssim 10^{19} \unitNHI$, and (4) \dvpeak does not depend on \sigsrc.

\subsection{Summary of Model S Results}

In our smooth medium model (Model S), we explore how various observables (surface brightness, velocity profile, and polarization) depend on the total \hi column density, the most dominant parameter.
As \NHI increases, we find that
%:
%\begin{enumerate}[leftmargin=+0.5cm,itemsep=0pt]
%\item[(1)]
(1) the surface brightness becomes more extended and flattened (Figure~\ref{fig:sb_NH_s});
%\item[(2)]
(2) the velocity offset (\dvpeak) and the line width of the \lya spectrum increase (Figure~\ref{fig:spec_NH_s});
%\item[(3)]
(3) however, the polarization behavior is more complex and does not monotonically vary as a function of \NHI (Figure~\ref{fig:pol_NH_s}).
Furthermore, the low and high column density cases show the different properties and dependence on other parameters.
In the low column density regime ($\NHI \leq 10^{19} \unitNHI$), the velocity offset of the line peak does not depend on the expansion velocity \vexp, the surface brightness strongly depends on the type of embedded source, and the polarization decreases due to core scattering.
In contrast, at high column density (\NHI $=10^{21} \unitNHI$), the properties of \lya halo depend on the kinematics and density profiles of the \hi halo, regardless of the input sources. As \vexp decreases, the surface brightness profile becomes more extended, the degree of polarization increases, and the velocity offset decreases.
In the case of the polarization,
The contributions of core, single-wing, and multiple-wing scattering determine the overall \pol and its gradient
(Figures~\ref{fig:single_multi} and \ref{fig:Pobs_s} for the schematic illustration and predicted $P_{obs}$, respectively).

\begin{figure}[t!]
	\includegraphics[width=85mm]{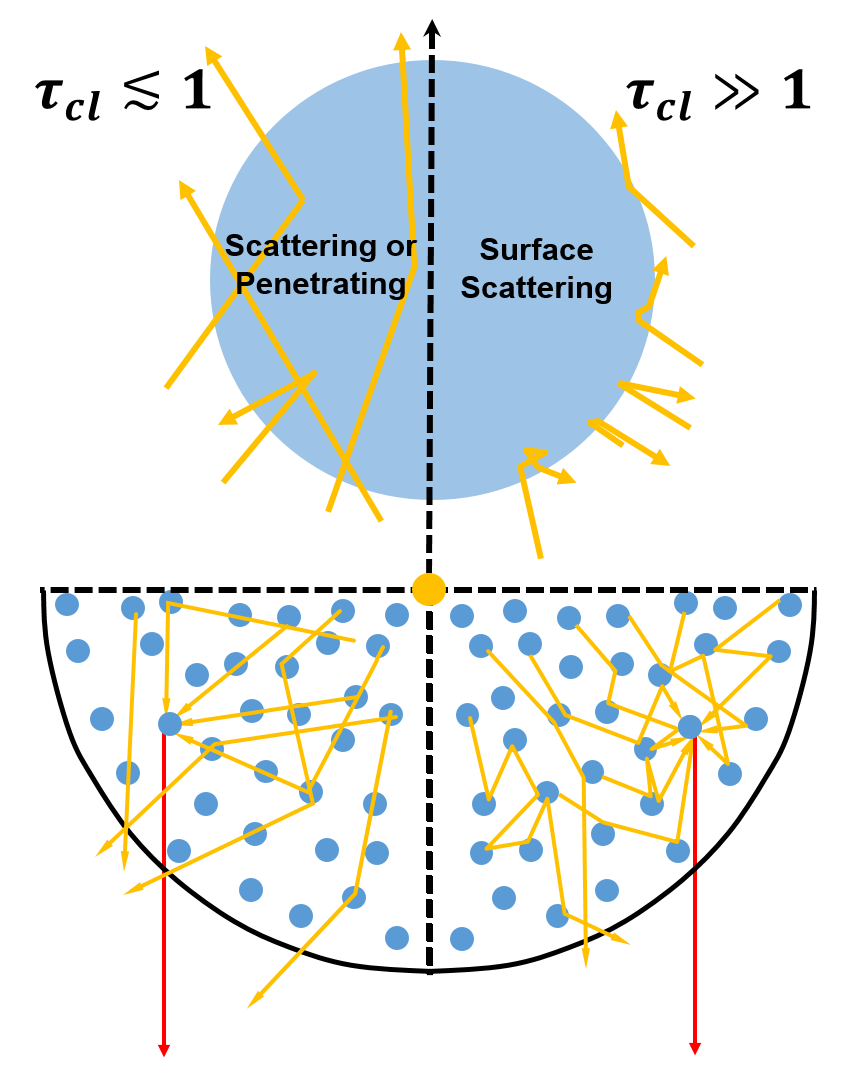}
	\caption{
        Schematic illustration for the clumpy model (Model C), where the clump optical depth is $\tau_{cl} \lesssim 1$ (left side) and $\gg 1$ (right side). The upper and bottom panels represent the behavior of scattered \lya with respect to a single \hi clump and within the entire \hi halo, respectively. 
        The yellow arrows are paths of scattered \lya photons. 
        The red arrows show the escaping direction from the last scattering clumps.
        For $\tau_{cl} \lesssim 1$, the photon can penetrate the clump or is scattered inside the clump.
        For $\tau_{cl} \gg 1$, scattering occurs at the surface of the clump (``surface scattering''). 
        At the last scattering point, the incident radiation field (yellow arrows) in the $\tau_{cl} \gg 1$ case is more isotropic than for $\tau_{cl} \lesssim 1$, because a large number of surface scatterings can produce photons traveling in the backward direction.
        Thus, surface scattering decreases the degree of polarization for $\tau_{cl} \gg 1$.
        }
	\label{fig:surface_scattering}
\end{figure}

\begin{figure*}[ht!]
	\includegraphics[width=\textwidth]{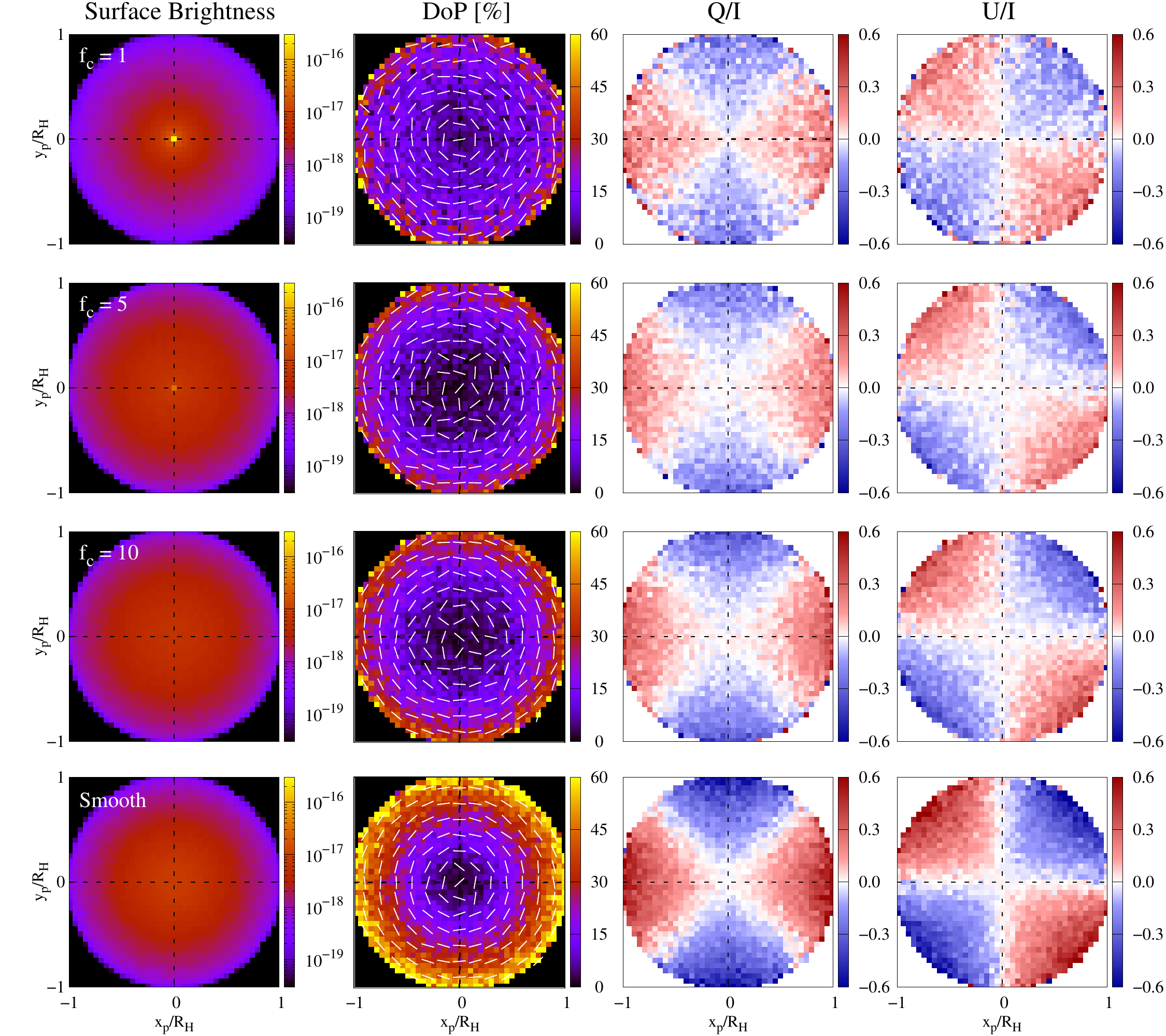}
	\caption{
		Projected surface brightness and polarization maps of the clumpy medium (Model C) for $f_c = 1$ (first), 5 (second), and 10 (third), as well as for the smooth medium (fourth row) at $\NHI = 10^{21} \unitNHI$ for $\vexp = 400$ \kms and $\sigsrc = 100$ \kms (SFG).
		This figure is equivalent to Figure~\ref{fig:image} for Model S.
		The fourth row in this figure corresponds to the fourth row of Figure~\ref{fig:image}.
		When $f_c \ge 5$, the surface brightness is identical to that of smooth medium.
		The polarization decreases with decreasing $f_c$ due to surface scattering, as explained in 
		Section \ref{sec:result_c} and Figure~\ref{fig:surface_scattering}.
		A clumpy medium with $f_c = 5$ is enough to describe the spatial diffusion in the smooth medium at the same \NHI;
		the left panel of the second row shows a surface brightness map like the smooth medium in the fourth row.
		At $f_c = 1$, a bright core exists, and the surface brightness is clearly less extended than that of the smooth medium model.
	}
	\label{fig:image_c}
\end{figure*}

\begin{figure*}[ht!]
\centering
	\includegraphics[angle=90,origin=c,width=152mm]{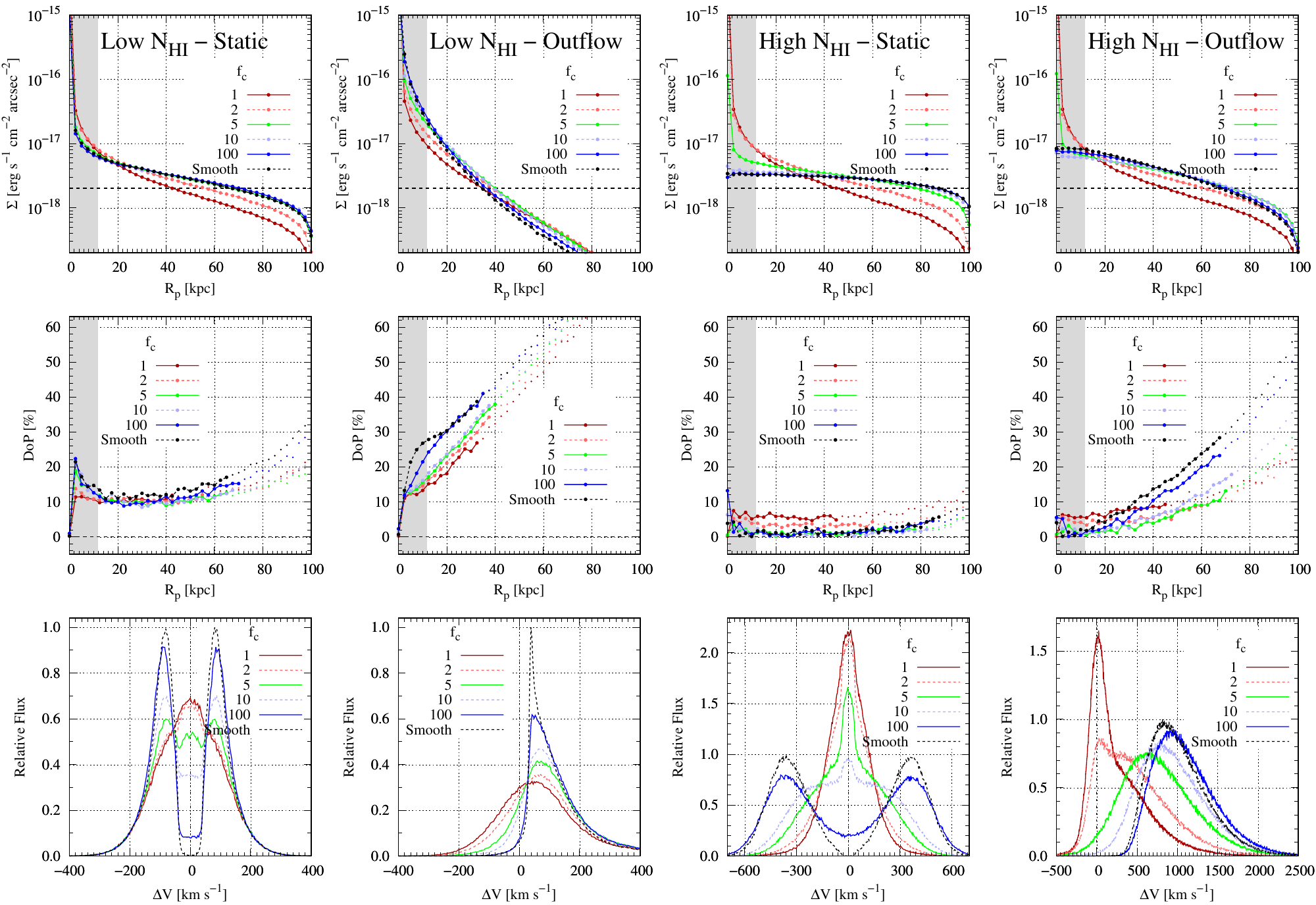}
	\caption{
  Surface brightness \SB, polarization \pol, and integrated spectra \spec (from top to bottom) of the clumpy model (Model C) with a covering factor $f_c = 1-100$ compared with the smooth model (Model S) with uniform density (i.e., $R_e/R_H = \infty$) at the same \NHI and \vexp. The central source is an SFG with $\sigsrc = 100 \kms$. ``Low'' and ``High'' \NHI cases represent $\NHI = 10^{19}$ and $10^{21} \unitNHI$, respectively. The ``Static'' and ``Outflow'' cases represent $\vexp = 0$ and 400\,\kms, respectively. These cases are identical to those in Figure~\ref{fig:sb_NH_s} and \ref{fig:sb_vexp_s}. High \NHI cases show stronger dependence on $f_c$,
  because the effect of surface scattering is more significant in the high \NHI regime at the same $f_c$. Polarization decreases with decreasing $f_c$ due to surface scattering (see Section \ref{sec:clumpy_high_fc} and Figure~\ref{fig:image_c}). The results for $f_c = 1-2$, especially \spec, are distinct from those of Model S: (1) spectral peaks are located close to the systemic velocity, and (2) bright cores and polarization jumps exist at the same time.
	}
	\label{fig:compare_SFG}
\end{figure*}

\begin{figure*}[ht!]
\centering
	\includegraphics[angle=90,origin=c,width=160mm]{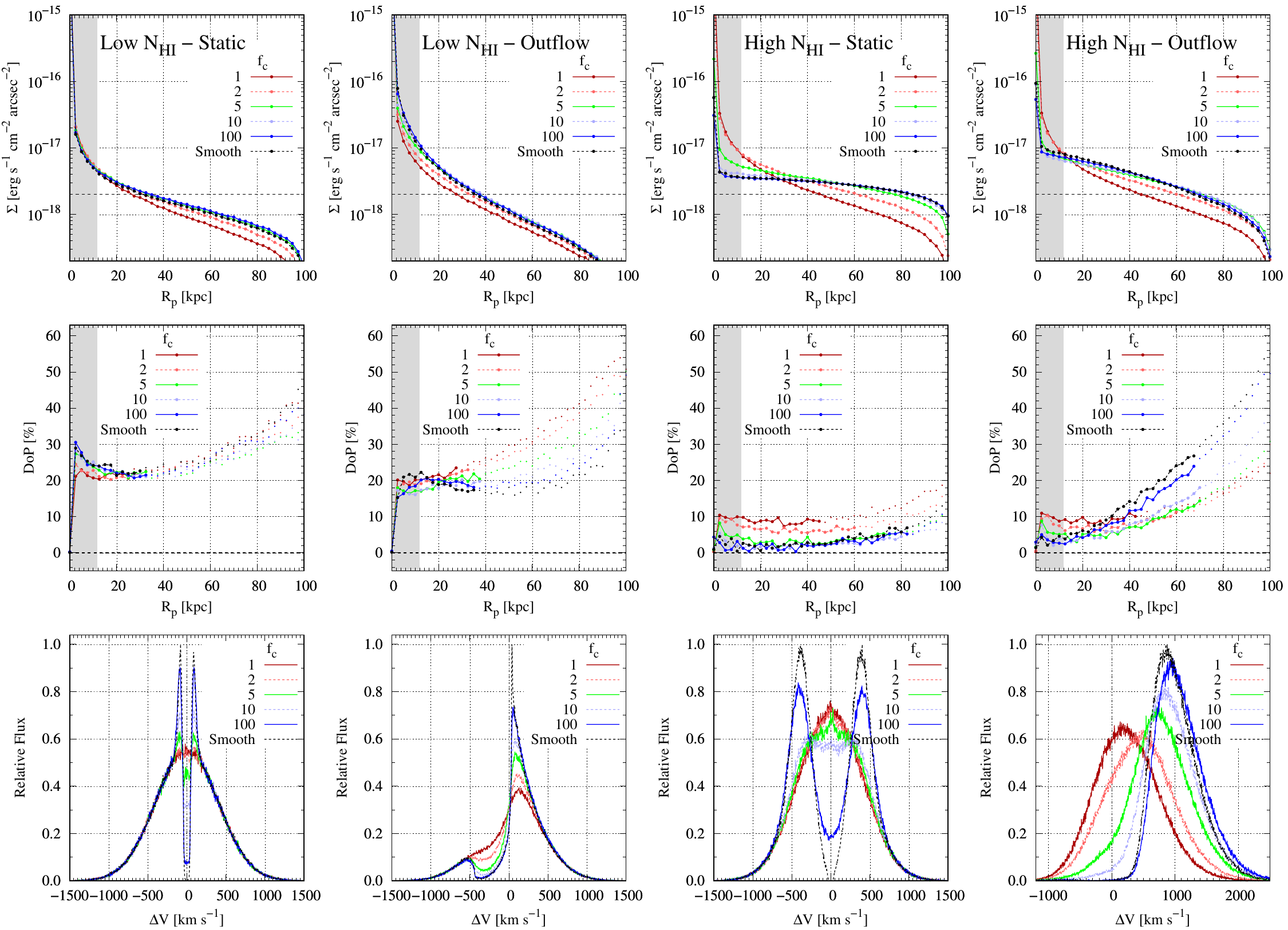}
	\caption{
        \SB, \pol, and \spec of the clumpy model (Model C) compared with Model S for the AGN case (\sigsrc = 400 \kms) with other parameters identical to Figure~\ref{fig:compare_SFG}. When $f_c$ decreases, the behaviors of \SB, \pol, and \spec are similar to those of the SFG case in Figure~\ref{fig:compare_SFG}.
        }
	\label{fig:compare_AGN}
\end{figure*}

\begin{comment}

\begin{figure*}[ht!]
\centering
	\includegraphics[angle=90,origin=c,width=160mm]{r_cl.pdf}
	\caption{
    \SB, \pol, and \spec (from top to bottom) of Model C for a range of the clump size $r_{cl}$.
    The line colors represent $\NHIcl = 10^{16}$ (red), $10^{17}$ (blue), $10^{18}$ (green), and $10^{19}$ (yellow) \unitNHI.
    The line styles represent $r_{cl}$ = 10\,pc (solid line), 100\,pc (dashed line), and 1\,kpc (dotted line).
    \SB, \pol, and \spec do not depend on $r_{cl}$.
	}
	\label{fig:r_cl_c}
\end{figure*}

\end{comment}

%----------------------------------------------------------------------

\begin{table*}[]
\centering
\begin{tabular}{rl}
\hline
%    Figure Number    & Contents  \\ \hline
 & For Comparison between Model C with $f_c = 1-100$ and Model S at the same \NHI  \\ \hline
Figure~\ref{fig:compare_SFG} & \SB, \pol, and \spec of Model C with \sigsrc = 100\,\kms (SFG case) 
\\
\ref{fig:compare_AGN} & \SB, \pol, and \spec of Model C with \sigsrc = 400\,\kms (AGN case) 
%\\
%\ref{fig:r_cl_c} & \SB, \pol, and \spec of Model C for the clump radius $r_{cl} = $ 10 pc, 100 pc, and 1 kpc
\\
\ref{fig:size_c} & $R_{obs}$ of Model C for $f_c = 1$, 2, 5, 10, and 100 and Model S
\\
\ref{fig:dop_c} & $P_{obs}$
\\
\ref{fig:peak_c} & $\Delta V_{peak}$
\\ \hline
 & For Low $f_c$ (1 and 2) and High \NHI ($10^{20}$ and $10^{21} \unitNHI$) Model C \\ \hline
Figure~\ref{fig:low_fc_sigma_emit} & \SB, \pol, \spec for \sigsrc = 100 -- 400\,\kms 
\\
\ref{fig:low_fc_vexp_SFG} & \SB, \pol, \spec of SFG case for \vexp = 0 -- 400\,\kms
\\
\ref{fig:low_fc_vexp_AGN} & \SB, \pol, \spec of AGN case for \vexp = 0 -- 400\,\kms
\\[0.2em]\hline
\end{tabular}
 	\caption{Figures showing results for Model C.}
	\label{table:figures_c}
\end{table*}

\section{Clumpy Medium (Model C) Results}\label{sec:result_c}

\subsection{\lya Radiative Transfer in Clumpy Medium}
\label{sec:lya_clumpy}

\lya radiative transfer in a clumpy medium has been studied extensively by \cite{gronke16,gronke17}. They show that, as the covering factor $f_c$ increases, \lya spectra emerging from a clumpy medium approach those from a continuous medium at the same total \hi column density $\NHI = f_c \NHIcl$, where $\NHIcl$ is the column density of a individual clump.
While \cite{dijkstra12} and \cite{trebitsch16} investigated the surface brightness and polarization in a multi-phase medium, they explored only a limited parameter space. Here we investigate the surface brightness profile \SB, polarization \pol, and spectrum \spec of \lya escaping from a clumpy medium for a wide range of parameters (see Table~\ref{table:model_s}).

Before presenting our results, we briefly explain ``surface scattering,'' a critical concept in understanding \lya radiative transfer in a clumpy medium. \cite{neufeld91} first introduced surface scattering in \lya RT. \cite{hansen06} and \cite{duval14} confirmed that surface scattering can help \lya photons escape more easily from a clumpy medium, thus increasing the escaping fraction of \lya.
Figure~\ref{fig:surface_scattering} is a schematic illustration of the behavior of \lya in a clumpy halo. If the wavelength of the incident photon is close to the line center of a clump in motion, it is hard for the photon to penetrate into the clump due to the high optical depth. This photon experiences several scatterings, mainly on the clump surface, leaves the clump, and propagates to another clump. The left and right panels represent the case where the optical depth of the clump $\tau_{cl}$ is $\lesssim 1$ and $\gg 1$ to the incident photon, respectively. If $\tau_{cl} \gg 1$, surface scatterings mainly dominate, potentially allowing the photons to be spatially diffused through a smaller number of scatterings than in a continuous medium.

To illustrate how the spatial diffusion and surface scattering depend on the clumpiness, we compare \SB and \pol for the smooth medium (Model S) with those for a clumpy medium (Model C) with covering factor $f_c = 1$, 5, and 10  (Figure~\ref{fig:image_c}). In the figure, we focus on the case with $\vexp = 400\, \kms$ and $\NHI = 10^{21} \unitNHI$. The last row is identical to that in Figure~\ref{fig:image}.

The projected surface brightness profiles with  $f_c \ge 5$ are almost identical to Model S. However, the polarizations are weaker than in Model S due to surface scattering. If most photons experience surface scattering, the incident radiation field at the last scattering position becomes more isotropic (see the right panel of Figure~\ref{fig:surface_scattering}). Thus,  polarization can decrease with decreasing $f_c$ due to surface scattering. In a clumpy medium, the polarization is more affected by surface scattering than is the surface brightness profile.

\subsection{Comparison between Models C and S}\label{sec:compare_model_c_s}

In this section, we systematically compare the surface brightness, polarization, and \lya spectrum of the clumpy medium (Model C) with the smooth medium case (Model S).  Given that most of the physics is similar for both Models, we focus only on the notable differences here. Note that due to computational limitations, we simulate only a uniform distribution of clumps in Model C. Therefore, for the purposes of comparison, we assume a Model S with a constant \hi number density (i.e., $R_e/R_H = \infty$). We also compare the Models at the same total \hi column density (\NHI).

In Figures~\ref{fig:compare_SFG} and \ref{fig:compare_AGN}, we compare Models S and C for the SFG and AGN cases, respectively. We show \SB, \pol, and \spec for Model S and for Model C with $f_c = 1-100$. For each SFG and AGN case, we show four combinations of two \NHI and two \vexp values: \NHI = $10^{19} \unitNHI$ (low \NHI) and $10^{21} \unitNHI$ (high \NHI); \vexp = 0 \kms (static) and 400 \kms (outflow). 
In Appendix~\ref{sec:clump_size}, we show that the choice of clump size 
$r_{cl}$ (10\,pc, 100\,pc, or 1\,kpc) does not affect our results (Figure~\ref{fig:r_cl_c}), so we adopt $r_{cl} = 100$ pc throughout.
%In addition, to investigate the dependence on the clump size $r_c$, we show the results for three clump sizes that are spatially unresolved from ground: $r_{cl}$ = 10~pc, 100~pc, and 1~kpc (Figure~\ref{fig:r_cl_c} {\bf in Appendix~\ref{sec:clump_size}}).

Our findings from the clumpy vs.~smooth medium comparison are summarized below and discussed further in the following sections.
\begin{enumerate}[leftmargin=+0.5cm,itemsep=0pt]

\item[$\bullet$]
With increasing covering factor ($f_c$), the surface brightness and degree of polarization profiles, as well as \lya spectra, of Model C converge to those of Model S. In particular, when $f_c \geq 5$, the surface brightness profiles are almost identical to those of Model S.

\item[$\bullet$]
The clumpy medium with $f_c = 1$ and 2 shows a unique behavior different from that of Model S.

\item[$\bullet$]
In the static clumpy medium, the \lya spectrum can show a peak at the systemic velocity or a non-zero central dip despite the high total optical depth at the line center.

\item[$\bullet$]
In the high \NHI--outflow case with $5 \leq f_c \leq 100$, the overall degree of polarization decreases with decreasing $f_c$. 

\item[$\bullet$]
The dependence on the clump radius $r_{cl}$ is negligible as long as the clumps are much smaller than the halo ($r_{cl} \ll R_H$). 

\end{enumerate}

\subsubsection{Clumpy Medium with High $f_c$ $\geq 5$}
\label{sec:clumpy_high_fc}

In Figures~\ref{fig:compare_SFG} and \ref{fig:compare_AGN}, we show that the results of Model C converge to those of Model S as $f_c$ increases from 1 to 100. As in \citet{gronke16,gronke17}, we confirm that the spectra at $f_c = 100$ are almost identical to those of Model S. In our simulation, \SB, \pol, and \spec in the high $f_c$ regime are also almost identical to those of Model S. 
As $f_c$ increases, the halo consists of larger number of clumps ($N_{cl} \sim f_c R_H^2 / r_{cl}^2 = 10^6 - 10^8$; $f_c = 1 - 100$) at fixed $r_{cl} = 100$\,pc, so that the medium becomes ``foggy'' and indistinguishable from the smooth medium case.

The surface brightness profiles are the least sensitive to $f_c$. In the top panels of Figures~\ref{fig:compare_SFG} and \ref{fig:compare_AGN}, 
\SB, even at $f_c = 5$ ({\it green}), is already indistinguishable from Model S ({\it black}). However, in the middle panels, the \lya spectra with $f_c = 5$ differ from  Model S, especially in the static medium. In addition, the overall polarization at this $f_c$ in the outflow cases is weaker than the polarization of Model S.
We conclude that the dependence on $f_c$ is more evident for the spectrum and the polarization than for the surface brightness. At $f_c \geq 5$, the spatial diffusion in the clumpy medium is enough to generate an extended \lya halo like the smooth medium.

\subsubsection{Clumpy Medium with Low $f_c = 1-2$}
\label{sec:compare_low_fc}

In the low covering factor regime ($f_c$ = 1 and 2), \SB, \pol and \spec are all peculiar compared to the large $f_c$ cases and Model S. As shown in in Figures~\ref{fig:compare_SFG} and \ref{fig:compare_AGN}, the spectra  have peaks near the systemic velocity; \SB becomes more concentrated or develops a bright core; thus, the polarization jump is always present.
These features are analogous to the properties of the single wing scattering case of Model S (Figure~\ref{fig:single_multi}).

At low $f_c$, the photons escape after interacting with only one or two clumps. The photons can escape from the inner halo after interacting with slowly moving clumps there. The wavelength change is negligible because scatterings cause a small line shift at the $\sim v_{th}$ scale.
As a result, low $f_c$ cases have bright cores, spectral peaks near the systemic velocity, and polarization jumps like Model S at low $\NHI \leq 10^{19} \unitNHI$. We revisit low $f_c$ case in more detail in Section \ref{sec:low_fc}.

\subsubsection{Formation of \lya Spectrum}\label{sec:lya_spec_c}

In contrast to \SB, the \lya spectra are most sensitive to the covering factor.  The most distinct feature of the clumpy model is that the resulting spectra can have \lya photons at the systemic velocity of the halo gas. A steep dip at the systemic velocity is a common feature of observed high-resolution \lya spectra \cite[e.g.,][]{yang14b,fabrizio19}. Here we explore the simulated spectra as a function of \NHI and \vexp, while varying $f_c$. 

{\it Static medium ---}
\lya photons at $\Delta v \sim 0\kms$ appear as a spectral peak or weak central dip feature depending on $f_c$. 
In Model S, the spectra in the static medium show the double peaks completely separated by the null flux at the middle. The spectrum at the systemic velocity always drops to zero due to the extremely high optical depth. However, for the static cases in Figures~\ref{fig:compare_SFG} and \ref{fig:compare_AGN}, the emergent spectra show central peaks at low $f_c \leq 5$, while developing dips with non-zero flux as $f_c$ increases.

%% Explanation:
This behavior arises because, in a medium with static clumps, surface scatterings mostly occur near the systemic velocity. The reason is due to the high optical depth of the clumps; photons can experience only several scatterings on the clump surface. Thus, only slight wavelength changes occur between interactions with clumps.  Photons at the systemic velocity  continuously experience surface scattering before escaping the system, while maintaining their initial wavelength, leading to peaks or non-zero fluxes at the systemic velocity.
In summary, in the smooth medium, the central dip originates from the high optical depth of the \hi\ {\it halo} that photons must pass through. In contrast, in the clumpy medium, the central peak arises due to surface scattering by each \hi\ {\it clump} with sufficient optical depth to prevent photons from penetrating it.

In the static medium with $f_c \leq 5$, 
%we find that
high \NHI cases show a {\it stronger} peak at the line center than low \NHI cases (first and third bottom panels in Figure \ref{fig:compare_SFG}). The effect of  surface scattering becomes stronger with increasing $\NHI$ at the same $f_c$. This seemingly counter-intuitive result is because the velocity range of photons experiencing surface scattering broadens with increasing \NHIcl ($\NHI \propto \NHIcl$ at a fixed $f_c$). The higher the \NHIcl of the clumps, the more the photons initially near the systemic velocity maintain their wavelength through the surface scattering.
Similarly, \cite{gronke16} find that spectra from a clumpy medium at higher \NHIcl require higher values of $f_c$ to resemble spectra from a smooth medium.

{\it Outflow medium ---}
As already mentioned in Section \ref{sec:clumpy_high_fc}, as $f_c$ increases ($f_c \geq 5$), the spectra of the expanding medium approach those of Model S. In the bottom panels of Figures~\ref{fig:compare_SFG} and \ref{fig:compare_AGN}, the spectra at $f_c \geq 5$ for the outflow cases are asymmetric with redshifted peaks and extended wings toward the red like Model S.
In the static clumpy medium, surface scattering mostly occurs near the systemic velocity. In the clumpy outflow medium, however, a blueward photon can also experience surface scattering; thus, surface scattering affects \lya radiative transfer at a wide range of velocities. As a result, the spectra for the outflow medium more easily converge to  Model S than those for the static medium at the same $f_c$.

\subsubsection{Polarization Behavior}\label{sec:pol_c}

{\it High \NHI case --- }
The degree of polarization decreases with decreasing $f_c$ over the range 5--100; the dependence on $f_c$ is strongest in high \NHI--Outflow cases (last column of Figures~\ref{fig:compare_SFG} and \ref{fig:compare_AGN}). 
The polarization decrease  occurs because the incident radiation field at the last scattering position becomes more isotropic due to the surface scattering. 
We illustrate this effect in Figure~\ref{fig:surface_scattering}, where the red and orange arrows represent the escape of photons from the last scattering and the path of the incident radiation field, respectively. 
For a fixed total \NHI, a smaller $f_c$ represents a larger column density for individual clumps ($\NHIcl$), leading to a higher optical depth of clumps  ($\tau_{cl} \gg 1$) and making it more likely that photons are scattered from their denser surfaces.
Therefore, the radiation field around the last scattering atoms (orange arrows) becomes more isotropic, and the weaker polarization emerges.
On the other hand, in the larger $f_c$, smaller $\NHIcl$, and $\tau_{cl} \leq 1$ case, surface scattering does not occur over the wide velocity range. Therefore, the radiation field becomes more anisotropic, and the polarization increases.
In high \NHI--Static cases, the radiation field of Model S is substantially isotropic without the surface scattering, and \pol is low at $\lesssim 10 \%$; thus, the variation of \pol by $f_c$ is negligible.

{\it Low \NHI case ---}
For the low \NHI--outflow case, as $f_c$ decreases over the range 5--100, the variation of \pol depends on \sigsrc. 
In the second column of Figure~\ref{fig:compare_SFG} for SFG cases, 
\pol (as well as the polarization of the high \NHI--outflow case discussed above) decreases with decreasing $f_c$ due to surface scattering.
However, in the second column of Figure~\ref{fig:compare_AGN} for AGN cases,
\pol increases with decreasing $f_c$.
As noted in Section~\ref{sec:result_a}, the polarization of the low \NHI--outflow case with an AGN-type source decreases with increasing \vexp due to core scattering.
The strong outflow decreases the layer of \hi halo having similar outflow velocity. If the moving layer is much thin and optically thin, the blueward photons are able to escape through 
%only 
core scattering alone. 
In the clumpy medium, the increase of \NHIcl by small $f_c$ reduces this contribution of core scattering, and wing scattering is more likely to occur; thus, the degree of polarization increases despite the radiation field becoming more isotropic through surface scattering.

\begin{figure*}[ht!]
	\centering
	\includegraphics[width=\textwidth]{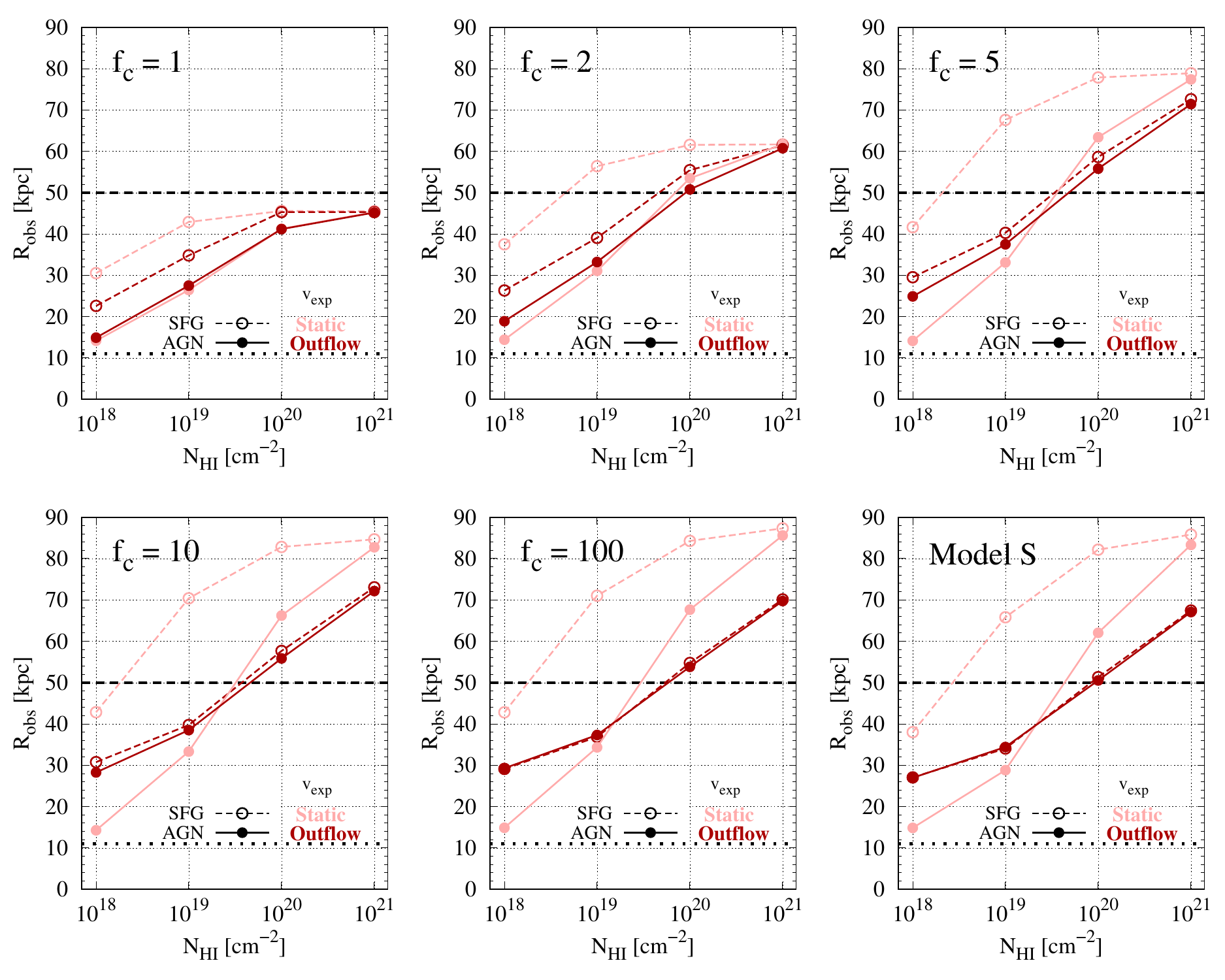}
	\caption{
    Observable halo size $R_{obs}$ of Model C for
    $f_c = 1$, 2, 5, 10, and 100, and of Model S.
    The solid and dashed lines represent the SFG (\sigsrc = 100\,\kms) and AGN (400 \kms) cases, respectively.
    The pink and red line colors are for the Static (\vexp = 0 \kms) and Outflow (\vexp = 400 \kms) cases, respectively.
    The right bottom panel for Model S shows $R_{obs}$ with $R_e/R_H = \infty$ from Figure~\ref{fig:size_s}.
    As introduced in Figure~\ref{fig:size_s}, the dotted and dashed horizontal lines represent the radius of the \lya halo considering only seeing (without scattering) $R_{obs} = 12$\,kpc and the typical size of giant \lya nebulae $R_{obs} =50$\,kpc.
    When $f_c \geq 2$ and $\NHI = 10^{20} \unitNHI$, $R_{obs}$ is always over 50~kpc.
    The trends as a function of \NHI, \vexp, and $\sigsrc$ of Model C with $f_c \geq 5$ are similar to those of Model S. 
    $R_{obs}$ decreases with decreasing covering factor, especially for $f_c < 5$.
    In this low $f_c$ regime, the dependence on \vexp becomes weaker and is negligible for the AGN cases.
    }
	\label{fig:size_c}
\end{figure*}

\begin{figure*}[ht!]
	\epsscale{1.1}
	\plotone{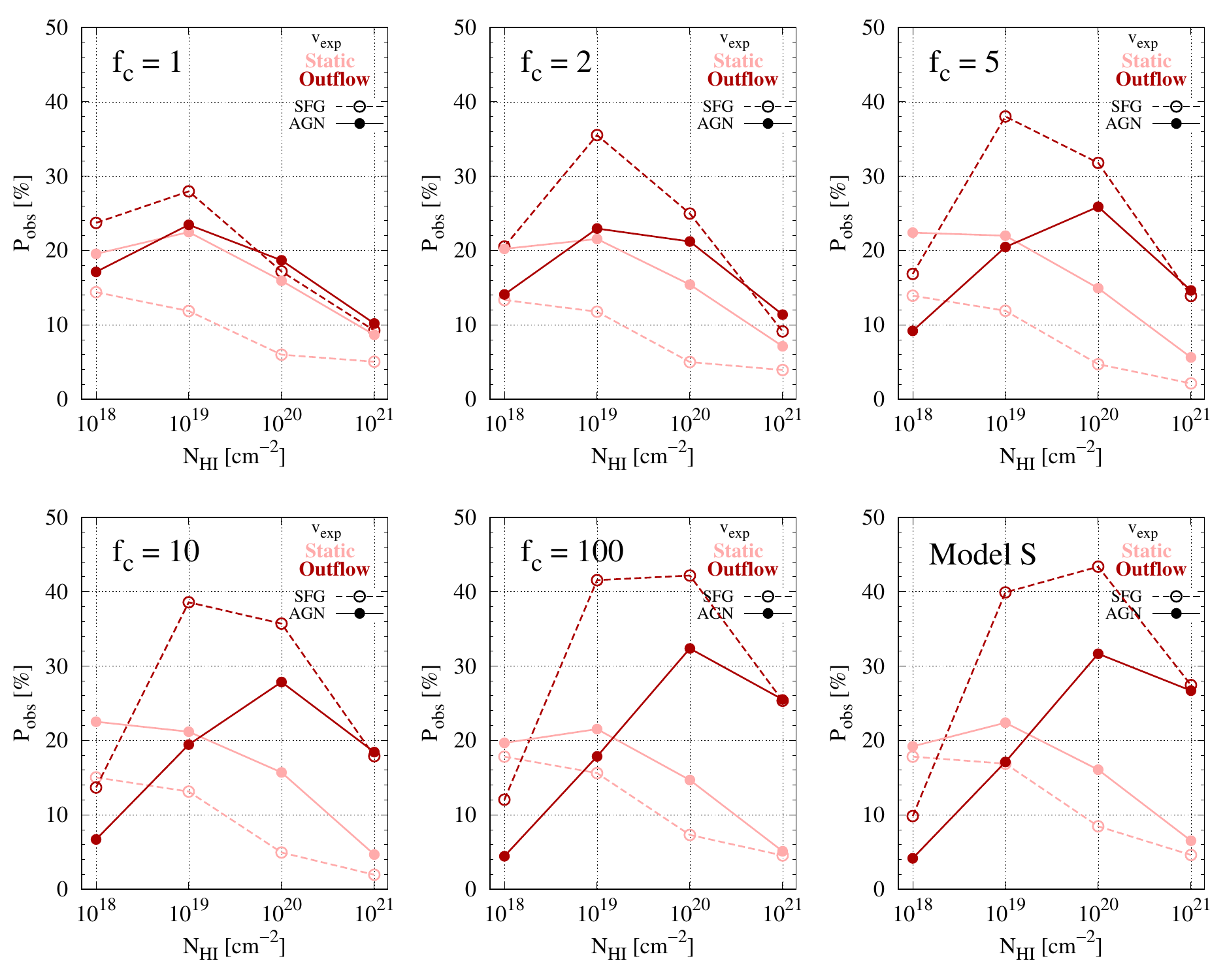}
      % \plotone{static_par_dop.pdf}
	\caption{
	Degree of polarization measured at $R_{obs}$ ($P_{obs}$) of Model S and of Model C with $f_c = 1$, 2, 5, 10, 100. 
	In each panel, we show the combinations of two outflow speeds and two source types: \vexp = 0 \kms (Static) and 400 \kms (Outflow); \sigsrc = 100 \kms (SFG) and \sigsrc = 400 \kms (AGN).
      % The parameters in each panels are same to those of Figure~\ref{fig:size_c}.
        The trends of $P_{obs}$ in Model C as functions of \NHI, \vexp, and \sigsrc when $f_c \geq 5$ are similar to those of Model S.
	$P_{obs}$ decreases with decreasing $f_c$ due to the effect of surface scattering (see the projected image in Figure~\ref{fig:image_c}).
	This decreasing trend of $P_{obs}$ becomes stronger with increasing $\NHI$.
	}
	\label{fig:dop_c}
\end{figure*}

\begin{figure*}[ht!]
	\epsscale{1.1}
	\plotone{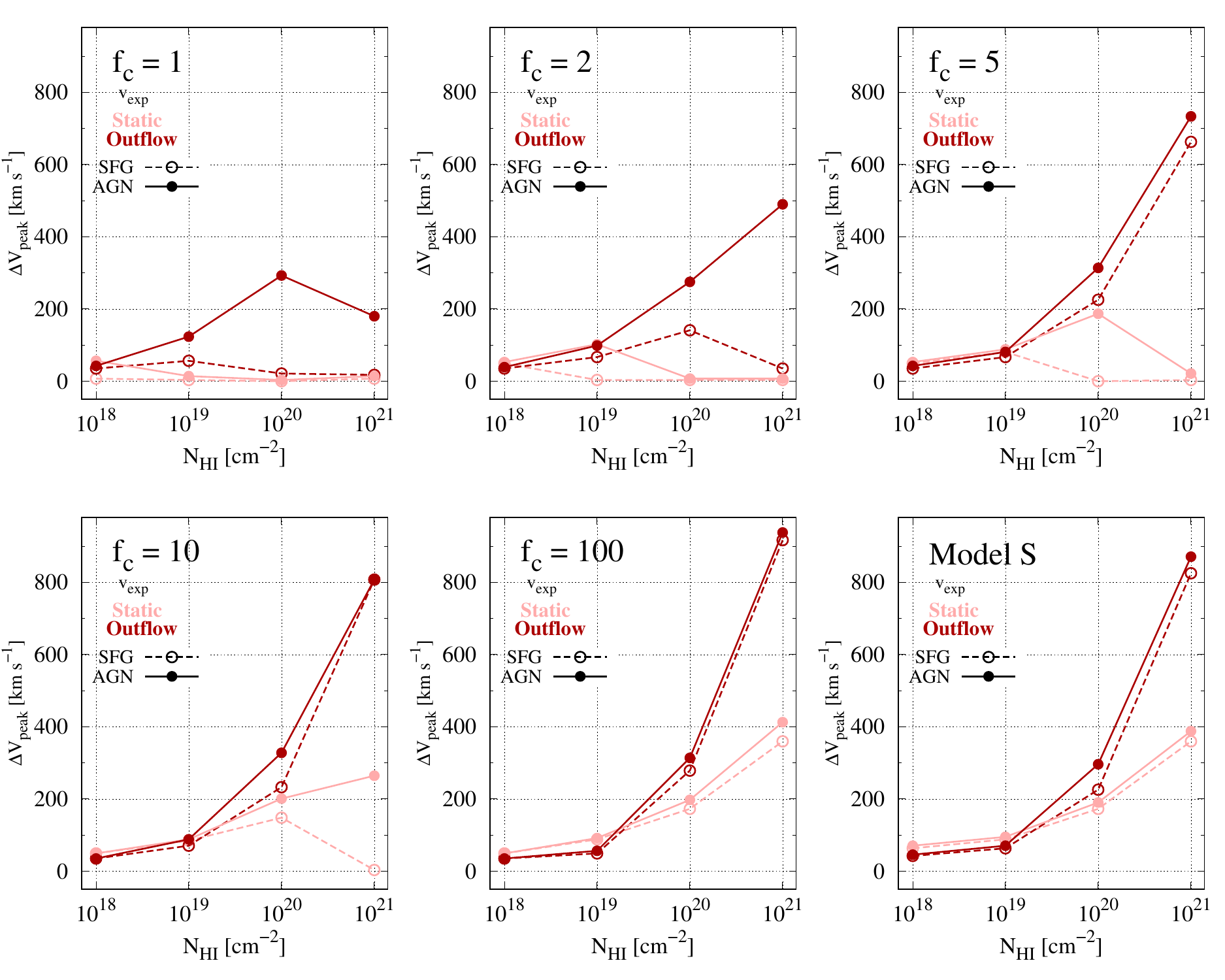}
	\caption{
        Doppler shift of the spectral peak (\dvpeak) of Model C and Model S. Parameters in each panel are identical to those of Figure~\ref{fig:size_c}.
        In each panel, we show the combinations of two outflow speeds and two source types: \vexp = 0 \kms (Static) and 400 \kms (Outflow); \sigsrc = 100 \kms (SFG) and \sigsrc = 400 \kms (AGN).
        In the outflow cases, the trends as a function of \NHI, \vexp, and $\sigsrc$ of Model C with $f_c = 5 - 100$ are similar to those of Model S, while the dependence at $f_c = 1 - 2$ is distinct.
        In $f_c \leq 10$, as shown in the bottom panels for \spec in Figures~\ref{fig:compare_SFG} and \ref{fig:compare_AGN}, the static cases show a central peak at the systemic velocity (\dvpeak = 0\,\kms) due to surface scattering.
        The top left and middle panels ($f_c = 1$ and 2) indicate a strong dependence on \sigsrc at $\NHI \geq 10^{20} \unitNHI$; unlike Model S, \dvpeak increases with increasing $\sigsrc$. 
        }
	\label{fig:peak_c}
\end{figure*}

\subsection{Observables ($R_{obs}$, $P_{obs}$, and $\Delta V_{peak}$)}

In this section, we study the dependence of the observables on model parameters. As for Model S in Section \ref{sec:result_a}, we measure three observables for Model C: the observable radius ($R_{obs}$), the degree of polarization at $R_{obs}$ ($P_{obs}$), and the Doppler shift of the spectral peak (\dvpeak). 
Figures~\ref{fig:size_c}, \ref{fig:dop_c}, and \ref{fig:peak_c} show the variation of $R_{obs}$, $P_{obs}$, and \dvpeak with \NHI, respectively, as well as with other three parameters: \sigsrc, \vexp, and $f_c$. Each panel in the figures shows either Model C with $f_c = 1$, 2, 5, 10, or 100 or Model S. 
The Model S panels are identical to the uniform density case ($R_e/R_H = \infty$) in Figures~\ref{fig:size_s} ($R_{obs}$), \ref{fig:Pobs_s} ($P_{obs}$), and \ref{fig:peak_s} ($\Delta V$). Given that observables are most sensitive to the total column density, the $x$-axis is \NHI = $f_c \NHIcl$.  The colors and line shapes represent $\vexp = 0$ (Static) and $400 \kms$ (Outflow), and \sigsrc  = 100 (SFG) and 400 (AGN) \kms, respectively.

Figures~\ref{fig:size_c}, \ref{fig:dop_c}, and \ref{fig:peak_c} indicate that, at large $f_c \geq 5$, the trends of the three observables for Model C are similar to those of Model S, although the values themselves are not identical for certain parameters. This result is expected, because the clumpy models converge to the smooth model at large $f_c$. 
In Figure~\ref{fig:size_c}, $R_{obs}$ for both Model C with $f_c$ = 5 -- 100 and Model S increases with 
increasing \NHI (along the $x$-axis), 
decreasing \vexp (line color), and 
decreasing \sigsrc (line type). 
In Figure~\ref{fig:dop_c}, $P_{obs}$ for Model C is generally smaller, but follows similar trends, as that for Model S.
In Figure~\ref{fig:peak_c}, the trends of \dvpeak for Model C with $f_c \geq 5$ are almost identical to Model S. The only exception is that \dvpeak of the Static case sometimes drops to zero at $f_c$ = 5 and 10, a manifestation of the central peak of the \lya spectrum originating from surface scattering (see Section~\ref{sec:lya_spec_c}).

In Section~\ref{sec:compare_low_fc}, we mention that \SB, \pol, and \spec for Model C with $f_c = 1$ and 2 are different from the high $f_c$ cases and from Model S. The bright core, the polarization jump, and the blueward spectrum always exist in low $f_c$ cases.
A summary of our findings in the low covering factor regime is:

\begin{enumerate}[leftmargin=+0.5cm,itemsep=0pt]

\item[$\bullet$]
At $f_c = 1$, a \lya halo cannot be observed as a giant \lya nebulae (i.e., extended over $\sim 100$ kpc), regardless of other parameters.

\item[$\bullet$]
At $f_c$ = 1 and 2, the dependence of $R_{obs}$ and $P_{obs}$ on \vexp becomes weaker as \NHI increases.

\item[$\bullet$]
When $f_c \leq 2$ and \NHI$\geq 10^{20} \unitNHI$, $\Delta V_{peak}$ increases with increasing \sigsrc. We do not see this dependence for Model C with high $f_c \geq 5$ and for Model S.

\end{enumerate}

In Figure~\ref{fig:size_c}, we find that a \lya halo with $f_c = 1$ cannot be observed as extended over $\sim$ 100~kpc scale in diameter. For Model S, the observable size of the \lya halo is $\sim$100 kpc as long as $\NHI \geq 10^{20} \unitNHI$. The top left panel shows that $R_{obs}$ with $f_c = 1$ is only $\sim$40 kpc, even at high \NHI ($10^{20-21} \unitNHI$). 
For Model C with $f_c = 1$, the photons emitted from the source can escape the system after interacting with only one or two clumps or even without scattering. Furthermore, the last interaction with a clump can occur in the inner part of the halo; thus, \SB is less extended than for Model C at high $f_c$ and for Model S. When $f_c \geq 2$ at high \NHI, a \lya halo can be observed as highly extended regardless of \vexp, and \sigsrc.

In the $f_c$ = 1 and 2 panels of Figures~\ref{fig:size_c} and  \ref{fig:dop_c}, $R_{obs}$ and $P_{obs}$ do not strongly depend on \vexp. Furthermore, at $\NHI = 10^{21} \unitNHI$, the dependence on \vexp is negligible.
In this low $f_c$ regime (e.g., \NHIcl $\sim \NHI$), where an initial photon first interacts with a  clump determines spatial diffusion because the photon escapes after only one or two scatterings. 
Because \NHIcl is very large ($\sim 10^{21} \unitNHI$), photons incident upon a clump are always scattered regardless of the incident wavelength. As a result, the spatial information parameters $R_{obs}$ and $P_{obs}$ do not depend on \vexp and \sigsrc. For a more detailed analysis of spatial diffusion at low $f_c$ and high \NHI, we investigate the \SB and \pol in Section~\ref{sec:low_fc}.

At $f_c =$ 1 and 2, $\Delta V_{peak}$ at $\NHI \geq 10^{20} \unitNHI$ depends on the width of the emitted \lya emission (\sigsrc). Recall that the shift of the spectral peak does not depend on \sigsrc for Model S. 
Thus, in the clumpy medium with $f_c =$ 1 and 2, the information on the \lya sources is imprinted in the spectrum. 
Note that this holds only for $\NHI \geq 10^{20} \unitNHI$. 
Figure~\ref{fig:peak_c} shows that \dvpeak does not depend on \sigsrc for $\NHI \leq 10^{19} \unitNHI$; there is no or little spread between the SFG and AGN cases with the same line colors. 
Nevertheless, in the high \NHI regime, \dvpeak for the outflow cases increases as \sigsrc increases, i.e., the value for the AGN case (solid) is higher than for the SFG  (dashed). 
We investigate the profiles of \lya spectra in this low $f_c$ and high \NHI regime in following section.

\begin{figure*}[ht!]
	\epsscale{1.16}
	\plotone{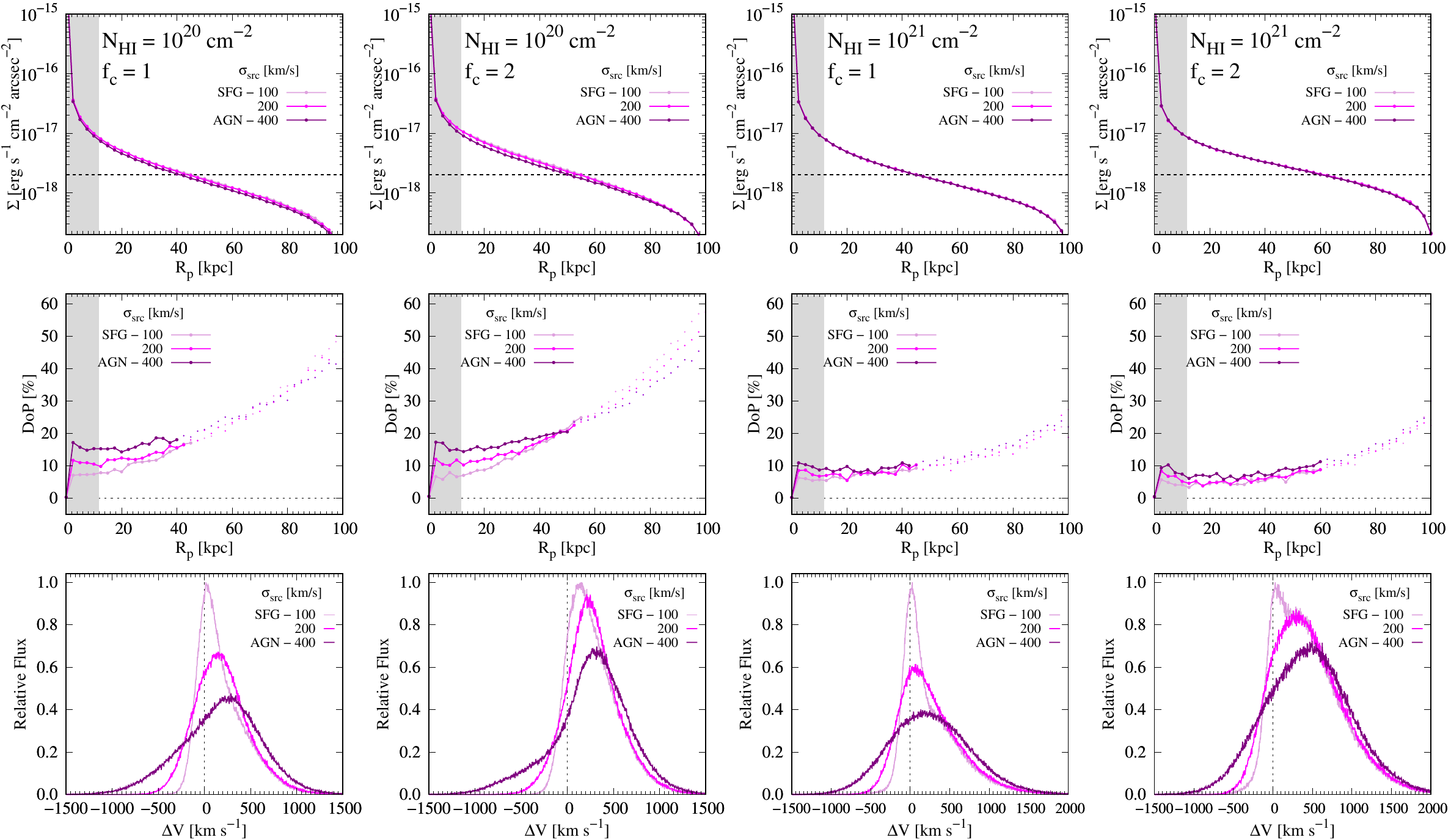}
	\caption{
        \SB, \pol, \spec (from top to bottom) of Model C with $f_c = 1-2$, high \NHI ($10^{20-21} \unitNHI$), and a strong outflow (\vexp = 400\,\kms) for a range of \sigsrc.
        Each column shows four combinations of \NHI = $10^{20}$, $10^{21}$ \unitNHI and $f_c$ = 1, 2. The line colors represent \sigsrc.
        When $\sigsrc$ increases, the spectral peak becomes more redshifted, while \SB remains virtually the same.
        The surface brightness profile does not vary, because, at this large \NHIcl ($\gtrsim 10^{20} \unitNHI$), an initial photon can always interact with a clump regardless of its incident wavelength.
        The dependence on \sigsrc arises because photons escape after only 1 -- 2 interactions with clumps, and these interactions are not sufficient for the halo kinematics to be imprinted on the spectrum. Thus, the escaping spectrum maintains the initial profile, i.e., the profiles simply become broader with increasing \sigsrc.
	}
	\label{fig:low_fc_sigma_emit}
\end{figure*}

\begin{figure*}[ht!]
	\epsscale{1.16}
	\plotone{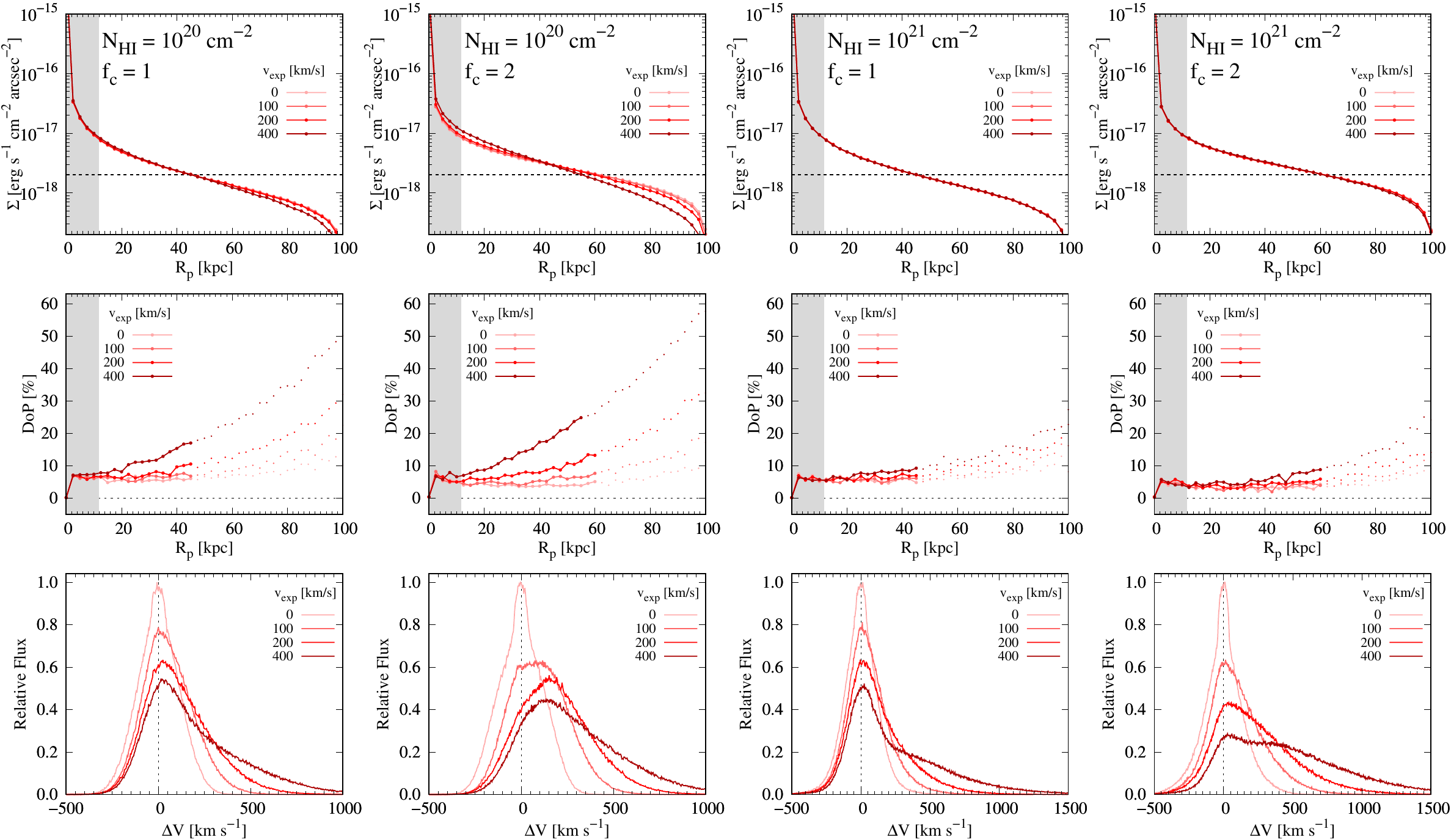}
	\caption{
        \SB, \pol, \spec (from top to bottom) of Model C for low $f_c$--high \NHI cases, with a fixed \sigsrc (100 \kms; SFG) for a range of \vexp.
        The line colors represent \vexp.
        When \vexp increases, the spectrum becomes asymmetric and extends more redward. However, note that unlike the cases with $f_c \geq 5$, the spectral peak is not shifted. 
        The initial spectra of SFG cases are narrow enough to maintain the spectral peak at the systemic velocity even after interactions with outflowing clumps.
	}
	\label{fig:low_fc_vexp_SFG}
\end{figure*}

\begin{figure*}[ht!]
	\epsscale{1.16}
	\plotone{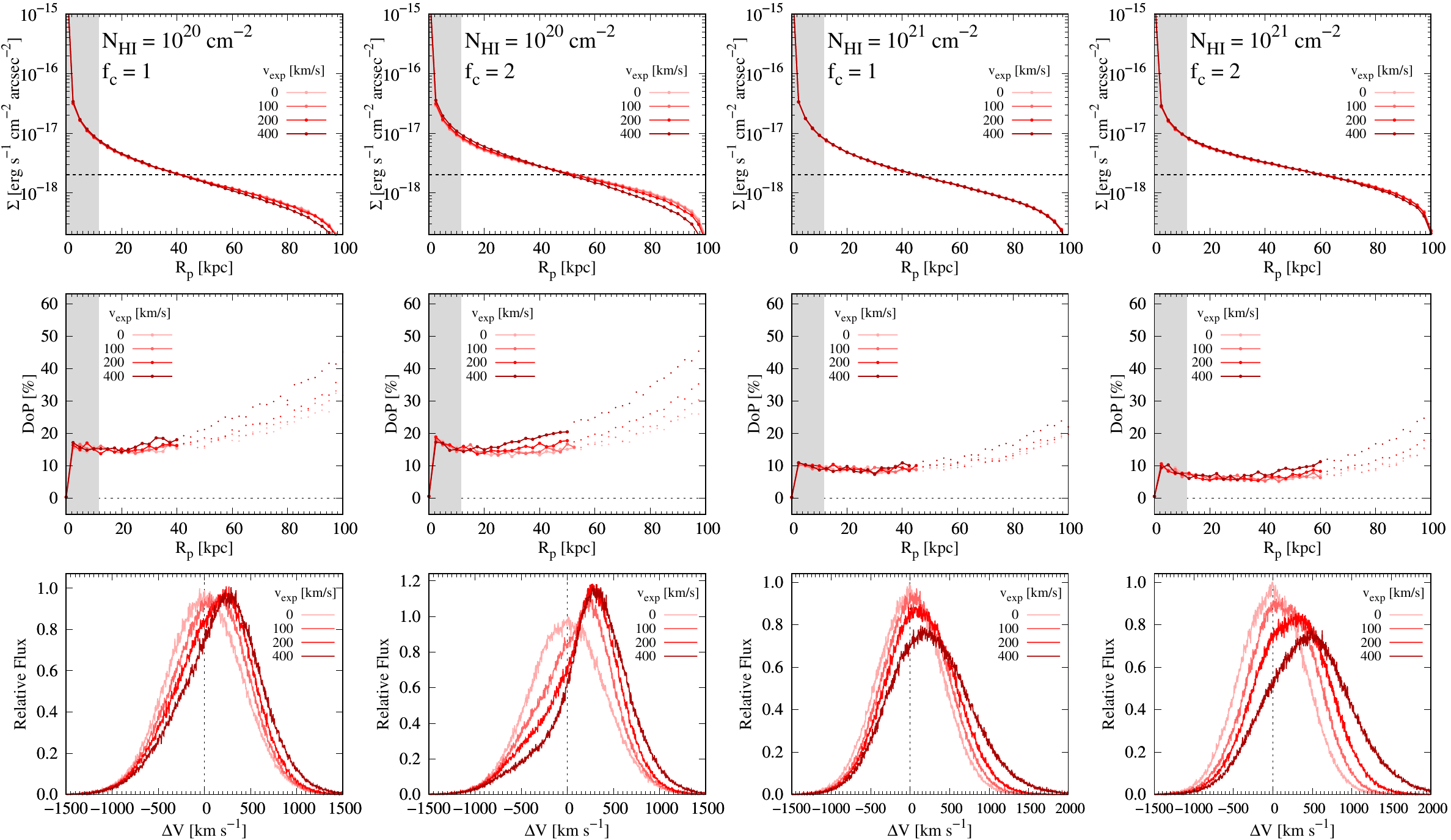}
	\caption{
        \SB, \pol, \spec (from top to bottom) of Model C for low $f_c$--high \NHI cases, with a fixed \sigsrc (400 \kms; AGN) for a range of \vexp.
        The line colors represent \vexp.	
        When \vexp increases, both \SB and \pol do not change, while the nearly symmetric spectrum becomes more redshifted.
        In this range of \vexp, one or two interactions with clumps are not enough to generate the red extended wing.
	}
	\label{fig:low_fc_vexp_AGN}
\end{figure*}

\subsection{Low $f_c$--High \NHI Clumpy Medium}
\label{sec:low_fc}

In this section, we focus on the low $f_c$--high \NHI case of Model C, because it exhibits observational features that cannot be produced by Model S. 
Almost symmetric broad \lya profiles and/or \lya spectra with peaks at the systemic velocity have been observed in several \lya blobs \citep{dey05, yang14a, yang14b,fabrizio19, li21}. These \lya spectral features are difficult to reproduce using \lya radiative transfer in a smooth medium. Note that to use Model S to explain the spatial diffusion of \lya over a 100\,kpc scale,  high \hi column density (\NHI $\gtrsim 10^{20} \unitNHI$) is required. If the scattering medium is outflowing, the \lya spectrum in this \NHI regime should be asymmetric to the red, and the spectral peak should move redward by at least $200\kms$. 
In our model library, the only case where a Gaussian-like profile centered on a systemic velocity can emerge is when \lya photons escape directly from the source through an optically thin continuous \hi medium. However, in this case, \lya is unpolarized, and the surface brightness cannot be significantly extended. 

The clumpy medium with low covering factor ($f_c$ = 1 -- 2) and high \NHI (Figures~\ref{fig:compare_SFG} and \ref{fig:compare_AGN}) can explain these peculiar spectral features, i.e., almost symmetric broad \lya profiles centered at the systemic velocity. In Model C with low $f_c$ and high \NHI, the \lya line can be symmetric while the \lya halo is spatially extended through scattering, and high polarization up to 10\% can be achieved.

A clumpy medium has already been considered to explain the spectra of a giant \lya nebula. Using the ratio of \lya to H$\beta$ in SSA22-LAB1, \cite{li21} suggest that the extended \lya originates from the photo-ionization plus scattering with atomic hydrogen, rather than from collisional excitation.
While there is evidence for scattering from the concentric polarization pattern \citep{hayes11}, they claim that SSA22-LAB1's \lya spectra cannot be explained by scattering in a smooth halo.
Because the \lya spectra in several regions are symmetric and some spectral peaks are located at the systemic velocity, they invoke the clumpy shell model of \cite{gronke17}. However, they adopt \lya RT for only the \lya spectra and do not consider the spatial diffusion and the polarization as we do here.

In Figures~\ref{fig:low_fc_sigma_emit}, \ref{fig:low_fc_vexp_SFG}, and \ref{fig:low_fc_vexp_AGN}, we once again show surface brightness (top), degree of polarization (middle), and spectral line (bottom) profiles. 
We consider two \NHI ($ 10^{20}$ and $10^{21} \unitNHI$) and two $f_c$ (1 and 2) values. Figure~\ref{fig:low_fc_sigma_emit} shows the results for \sigsrc = 100, 200, and 400 \kms at fixed \vexp = 400 \kms.
Figures~\ref{fig:low_fc_vexp_SFG} and \ref{fig:low_fc_vexp_AGN} show the results for \vexp = $100-400 \kms$
for the SFG and AGN cases, respectively.

Our findings for the low $f_c$--high \NHI medium are:
\begin{enumerate}[leftmargin=+0.5cm,itemsep=0pt]

\item[$\bullet$]
The escaping \lya spectrum is similar to the intrinsic spectrum despite the high \NHI.

\item[$\bullet$]
When \vexp increases, the red wing is more extended in the SFG cases, and the spectral peak is more redshifted in AGN cases.

\item[$\bullet$]
The \lya spectrum is more sensitive to the model parameters than either the surface brightness or  polarization.

\end{enumerate}

\subsubsection{Formation of \lya Spectrum} 

The key feature of the low $f_c$--high \NHI case is that the integrated spectrum is similar to the intrinsic \lya source profile; thus, source information is still preserved in the \lya spectrum. On the other hand, recall that in the high \NHI ($\geq 10^{20} \unitNHI$) regime of Model S (bottom panels of Figures~\ref{fig:compare_SFG} and \ref{fig:compare_AGN}), the \lya lines in the outflow medium always show asymmetry to the red. \lya photons entirely move to the redward of the systemic velocity through multiple wing scatterings in the outflow medium.
However, in the clumpy medium, the photons interact with a clump one or two times before escaping; thus, the frequency change due to moving clumps occurs one or two times. For example, the photon escapes after interacting with slowly moving clumps in the inner halo. The initial and escaping wavelength are not much different despite high \NHI. Consequently, the blueward photon survives, keeping its initial wavelength.  In Figure~\ref{fig:low_fc_sigma_emit}, the \lya photons appear in the blueward of the systemic velocity.

Because of the insufficient interaction with clumps, \lya profiles depend on the \lya source velocity width (\sigsrc) in the outflow medium. In the SFG cases, \lya spectra show central peaks with extended red wings, while symmetric profiles with redshifted peaks emerge in the AGN cases. In Figure~\ref{fig:low_fc_sigma_emit}, at the fixed \vexp = 400\kms, the spectral peaks move to the redward with increasing \sigsrc. 
Regardless of the initial wavelength, all photons experience one or two interactions with clumps due to high \NHIcl. Because the number of initial photons near the systemic velocity of the small \sigsrc case (SFG) is higher than the large \sigsrc case (AGN), the central peak can survive in the SFG case.
Note that for Model S, there is virtually no difference in profile shapes at this high \NHI (Figures~ \ref{fig:spec_vexp_s} and \ref{fig:spec_A_s}), because all \lya photons experience an enormous number of scatterings, thus losing the memory of the initial velocity information. 

%% spectral difference by the source type is not conspicuous in Model S at this high \NHI.
%% The spectral peak of the \lya spectra is closer to the systemic velocity than those of AGN case.

%% The extended red wing and the peak of the \lya spectrum represent the outflow velocity \vexp and the intrinsic \lya emission $\sigma_{src}$, respectively.

Although more weakly than for \sigsrc, the outflow information is encoded in different spectral features. 
In the SFG case, outflow information is in the extended red wing, whereas,
in the AGN case, it is reflected in the redshift of the peak.
In the bottom panels of Figure~\ref{fig:low_fc_vexp_SFG}, the \lya spectra of the SFG cases show a spectral peak at the systemic velocity and a red wing that becomes more extended with increasing \vexp.
In Figure~\ref{fig:low_fc_vexp_AGN}, when \vexp increases, the \lya profiles for the AGN case remain roughly Gaussian, and the spectral peaks move to the redward.

\subsubsection{Surface Brightness Profile with Bright Core}
\label{sec:low_fc_SB}

In the low $f_c$--high \NHI case, the dependence of the surface brightness profile on \sigsrc and \vexp is negligible (Figures~\ref{fig:low_fc_vexp_SFG} and \ref{fig:low_fc_vexp_AGN}), and \SB has bright core within a diffuse halo (Figure~\ref{fig:low_fc_sigma_emit}). The polarization jump feature occurs at the center.

The bright core and the polarization jump arise due to directly escaping photons, like the single scattering case of Model S (Figure~\ref{fig:single_multi}). In other words, photons escape the system without interacting with any clump. When $f_c \gg 1$, clumps cover the entire sky from the point of view at the center. However, at $f_c \sim 1-2$, tiny holes can exist for certain sightlines, allowing some photons to leak without first interacting with clumps. Furthermore, a photon grazing a clump maintains its incident wavelength and direction, just like a directly escaping photon.

The negligible dependence on \sigsrc and \vexp originate from the fact that the column densities of the clumps (\NHIcl $\gtrsim 10^{20} \unitNHI$) are so thick that they can scatter most incident photons of any wavelength. In general, \vexp and \sigsrc determine the incident wavelength of the photon when it encounters clumps.
In the low $\NHIcl \lesssim 10^{19} \unitNHI$ case, whether or not an incident photon interacts with a clump is determined by the photon's wavelength. However, when $\NHIcl$ is high enough ($\sim 10^{21} \unitNHI$), the clumps become optically thick regardless of the incident wavelength; the photons must be scattered by clumps. Therefore, in the top panels of Figures~\ref{fig:low_fc_sigma_emit}, \ref{fig:low_fc_vexp_SFG}, and \ref{fig:low_fc_vexp_AGN}, \SB does not depend on \vexp and \sigsrc.

\subsubsection{Polarization Behavior}

Unlike for the smooth medium or the clumpy medium with $f_c \gtrsim 5$, the polarization is not a good indicator of the \hi halo kinematics in a low $f_c$ clumpy medium with $\NHI = 10^{21} \unitNHI$.
In the right panels of Figure~\ref{fig:pol_vexp_s}, \pol for Model S where $\NHI = 10^{21} \unitNHI$ increases with increasing \vexp. Because Model C  with a high covering factor $(f_c \geq 5)$ follows the trend of Model S, the high $f_c$ case with $\NHI \geq 10^{21} \unitNHI$ also shows strong dependence of $P_{obs}$ on \vexp (Figure~\ref{fig:dop_c}).
However, the third and fourth columns ($\NHI = 10^{21} \unitNHI$) of Figures~\ref{fig:low_fc_sigma_emit}, \ref{fig:low_fc_vexp_SFG}, and \ref{fig:low_fc_vexp_AGN} show that \pol does not depend on \sigsrc and \vexp; \pol remains flat after the polarization jump, and the overall degree of polarization is smaller than $\sim 10\%$.

%% YY: Explanation
The negligible \pol dependence on \sigsrc and \vexp, as well as the negligible \SB dependence, originates from the high \NHIcl. As stated in \S \ref{sec:low_fc_SB}, for $\NHIcl \gtrsim 10^{20} \unitNHI$, whether the photon interacts with the clumps does not depend on the incident wavelength, which is determined by \sigsrc and \vexp.
When the photon interacts with higher $\NHIcl \sim 10^{21} \unitNHI$ clumps, the surface scattering always occurs for any wavelength and causes the radiation field to become isotropic.  Thus, the overall \pol at $\NHI =10^{21} \unitNHI$ is as low as $\sim 10\%$.
This point is illustrated by the schematic diagram (Figure~\ref{fig:surface_scattering}).
When $\NHIcl \sim 10^{21} \unitNHI$, surface scattering dominates the polarization behavior.

The polarization behavior of the SFG case at $\NHI = 10^{20} \unitNHI$ is similar to that of the single scattering case for Model S.
In Figure~\ref{fig:low_fc_vexp_SFG}, the overall degree of polarization for the SFG case at $\NHI = 10^{20} \unitNHI$ increases with increasing \vexp.
The interaction with clumps of $\NHIcl \sim 10^{21} \unitNHI$ is more likely to be surface wing scattering.
However, when $\NHIcl \sim 10^{20} \unitNHI$, the photons with wavelength far from the line center of a clump can be either scattered inside of clumps or penetrate clumps.
In other words, the optical depth for clumps $\tau_{cl}$ can be $\lesssim 1$. 
If the single wing scattering contribution increases due to a strong outflow, the polarization increases radially outward up to $10\%$. In Figure~\ref{fig:low_fc_vexp_AGN}, \pol in the AGN case at $\NHI = 10^{20} \unitNHI$ does not depend on \vexp. Because surface scattering dominates the polarization behavior of the blueward photons, \pol maintains a flat shape after the polarization jump. 
Only when $\NHI = 10^{20} \unitNHI$ and the source is an SFG, does the polarization depend on \vexp and increase with increasing \vexp.

\subsection{Comparison with Observations} \label{sec:compare_observation}

The clumpy medium model with low $f_c$ and high \NHI (\S~\ref{sec:low_fc}) can explain certain observed peculiar spectra of \lya blobs. For example, \cite{yang14b} report that the peak of the \lya spectrum of LABd05 blob is located at the systemic velocity pinned by the CO molecular line observation. At first, this spectral feature was interpreted as the lack of a strong outflow and/or the presence of photo-ionization. However, according to our clumpy model, this feature might indicate that this CGM has a low covering factor ($f_c$ = 1 -- 2) and high column density ($\NHI \geq 10^{20} \unitNHI$), and that 
the system has a large \sigsrc (the AGN case).
%%
%%\yy{(Can we constrain other parameters, too? Did we ever try actual fitting?)}
%%
Furthermore, \cite{kim20} map \lya polarization across LABd05, where the highest degree of polarization ($\sim 10\%-20\%$) is found at $r$ $\sim$ 40 kpc from the AGN. This level of polarization is consistent with our models in Figure \ref{fig:low_fc_vexp_AGN}.

Thus, the clumpy medium model with a range of covering factors ($f_c$ = 1 -- 100) is essential to understand the observed properties of extended \lya halos produced fully or in part by scattering. If the \lya spectrum of a system is symmetric around the systemic velocity, and the spectrum extends to the blueside of the systemic velocity, a clumpy medium model with low $f_c$ must be adopted.
In this clumpy medium, the profiles of the surface brightness and the degree of polarization are not sensitive to \vexp and \sigsrc. 
The emergent \lya line profile is not much different from the intrinsic \lya profile of the central source. But, the \lya line profile can provide key information about the \hi halo and the source. The position of the spectral peak and the extended red wing constrain the source type (SFG vs.~AGN) and the gas kinematics, respectively. In this experiment, we emphasize that accurate measurement of the systemic velocity is key in applying the low $f_c$--high \NHI clumpy medium model, as advocated by \citet{yang11,yang14a}.

\subsection{Summary of Model C results}
In this section, we have compared the results of \lya radiative transfer modeling in a clumpy medium (Model C) to a smooth medium (Model S). 
%%
%We find that
Our findings are: 
%\begin{enumerate}[leftmargin=+0.5cm,itemsep=0pt]
%\item[(1)]
(1) the surface brightness profile of Model C with covering factor $f_c \geq 5$ is identical to that of Model S at the same \NHI (Figures~\ref{fig:compare_SFG} and \ref{fig:compare_AGN});
%\item[(2)]
(2) surface scatterings decrease the overall polarization and the spectral line broadening (Figures~\ref{fig:compare_SFG} and \ref{fig:compare_AGN});
%\item[(3)]
(4) the clump size does not affect the \lya radiative transfer results if the volume filling factor is small (Figure~\ref{fig:r_cl_c} {\bf in Appendix~\ref{sec:clump_size}});
%\item[(4)]
(4) the trends of $R_{obs}$, $P_{obs}$, and $\Delta V_{peak}$ for Model C with $f_c \geq 5$ are similar to the those of Model S at the same \NHI (Figures~\ref{fig:size_c}-\ref{fig:peak_c});
%\item[(5)]
(5) the spectra of Model C with $f_c = 1-2$ are entirely different from those of Model S. (Figures~\ref{fig:low_fc_vexp_SFG}-\ref{fig:low_fc_sigma_emit}).
%\end{enumerate}
The effect of surface scattering becomes strong in the high $\NHIcl$ regime (i.e., in the low $f_c$ regime at the same \NHI).
In the low $f_c$ - high \NHI case (Section~\ref{sec:low_fc}), \lya RT results are dominated by surface scattering.
%% We believe that it help studying symmetry spectra of \lya nebulae with its peak near the systemic velocity (see Section~\ref{sec:compare_observation}). 

\section{Conclusions}
\label{sec:summary}

Using new Monte-Carlo radiative transfer simulations, we investigate \lya scattering from a central source within a spherical \hi halo of $\sim$100\,kpc. To model polarization correctly, our code accounts for both resonance (core) and Rayleigh (wing) scattering using Stokes vectors \citep{seon22}. 
Because recent observations suggest that the CGM is clumpy \cite[e.g.,][]{hennawi13,fabrizio15}, we simulate two types of spherical \hi halo around the central \lya point source---a smooth continuous medium (Model S) and a clumpy medium (Model C). The clumps in the halo are not spatially resolved ($r_{cl}$ $\ll$ $R_H$), and their distributions are parameterized with a covering factor ($f_c$).

Our large suite of \lya radiative transfer simulations allows us to consider a wide range of physical parameters for the first time. 
We examine both a static (\vexp = 0) and an outflowing scattering medium (\vexp = 100 -- 400 \kms). To represent various types of powering central sources, we explore intrinsic \lya line widths consistent with star-forming galaxies ($\sigsrc = 100 \kms$) up to AGN ($\sigsrc = 400 \kms$). 
For Model S, we also vary effective radius of \hi distribution ($R_e/R_H$ = 0.3 -- $\infty$) to test effect of gas concentration.
For Model C, we also explore clumpiness of medium (i.e., covering factor $f_C$ = 1 -- 100).
For these wide range of parameters, we alter the total column density ($10^{18-21} \unitNHI$) as a key parameter.
Our extensive model library can thus be used to isolate how scattering in the CGM affects the \lya polarization, spectrum, and surface brightness profile.

%%%%%%%%%%%%%%%%%%%%%%%%
%%%%%%%%%%%%%%%%%%%%%%%%

\vspace{0.2cm}
Our main conclusions are:

\begin{itemize}[leftmargin=+0.6cm,itemsep=0pt]

\item 
The simulated \lya halos divide into two general classes: 1) those with a bright core in the surface brightness profile and 2) those without. Halos with a bright core also have a polarization jump, a steep increase in the polarization radial profile just outside the core (see Figure~\ref{fig:single_multi}). 
Because this behavior arises from single wing scattering, the bright core and polarization jump mainly appear for low \NHI and originate from the direct escape of \lya photons from the central source. The core and jump disappear at high \NHI $=10^{21}$ \unitNHI, where multiple wing scatterings dominate the formation of the \lya halo. Figure~\ref{fig:compare_SFG} (top and middle panels) shows this coexistence of the bright core and polarization jump.

\item 
Unlike previous calculations where only radially increasing polarization profiles were predicted, we find that the radial polarization profiles are diverse. The radial profile \pol can have positive, flat, and even negative gradients (Section~\ref{sec:result_dop_s} and Figure~\ref{fig:pol_NH_s}). This diversity arises from the range of intrinsic line widths (\sigsrc) for different types of \lya sources, especially when \sigsrc is comparable to the outflow speed of the halo (\vexp). 
If \sigsrc is much smaller than the outflow speed, a typical positive gradient is produced. On the other hand, if the polarization behavior is dominated by single-wing scattering near the center and by core scattering in the outer halo, the gradient becomes negative (see the outflow--AGN case at $\NHI = 10^{18-19} \unitNHI$ in Figure~\ref{fig:pol_NH_s}). 
We emphasize that resonance scattering must be considered in the \lya radiative transfer to compute polarization correctly, especially for \NHI $\lesssim 10^{19}$ \unitNHI.

\item
We test if scattering alone can produce halos that are spatially extended over $\sim$100\,kpc like \lya blobs or enormous \lya nebulae. We find that the column density of the scattering medium is the main driver of the observed size of \lya halos. 
In Model S, for a fixed \lya luminosity of $10^{44}$ \unitcgslum at $z=3$, a large column density of $\NHI \gtrsim 10^{20} \unitNHI$ is required to make \lya halos bigger than $R_{obs}$ $\gtrsim$ 50\,kpc (Section~\ref{sec:LAH_size} and Figure~\ref{fig:size_s}).
In Model C, for the same limit of \NHI as above, a covering factor of $f_c \geq 2$ is needed to produce LABs and ELANe (Section~\ref{sec:clumpy_high_fc} and Figure~\ref{fig:size_c}).

\item
The clumpy model with a low covering factor ($f_c \leq 2$) and a high \NHI shows very distinctive spectral features: the \lya line profiles are nearly symmetric, and their spectral peak is centered at the systemic velocity (Section~\ref{sec:low_fc} and Figures~\ref{fig:low_fc_sigma_emit}-\ref{fig:low_fc_vexp_AGN}). These features---which cannot be reproduced by the smooth model at all---might be able to explain the \lya spectra of some \lya blobs \cite[e.g.,][]{li21}.

\end{itemize}

In future work, we will compare our \lya RT library directly with existing \lya observations, including measurements of the surface brightness profile, spatially-resolved spectrum, and polarization profile for individual LABs. \lya nebulae with polarimetric observations such as LABd05 \citep{dey05, kim20} and SSA22-LAB1 \citep{steidel00, hayes11, li21} would be ideal targets for such a comprehensive analysis. As discussed in Section~\ref{sec:compare_observation}, the clumpy medium model is a promising tool to properly fit the data.

We will also extend our \lya RT model to include photo-ionization. To simulate the case where \lya is produced {\it in situ} in the halo by photo-ionization, we will distribute \lya emitting regions over the halo and analyze the predicted observables. Spatially extended halos of \ion{He}{2} and \ion{C}{4} have been discovered recently \citep{yang14a, cabot16, fabrizio19}, suggesting that the intrinsic \lya emission is in fact produced by halo gas. Our model will include not only \lya, but also these key emission lines to compare directly with the observations of the clumpy CGM \citep{hennawi13, fabrizio15}.

\section*{Acknowledgements}
Y.Y. and S.C. were supported by the Basic Science Research Program through the National Research Foundation of Korea (NRF) funded by the Ministry of Science, ICT \& Future Planning (NRF-2019R1A2C4069803).
K.S. and H.L. were supported by the National Research Foundation of Korea (NRF) grants funded by the Korea government (No. 2020R1A2C1005788 and No. 2018R1D1A1B07043944).
A.I.Z. acknowledges support from NSF AST-1715609. She also thanks the hospitality of the Columbia Astrophysics Laboratory at Columbia University, where some of this work was completed.

\appendix
\restartappendixnumbering

\section{Dependence on clump size}\label{sec:clump_size}

\begin{figure*}[ht!]
\centering
	\includegraphics[angle=90,origin=c,width=160mm]{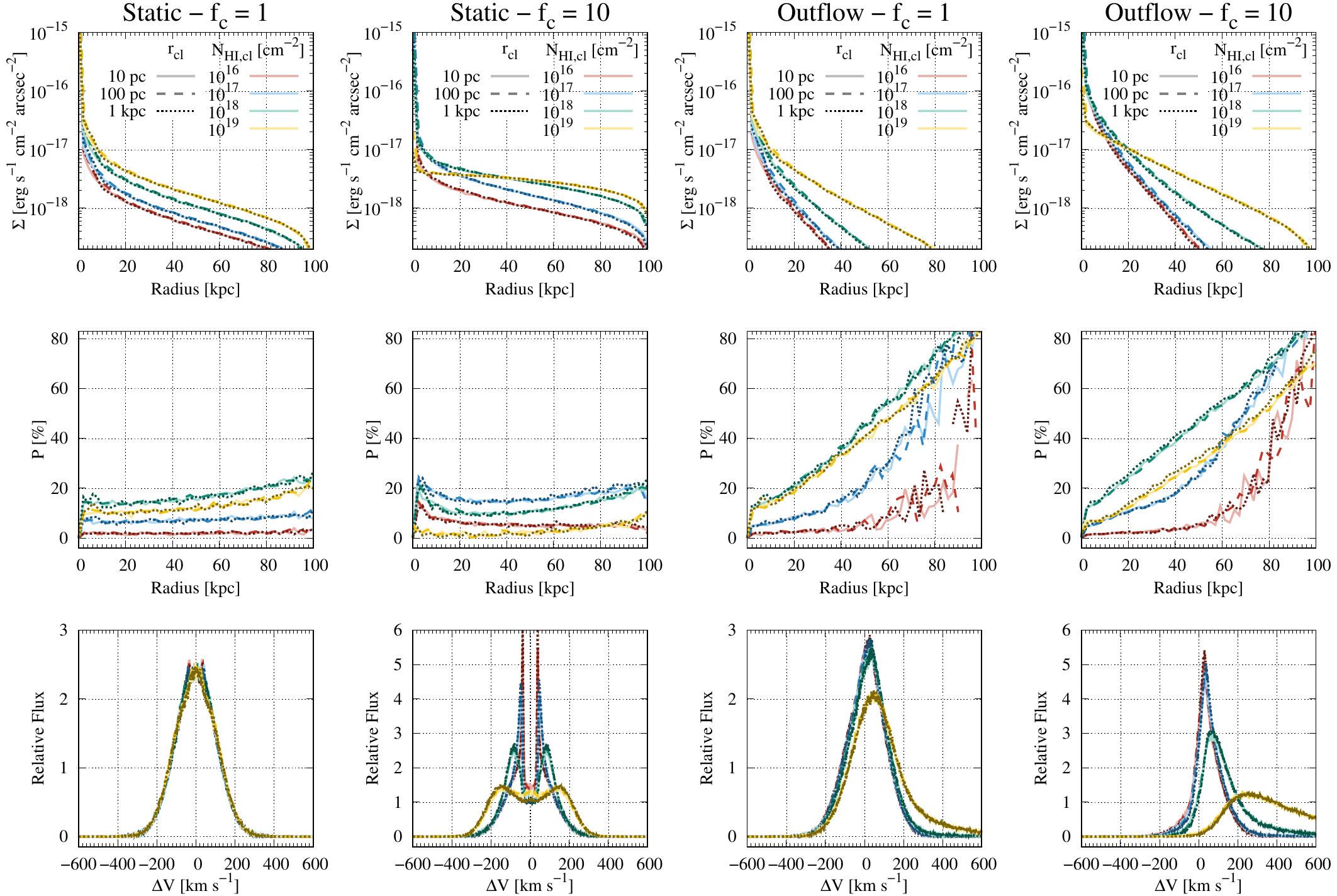}
	\caption{
    \SB, \pol, and \spec (from top to bottom) of Model C for a range of clump size $r_{cl}$.
    The line colors represent $\NHIcl = 10^{16}$ (red), $10^{17}$ (blue), $10^{18}$ (green), and $10^{19}$ (yellow) \unitNHI.
    The line styles represent $r_{cl}$ = 10\,pc (solid), 100\,pc (dashed), and 1\,kpc (dotted).
    \SB, \pol, and \spec do not depend on $r_{cl}$.
	}
	\label{fig:r_cl_c}
\end{figure*}

We explore the dependence of 
%Figure~\ref{fig:r_cl_c} shows 
\SB, \pol, and \spec on three values of clump radius $r_{cl}=$ 10\,pc (solid lines), 100\,pc (dashed), and 1\,kpc (dotted) in Figure~\ref{fig:r_cl_c}.
%to investigate the dependence on $r_{cl}$.
Because the simulated results for the three very different $r_{cl}$ values are identical, we conclude that
\lya radiative transfer in our model of a clumpy medium  (Model C) is unaffected by clump size.

This insensitivity to $r_{cl}$ would hold only for a small volume filling factor $f_V$ $\sim {f_c \, r_{cl}}/R_H$, the ratio between the total volume of all the clumps and the halo volume.
In our simulation, we adopt $r_{cl} \ll 0.1 R_H$ $\sim$ 10\,kpc, because ground-based observations have not resolved any clumps in \lya nebulae. In this case, the filling factor is much smaller than 1; $f_V \sim 0.1$ when $r_{cl} = 1$\,kpc and $f_c = 10$. Because there are so many clumps per sightline, each sightline has the same number of clumps to within the counting errors. Thus, there is also no dependence on the simulated sightline.

If $r_{cl}$ becomes large enough such that $f_V \sim 1$, clumps will start to overlap spatially and will no longer be uniformly distributed. In this case, the number of clumps along the sightline depends on the direction on the sky, and we expect that $r_{cl}$ will affect the \lya radiative transfer. We adopt the fixed value $r_{cl} = 100$\,pc throughout this paper.


\begin{thebibliography}{}
\bibliographystyle{aasjournal}
\bibitem[Ahn et al.(2002)]{ahn02} Ahn, S.-H., Lee, H.-W., Lee, H. M., 2002, ApJ, 567, 922 
\bibitem[Ahn \& Lee (2002)]{ahn02a} Ahn, S.-H., Lee, H.-W., 2002, JKAS, 35, 175 
\bibitem[Ahn \& Lee (2003)]{ahn03} Ahn, S.-H., Lee, H.-W., Lee, H. M., 2003, MNRAS, 340, 863 
\bibitem[Ao et al.(2020)]{ao20} Ao, Y., Zheng, Z., Henkel, C., Nie, S., Beelen, A., Cen, R., Dijkstra, M., 2020, Nature Astronomy, 47
\bibitem[Arrigoni Battaia et al.(2019)]{fabrizio19} Arrigoni Battaia, F., Hennawi, J. F., Prochaska, J. X., O\~norbe, J., Farina, E. P., Cantalupo, S., Lusso, E., 2019, MNRAS, 482, 3162
\bibitem[Arrigoni Battaia et al.(2015)]{fabrizio15} Arrigoni Battaia, F., Hennawi, J. F., Prochaska, J. X., Cantalupo, S., 2015, ApJ, 809, 163
\bibitem[B{\u{a}}descu et al.(2017)]{Badescu2017} B{\u{a}}descu, T., Yang, Y., Bertoldi, F., et al.\ 2017, \apj, 845, 172. doi:10.3847/1538-4357/aa8220
\bibitem[Borisova et al.(2016)]{borisova16} Borisova, E., Cantalupo, S., Lilly, S. J., et al., 2016, ApJ, 831, 39
\bibitem[Cabot et al.(2016)]{cabot16} Cabot, S. H. C., Cen, R., Zheng, Z., 2016, MNRAS, 462, 1076
\bibitem[Cai et al.(2017)]{cai17} Cai, Z., Fan X., Yang, Y., Bian, F.,, et al., 2017, ApJ, 837, 71
\bibitem[Chandrasekhar (1960)]{chandrasekhar60} Chandrasekhar, 1960, Radiative Transfer (Reading : New York Dover)
\bibitem[Chang et al.(2017)]{chang17} Chang, S.-J., Lee, H.-W., Yang, Y., 2017, MNRAS, 464, 5018
\bibitem[Chang \& Lee (2020)]{chang20} Chang, S.-J., Lee, H.-W., 2020, JKAS, 53, 169
\bibitem[Daddi et al.(2020)]{daddi20} Daddi, E., Valentino, F., Rich, R. M., Neill, J. D.,  2020, arXiv:2006.11089
\bibitem[Dey et al.(2005)]{dey05} Dey, A., Bian, C., Soifer, B.~T., et al.\ 2005, \apj, 629, 654
\bibitem[Dijkstra \& Loeb (2008)]{dijkstra08} Dijkstra, M., Loeb, A., 2008, MNRAS, 386, 492
\bibitem[Dijkstra \& Kramer (2012)]{dijkstra12} Dijkstra, M., Kramer, R., 2012, MNRAS, 424, 1672
\bibitem[Duval et al.(2014)]{duval14} Duval, F., Schaerer, D., \"Ostlin, G., Laursen, P., 2014, A\&A, 562, 52
\bibitem[Eide et al.(2018)]{eide18} Eide, M. B., Gronke, M., Dijkstra, M., Hayes M., 2018, ApJ, 856, 156
\bibitem[Gawiser et al.(2007)]{gawiser07} Gawiser, E., Francke, H., Lai, K., Schawinski, K., et al., 2007, ApJ, 671, 278
\bibitem[Geach et al.(2016)]{geach16} Geach, J. E., Narayanan, D., Matsuda, Y., et al., 2016, ApJ, 832, 37
\bibitem[Gronke et al.(2016)]{gronke16} Gronke, M., Dijkstra, M., McCourt, M., Oh, S. P., 2016, ApJL, 833, 26
\bibitem[Gronke et al.(2017)]{gronke17} Gronke, M., Dijkstra, M., McCourt, M., Oh, S. P., 2017, A\&A, 607, 71
\bibitem[Hansen \& Oh (2006)]{hansen06} Hansen, M., Oh, S. P., 2006, MNRAS, 367, 979
\bibitem[Hayes et al.(2011)]{hayes11} Hayes, M., Scarlata, C., Siana, B., 2011, Nature, 476, 304
\bibitem[Heckman et al. (1991)]{heckman91} Heckman, T. M., Lehnert, M. D., van Breugel, W., Miley, G. K., 1991, ApJ, 370, 78
\bibitem[Hennawi \& Prochaska (2013)]{hennawi13} Hennawi, J. F., Prochaska, J. X., 2013, ApJ, 766, 58
\bibitem[Hennawi et al.(2015)]{hennawi15} Hennawi, J. F., Prochaska, J. X., Cantalupo, S., Arrigoni-Battaia, F., 2015, Science, 348, 779
\bibitem[Keel et al. (1999)]{keel99} Keel, W. C., Cohen, S. H., Windhorst, R. A., Waddington, I., 1999, ApJ, 118, 2547
\bibitem[Kim et al.(2020)]{kim20} Kim, E., Yang, Y., Zabludoff, A., Smith, P., Jannuzi, B., Lee, M. G., Hwang, N., Part, B.-G., 2020, ApJ, 894, 33
\bibitem[Lee et al. (1994)]{lee94} Lee, H.-W., Blandford, R. D., Western, L., 1994, MNRAS, 267,  303
\bibitem[Li et al.(2021)]{li21} Li, Z., Steidel, C. C., Gronke, M., Chen, Y., 2021, 2021, MNRAS, 502, 2389
\bibitem[Mas-Ribas \& Chang (2020)]{mas-ribas20} Mas-Ribas, L., Chang, T.-C., 2020, Phys. Rev. D, 101, 083032
\bibitem[Matduda et al.(2004)]{matsuda04} Matsuda, Y., Yamada, T., Hayashino, T., et al. 2004, AJ, 128, 569
\bibitem[Neufeld (1990)]{neufeld90} Neufeld, D. A., 1990, ApJ, 350, 216
\bibitem[Neufeld (1991)]{neufeld91} Neufeld, D. A., 1991, ApJL, 370, L85
\bibitem[Nilsson et al. (2006)]{nilsson06} Nilsson, K. K., Fynbo, J. P. U., Møller, P., Sommer-Larsen, J., Ledoux, C. , 2006, A\&A, 452, L23
\bibitem[Ouchi et al.(2008)]{ouchi08} Ouchi, M., Shimasaku, K., Akiyama, M., Simpson, C., et al., 2008, ApJS, 176, 301
\bibitem[Ouchi et al.(2018)]{ouchi18} Ouchi, M., Harikane, Y., Shibuya, T., Shimasaku, K., et al., 2018, PASJ, 70S, 13 
\bibitem[Prescott et al.(2009)]{prescott09} Prescott, M. K. M., Dey, A., Jannuzi B. T., 2009, ApJ, 702, 554
\bibitem[Rybicki \& Loeb (1999)]{rybicki99}Rybicki, G. B., Loeb, A., 1999, ApJL, 520, L79 
\bibitem[Seon \& Kim (2020)]{seon20} Seon, K.-I., Kim, C.-G., 2020, ApJS, 250, 35
\bibitem[Seon et al. (2022)]{seon22} Seon, K.-I., Song, H., Chang, S.-J., 2022, ApJS, 259, 3
\bibitem[Shukla et al. (2022)]{shukla22}  Shukla, G., Srianand, R., Gupta, N., Petitjean, P., Baker, A. J., Krogager, J.-K., Noterdaeme, P., 2022, MNRAS, 510, 786 
\bibitem[Sobral et al.(2017)]{sobral17} Sobral, D., Matthee, J., Best, P., Stroe, A., et al., 2017, MNRAS, 466, 1242
\bibitem[Stenflo (1980)]{stenflo80} Stenflo, J. O., 2006, A\&A, 84,  68
\bibitem[Steidel et al.(2000)]{steidel00} Steidel, C. C., Adelberger, K. L., Shapley, A. E., et al. 2000, ApJ, 532, 170
\bibitem[Trebitsch et al.(2016)]{trebitsch16} Trebitsch, M., Verhamme, A., Blaizot, J., Rosdahl, J., 2016, A\&A, 593, 122
\bibitem[Travascio et al.(2020)]{travascio20} Travascio, A., Zappacosta, L, Cantalupo, S., et al., 2020, A\&A, 635, 157
\bibitem[Yang et al.(2009)]{yang09} Yang, Y., Zabludoff, A., Tremonti, C., Eisenstein, D., Dav\'e, R. 2009, ApJ, 693, 1579
\bibitem[Yang et al.(2010)]{yang10} Yang, Y., Zabludoff, A., Eisenstein, D., Dav\'e, R. 2010, ApJ, 719, 1654
\bibitem[Yang et al.(2011)]{yang11} Yang, Y., Zabludoff, A., Jahnke, K., Dav\'e, R., Schectman, S. A., kelson, D. D., 2011, ApJ, 735, 87
\bibitem[Yang et al.(2014a)]{yang14a} Yang, Y., Walter, F., Decarli, R., Bertoldi, F., Weiss, A., Dey, A., 2014, ApJ, 787, 171
\bibitem[Yang et al.(2014b)]{yang14b} Yang, Y., Zabludoff, A., Jahnke, K., Dave, R., 2014, ApJ, 797, 114
\bibitem[You et al.(2017)]{you17} You, C., Zabludoff, A., Smith, P., et al., 2017, ApJ, 834, 182
\bibitem[Umehata et al.(2021)]{Umehata21} Umehata, H., Smail, I., Steidel, C.~C., et al.\ 2021, \apj, 918, 69
\bibitem[Verhamme et al.(2006)]{verhamme06} Verhamme, A., Schaerer, D., Maselli, A., 2006, A\&A, 460, 397
\bibitem[Villar-Mart\'in et al.(2007)]{villar07}  Villar-Mart\'in, M., S\'aáchez, S. F., Humphrey, A., Dijkstra, M., di Serego Alighieri, S., De Breuck, C., Gonz\'alez Delgado, R., 2007, MNRAS, 378, 416
\bibitem[Zheng \& Miralda-Escud\'e (2002)]{zheng02} Zheng, Z., Miralda-Escud\'e, J., 2002, ApJ, 578, 33

\end{thebibliography}
\end{document}